\documentclass[11pt,showpacs,preprintnumbers,superscriptaddress,amsmath,amssymb,nofootinbib]{revtex4}

\usepackage{graphicx}
\usepackage{dcolumn}
\usepackage{bm}
\usepackage{amssymb}
\usepackage{amsmath}
\usepackage{epsfig}    
\usepackage{color}    

\def\beq{\begin{equation}}
\def\eeq{\end{equation}}
\def\bea{\begin{array}}
\def\eea{\end{array}}
\def\be{\begin{equation}}
\def\ee{\end{equation}}
\def\ba{\begin{eqnarray}}
\def\ea{\end{eqnarray}}

\def\to{\rightarrow}

\def\[{\left[}
\def\]{\right]}
\def\({\left(}
\def\){\right)}

\def\sm0{{\widetilde{m}_0}}

\def\U1em{{U(1)_{\rm em}}}
\def\to{\rightarrow}

\def\sq2{\sqrt{2}}

\def\ee{e^+e^-}

\def\End{\end{document}}

\def\fsl#1{\setbox0=\hbox{$#1$}                 
   \dimen0=\wd0                                 
   \setbox1=\hbox{/} \dimen1=\wd1               
   \ifdim\dimen0>\dimen1                        
      \rlap{\hbox to \dimen0{\hfil/\hfil}}      
      #1                                        
   \else                                        
      \rlap{\hbox to \dimen1{\hfil$#1$\hfil}}   
      /                                         
   \fi}

\begin{document} 

\title{Radiative corrections to the Higgs boson couplings in the triplet model}
\preprint{KANAZAWA-12-09, UT-HET 076}
\author{Mayumi Aoki}
\email{mayumi@hep.s.kanazawa-u.ac.jp}
\affiliation{Institute for Theoretical Physics, Kanazawa University, \\Kanazawa 920-1192, Japan}
\affiliation{Max-Planck-Institut f\"ur Kernphysik, \\Saupfercheckweg 1, 69117 Heidelberg, Germany}
\author{Shinya Kanemura}
\email{kanemu@sci.u-toyama.ac.jp}
\affiliation{Department of Physics, University of Toyama, \\3190 Gofuku, Toyama 930-8555, Japan}
\author{Mariko Kikuchi}
\email{kikuchi@jodo.sci.u-toyama.ac.jp}
\affiliation{Department of Physics, University of Toyama, \\3190 Gofuku, Toyama 930-8555, Japan}
\author{Kei Yagyu}
\email{keiyagyu@ncu.edu.tw}
\affiliation{Department of Physics, National Central University, \\Chungli 32001, Taiwan}
\begin{abstract}

We calculate a full set of one-loop corrections to 
the Higgs boson coupling constants as well as 
the electroweak parameters.  
%
We compute the decay rate of the standard model (SM)-like Higgs boson ($h$) into diphoton. 
Renormalized Higgs couplings with the weak gauge bosons $hVV$ ($V=W$ and $Z$) and 
the trilinear coupling $hhh$ are also calculated  at the one-loop level in the on-shell scheme. 
Magnitudes of the deviations in these quantities are evaluated in the parameter regions 
where the unitarity and vacuum 
stability bounds are satisfied and the predicted W boson mass at the one-loop level 
is consistent with the data.
We find that there are strong correlations among deviations in the Higgs boson couplings $h\gamma\gamma$, $hVV$ and $hhh$. 
For example, if the event number of the $pp\to h\to\gamma\gamma$ channel deviates by $+30\%$ ($-40\%$)
from the SM prediction, deviations in the one-loop corrected $hVV$ and $hhh$ vertices
are predicted about $-0.1\%$ ($-2\%$) and $-10\%$ $(+150\%)$, respectively. 
The model can be discriminated from the other models
by measuring these coupling constants accurately at future collider experiments.

\end{abstract}

\date{\today}
\maketitle
\newpage
\section{introduction}


The new boson with the mass of around 126 GeV has been discovered 
in the $h\to \gamma\gamma$, $h\to ZZ^*\to 4\ell$ and $h\to WW^*\to \ell\nu\ell\nu$ 
channels with 5.9$\sigma$ at the ATLAS~\cite{Higgs_ATLAS}
and with 5.0$\sigma$ at the CMS~\cite{Higgs_CMS}. 
%
%
The observed mass of 126 GeV is consistent with the precision data at the LEP/SLC experiments~\cite{LEP} at the quantum level 
assuming that it is the Higgs boson in the Standard Model (SM). 
At the LHC, the Higgs boson production and decay are consistent with the SM predictions at the 2$\sigma$ level by both ATLAS and 
CMS experiments. 
The particle is most likely the Higgs boson.  


However, it is not necessary that the particle is the Higgs boson of the SM. 
The SM-like Higgs boson can also be predicted in various extended Higgs sectors; e.g., 
the Higgs sector with additional SU(2) singlets, doublets and/or triplets.   
Such a non-minimal Higgs sector is introduced in various 
new physics models beyond the SM which are considered to solve the problems such as tiny neutrino masses, dark matter and/or 
the baryon asymmetry of the Universe. 
The deviations in coupling constants of the SM-like Higgs boson from the SM predictions may be detected 
at the LHC or at the future precision 
collider experiments such as the LHC at the integrated luminosity of 3000 fb$^{-1}$ and the International Linear Collider (ILC). 
Therefore, we can discriminate models of new physics by comparing the accurate predictions on the coupling constants associated with 
the SM-like Higgs boson with the future precision measurements, even if additional new particles will be unfound.


In this paper, 
we focus on the Higgs boson properties in the minimal Higgs triplet model (HTM), where  
tiny neutrino masses can be explained via the so-called type-II seesaw mechanism~\cite{typeII}. 
The Higgs sector of this model is composed from an SU(2) doublet Higgs field with the hypercharge $Y=1/2$ 
and a triplet field with $Y=1$. 
One of the striking features of this model is the prediction that 
the electroweak rho parameter $\rho$ deviates from unity at the tree level due to 
the non-zero vacuum expectation value (VEV) of the triplet field $v_\Delta$. 
As the experimental value of $\rho$ is nearly one, 
$v_\Delta$ should be much suppressed as compared to the VEV of the doublet. 
Approximately, $v_\Delta$ is constrained to be less than 
8 GeV from the rho parameter data~\cite{PDG}. 
Under the requirement of $v_\Delta\ll v_\phi$, where $v_\phi$ is the VEV of the doublet, 
the SM-like Higgs boson $h$ can be separated from the triplet-like Higgs boson; namely, 
a pair of the doubly-charged (singly-charged) Higgs boson $H^{\pm\pm}$ ($H^\pm$), the CP-odd Higgs boson $A$ 
and the CP-even Higgs boson $H$. 


In order to identify the HTM at collider experiments, 
detection of the triplet-like Higgs bosons, especially of the doubly-charged Higgs boson,  
is important.  
The decay property of the triplet-like Higgs bosons strongly depends on the mass spectrum among them and 
$v_\Delta$. 
When the triplet-like Higgs bosons are degenerate in mass or 
$H^{\pm\pm}$ is the lightest of all of them, the main decay mode of $H^{\pm\pm}$ is 
the same-sign dilepton (diboson) in the case where $v_\Delta$ is less (larger) than about 1 MeV. 
The scenario based on the same-sign dilepton decay of $H^{\pm\pm}$ has been studied in Refs.~\cite{lplp1,lplp2,Han,HTM_pheno_w1}. 
This scenario has already been strongly constrained by the LHC data. 
The current lower mass limit on $H^{\pm\pm}$ is about 400 GeV~\cite{lplp_LHC}. 
The scenario for the same sign diboson decay of $H^{\pm\pm}$ has been 
discussed in Refs.~\cite{Han,Chiang_Nomura_Tsumura}, and the discovery potential of $H^{\pm\pm}$ at the LHC 
has also been investigated in Ref.~\cite{Chiang_Nomura_Tsumura}.  

Phenomenology of the HTM can be drastically changed
when there is mass splitting among the triplet-like Higgs bosons and 
$H^{\pm\pm}$ is the heaviest of all of them~\cite{HTM_pheno_w1,HTM_pheno_w2,AKY}.  
In such a case, the cascade decay of $H^{\pm\pm}$ can be dominant instead of the same-sign dilepton or diboson decay; 
namely, $H^{\pm\pm}$ decays into $H^\pm W^\pm$. 
At the same time, $H^\pm$ can decay into a neutral Higgs boson ($H$ or $A$) associated with a $W^\pm$. 
$H$ and $A$ can mainly decay into $b\bar{b}$ for $v_\Delta\gtrsim$ 1 MeV and into neutrinos 
for $v_\Delta\lesssim$ 1 MeV. 
In the former case, the triplet-like Higgs bosons may be reconstructed by using the invariant mass as well as 
the transverse mass distributions at the LHC~\cite{AKY}. 


If the triplet-like Higgs bosons are light enough to be produced at the LHC, 
the direct detection can be an important probe of the HTM as already discussed. 
Even if they are too heavy to be directly detected, they can be indirectly tested by measuring 
the deviations from the SM in the Higgs boson couplings associated with $h$ such as 
the coupling constants with the weak gauge bosons $hVV$, the Yukawa couplings $hf\bar{f}$ and 
the triple Higgs boson coupling $hhh$. 
In Ref.~\cite{Higgs_couplings_LHC,Peskin}, 
accuracy of the Higgs boson coupling measurements has been 
discussed at the LHC. 
%
Assuming the 14 TeV collision energy and with 
the integrated luminosity to be 300 fb$^{-1}$, 
the deviations in $hZZ$, $hWW$ and $h\gamma\gamma$  
can be measured with about $10\%$ accuracy,  
and that of the Yukawa couplings can be measured with about $20\%$ for $ht\bar{t}$ and $hb\bar{b}$ 
and about $10\%$ for $h\tau^+\tau^-$. 
At the ILC with the 1 TeV collision energy and with 
the integrated luminosity to be 500 fb$^{-1}$, 
accuracy of the measured deviations in the Higgs couplings is expected to be 
less than about $1\%$ for $hWW$ and $hZZ$, about $5\%$ for $h\gamma\gamma$, $2$-$3\%$ for $hb\bar{b}$ and $h\tau^+\tau^-$ and 
$5$-$10\%$ for $ht\bar{t}$~\cite{Peskin}. 
The triple Higgs boson coupling $hhh$ is expected to be measured with about $20\%$ accuracy~\cite{Fujii}, 
assuming the collision energy and the integrated luminosity being 1 TeV and 2 ab$^{-1}$, respectively at the ILC. 

In this paper, we calculate the deviations in these Higgs boson couplings in the HTM at the one-loop level. 
In particular, we focus on the deviations in the $h\to \gamma\gamma$ decay rate, $hZZ$ and $hWW$ couplings and also 
the triple Higgs boson coupling $hhh$. 
The couplings of the triplet field with the quarks are induced by a small mixing angle which is proportional to $v_\Delta$, 
so that we do not discuss about the one-loop corrections to the Yukawa couplings. 
In order to calculate finite predictions of various observables, we need the renormalization of the model. 
The renormalization of the electroweak sector in the HTM 
is different from that in models with $\rho=1$ at the tree level.  
Four parameters are necessary to describe the electroweak parameters instead of three parameters such 
as $\alpha_{\text{em}}$, $G_F$ and $m_Z$. 
This means that one extra renormalization condition is required to determine the counter-term which corresponds
to the one extra input parameter. 
The renormalization scheme in models with $\rho\neq 1$ has been discussed 
by Blank and Hollik in Ref.~\cite{Blank_Hollik}, where 
the effective weak mixing angle is chosen as the extra input parameter in the model with the $Y=0$ triplet Higgs field. 
In Ref.~\cite{Kanemura-Yagyu}, this renormalization scheme has been applied to the HTM. 
The two different renormalization schemes in the model with the $Y=0$ triplet Higgs field have been discussed 
in Refs.~\cite{Chen-Dawson-Jackson}, 
where the triplet VEV is chosen to be the fourth input parameter as the other renormalization scheme from that proposed in Ref.~\cite{Blank_Hollik}. 

We impose in this paper the new renormalization scheme for the electroweak parameters, 
in which we require the no-mixing condition between the physical CP-odd Higgs boson $A$ and 
the Z boson as an additional renormalization condition. 
We then compare the previous renormalization scheme, which is used four inputs from the electroweak precision data, 
and the new scheme. 
%
In the former renormalization scheme, 
the decoupling limit cannot be taken even when the triplet-like Higgs bosons are taken to be quite heavy. 
On the other hand, in the latter scheme, we can take the decoupling limit when a coupling constant among 
the doublet-doublet-triplet term is taken to be a fixed value. 
We then discuss the renormalization of the Higgs potential in the HTM. 
The on-shell renormalization scheme for the parameters in the potential has been constructed in Ref.~\cite{AKKY} 
in the limit of $v_\Delta/v_\phi\to 0$. 
We prepare the counter-terms which are necessary to calculate the renormalized $hZZ$, $hWW$ and $hhh$ vertices. 

Finally, we evaluate possible deviations from the SM in these Higgs couplings under the 
allowed parameter regions by the electroweak precision data and the unitarity and vacuum stability bounds. 
We also examine the event number of the $pp\to h\to \gamma\gamma$ channel. 
We find that there are strong correlations among the deviations in $h\gamma\gamma$, $hVV$ and $hhh$. 
For example, if the event number of the $pp\to h\to\gamma\gamma$ channel is predicted as $+30\%$ ($-40\%$)
compared to the SM prediction, deviations in the one-loop corrected $hVV$ and $hhh$ vertices
are predicted about $-0.1\%$ ($-2\%$) and $-10\%$ $(+150\%)$, respectively. 
Such large deviations in $h\gamma\gamma$ and $hhh$ are consequence of the non-decoupling loop effect of extra Higgs bosons, 
similarly to the case of the two Higgs doublet model. 
In the two Higgs doublet model, the Higgs boson couplings $hZZ$~\cite{KOSY} and $hhh$~\cite{KOSY,Kiyoura} 
and the electroweak precision observables~\cite{David} have been calculated at the one-loop level. 
By measuring these coupling constants accurately at future colliders such as 
the LHC with 3000 fb$^{-1}$ and at the ILC, the HTM can be discriminated from the other models.


This paper is organized as follows. 
In Sec.~II, we define the Lagrangian in the HTM. We calculate the mass spectrum among the triplet-like Higgs bosons, 
and we briefly review the tree level relations among parameters. 
The unitarity and vacuum stability bounds are also discussed in the end of this section. 
In Sec.~III, the renormalization of the HTM is discussed based on the on-shell scheme, 
where we study the renormalization of the electroweak precision parameters and 
that of the parameters in the Higgs potential.  
In Sec.~IV, we first calculate the decay rate of $h\to \gamma\gamma$ process and 
the renormalized Higgs boson couplings $hZZ$, $hWW$ and $hhh$ at the one-loop level.  
Numerical results for the deviations in these coupling constants from the SM are shown in the allowed parameter regions 
by the theoretical bounds and the electroweak precision data. 
Conclusions are given in Sec.~VI. 

%
%
%
%
%
%
%

\section{Tree level formulae}
The scalar sector of the HTM is composed of the isospin doublet field $\Phi$ with 
hypercharge $Y=1/2$ and the triplet field $\Delta$ with $Y=1$. 
The relevant terms in the Lagrangian are given by 
\begin{align}
\mathcal{L}_{\text{HTM}}=\mathcal{L}_{\text{kin}}+\mathcal{L}_{Y}-V(\Phi,\Delta), 
\end{align}
where $\mathcal{L}_{\text{kin}}$, $\mathcal{L}_{Y}$ and $V(\Phi,\Delta)$ are 
the kinetic term, the Yukawa interaction and the Higgs potential, respectively. 

The kinetic term of the Higgs fields is given by 
\begin{align}
\mathcal{L}_{\text{kin}}&=(D_\mu \Phi)^\dagger (D^\mu \Phi)+\text{Tr}[(D_\mu \Delta)^\dagger (D^\mu \Delta)], \label{kinetic}
\end{align}
where the covariant derivatives are defined as
\begin{align}
D_\mu \Phi=\left(\partial_\mu+i\frac{g}{2}\tau^aW_\mu^a+i\frac{g'}{2}B_\mu\right)\Phi, \quad
D_\mu \Delta=\partial_\mu \Delta+i\frac{g}{2}[\tau^aW_\mu^a,\Delta]+ig'B_\mu\Delta. 
\end{align}
The Higgs fields can be parameterized by
\begin{align}
\Phi=\left[
\begin{array}{c}
\phi^+\\
\frac{1}{\sqrt{2}}(\phi+v_\phi+i\chi)
\end{array}\right],\quad \Delta =
\left[
\begin{array}{cc}
\frac{\Delta^+}{\sqrt{2}} & \Delta^{++}\\
\Delta^0 & -\frac{\Delta^+}{\sqrt{2}} 
\end{array}\right]\text{ with } \Delta^0=\frac{1}{\sqrt{2}}(\delta+v_\Delta+i\eta), 
\end{align}
where $v_\phi$ and $v_\Delta$ 
are the VEVs of the doublet Higgs field and the triplet Higgs field, respectively which satisfy 
$v^2\equiv v_\phi^2+2v_\Delta^2\simeq$ (246 GeV)$^2$. 
The masses of the W boson and the Z boson are obtained at the tree level as
\begin{align}
m_W^2 = \frac{g^2}{4}(v_\phi^2+2v_\Delta^2),\quad m_Z^2 =\frac{g^2}{4\cos^2\theta_W}(v_\phi^2+4v_\Delta^2). \label{mV}
\end{align}
The electroweak rho parameter can deviate from unity at the tree level; 
\begin{align}
\rho \equiv \frac{m_W^2}{m_Z^2\cos^2\theta_W}=\frac{1+\frac{2v_\Delta^2}{v_\phi^2}}{1+\frac{4v_\Delta^2}{v_\phi^2}}. \label{rho_triplet}
\end{align}
The experimental value of the rho parameter is quite close to unity; i.e., $\rho^{\text{exp}}=1.0008^{+0.0017}_{-0.0007}$~\cite{PDG}, 
so that 
$v_\Delta$ has to be less than about 8 GeV from the tree level formula given in Eq.~(\ref{rho_triplet}). 

The Yukawa interaction for neutrinos~\cite{typeII} is given by 
\begin{align}
\mathcal{L}_Y&=h_{ij}\overline{L_L^{ic}}i\tau_2\Delta L_L^j+\text{h.c.}, \label{nu_yukawa}
\end{align}
where $h_{ij}$ is the $3\times 3$ complex symmetric Yukawa matrix. 
Notice that the triplet field $\Delta$ carries the lepton number of $-2$.  
The mass matrix for the left-handed neutrinos is obtained as 
\begin{align}
(\mathcal{M}_\nu)_{ij}=\sqrt{2}h_{ij}v_\Delta.  \label{eq:mn}
\end{align}
Current neutrino oscillation data are given by the following central values~\cite{ndata}
\begin{align}
&\sin^2\theta_{12} = 0.31,\quad  \sin^2\theta_{23} = 0.39,\quad \sin^2\theta_{13} = 0.024, \\
&\Delta m_{21}^2 = 7.5\times 10^{-5} \text{ eV}^2,\quad \Delta m_{32}^2= 2.4\times 10^{-3} \text{ eV}^2,
\end{align}
can be explained in the HTM~\cite{lplp2,Han,HTM_pheno_w1}. 
It is seen from Eq.~(\ref{eq:mn}) that the neutrino mixing pattern is purely determined by the $h_{ij}$ matrix. 
Since the decay rate of $H^{\pm\pm}$ into the same-sign dilepton is proportional to $|h_{ij}|^2$, 
we can test the type-II seesaw mechanism by looking at the same-sign dilepton decay mode of $H^{\pm\pm}$~\cite{lplp2,Han,HTM_pheno_w1}.

The most general form of the Higgs potential under the gauge symmetry is given by 
\begin{align}
V(\Phi,\Delta)&=m^2\Phi^\dagger\Phi+M^2\text{Tr}(\Delta^\dagger\Delta)+\left[\mu \Phi^Ti\tau_2\Delta^\dagger \Phi+\text{h.c.}\right]\notag\\
&+\lambda_1(\Phi^\dagger\Phi)^2+\lambda_2\left[\text{Tr}(\Delta^\dagger\Delta)\right]^2+\lambda_3\text{Tr}[(\Delta^\dagger\Delta)^2]
+\lambda_4(\Phi^\dagger\Phi)\text{Tr}(\Delta^\dagger\Delta)+\lambda_5\Phi^\dagger\Delta\Delta^\dagger\Phi, \label{pot_htm}
\end{align}
where $m$ and $M$ are the dimension full real parameters, $\mu$ is the dimension full complex parameter 
which violates the lepton number, and 
$\lambda_1$-$\lambda_5$ are the coupling constants which are real. 
We here take $\mu$ to be real. 

The potential respects additional global symmetries in some limits. 
First, when the $\mu$ term is absent,  
there is the global $U(1)$ symmetry in the potential, which conserves the lepton number. 
As long as we assume that the lepton number is not spontaneously broken, 
the triplet field does not carry the VEV; i.e., $v_\Delta=0$. 
%
Next, when both the $\mu$ term and the $\lambda_5$ term are zero, 
an additional global $SU(2)$ symmetry appears. 
Under this $SU(2)$ symmetry, $\Phi$ and $\Delta$ can be transformed with the different $SU(2)$ phases.  
In this case, all the physical triplet-like Higgs bosons are degenerate in mass. 

The tadpoles for the $\phi$ and $\delta$ fields are obtained as 
\begin{align}
T_\Phi&=-v_\phi\left[m^2+v_\phi^2\lambda_1+\frac{v_\Delta^2}{2}(\lambda_4+\lambda_5)-\sqrt{2}\mu v_\Delta\right],\\
T_\Delta&=-v_\Delta\left[M^2+v_\Delta^2(\lambda_2+\lambda_3)+\frac{v_\phi^2}{2}(\lambda_4+\lambda_5)-M_\Delta^2\right], \label{vc}
\text{ with } M_\Delta^2\equiv \frac{v_\phi^2\mu}{\sqrt{2}v_\Delta}. 
\end{align}
Because the tadpoles must be vanished at the tree level ($T_\Phi=T_\Delta=0$), we can eliminate $m^2$ and $M^2$ in the potential. 
The mass matrices for the scalar bosons can be diagonalized by rotating the 
scalar fields as 
\begin{align}
\left(
\begin{array}{c}
\phi^\pm\\
\Delta^\pm
\end{array}\right)&=
\left(
\begin{array}{cc}
\cos \beta & -\sin\beta \\
\sin\beta   & \cos\beta
\end{array}
\right)
\left(
\begin{array}{c}
G^\pm\\
H^\pm
\end{array}\right),\quad 
\left(
\begin{array}{c}
\chi\\
\eta
\end{array}\right)=
\left(
\begin{array}{cc}
\cos \beta' & -\sin\beta' \\
\sin\beta'   & \cos\beta'
\end{array}
\right)
\left(
\begin{array}{c}
G^0\\
A
\end{array}\right),\notag\\
\left(
\begin{array}{c}
\phi\\
\delta
\end{array}\right)&=
\left(
\begin{array}{cc}
\cos \alpha & -\sin\alpha \\
\sin\alpha   & \cos\alpha
\end{array}
\right)
\left(
\begin{array}{c}
h\\
H
\end{array}\right), \label{mixing1}
\end{align}
with the mixing angles
\begin{align}
\tan\beta=\frac{\sqrt{2}v_\Delta}{v_\phi},\quad \tan\beta' = \frac{2v_\Delta}{v_\phi}, \quad
\tan2\alpha &=\frac{v_\Delta}{v_\phi}\frac{2v_\phi^2(\lambda_4+\lambda_5)-4M_\Delta^2}{2v_\phi^2\lambda_1-M_\Delta^2-2v_\Delta^2(\lambda_2+\lambda_3)}. \label{tan2a}
\end{align}
We note that the mixing angle of the charged scalar states $(\beta)$ and that of the CP-odd scalar states ($\beta'$) 
are different in the triplet model. 
In the two Higgs doublet model, corresponding two mixing angles are the same at the tree level. 
This is because the kinetic term of the two doublet fields 
can be rewritten in terms of so-called the Georgi basis, where only one of the doublets has a non-zero VEV in which 
the NG bosons are included. 
Original basis and the Georgi basis are related to a single angle. 
In the HTM, because $\Phi$ and $\Delta$ are the different representation of SU(2), 
the kinetic term given in Eq.~(\ref{kinetic}) cannot be rewritten in terms of the Georgi basis. 
Thus, the diagonalization of the mass matrices has to be done by each component scalar field, and 
mixing angles for the charged scalar states and the CP-odd scalar states are different in general.

In addition to the three NG bosons $G^\pm$ and $G^0$ which are absorbed by the longitudinal components 
of the $W$ boson and the $Z$ boson, 
there are seven physical mass eigenstates $H^{\pm\pm}$, $H^\pm$, $A$, $H$ and $h$. 
The masses of these physical states are expressed as 
\begin{align}
m_{H^{++}}^2&=M_\Delta^2-v_\Delta^2\lambda_3-\frac{\lambda_5}{2}v_\phi^2,\label{mhpp}\\
m_{H^+}^2&= \left(M_\Delta^2-\frac{\lambda_5}{4}v_\phi^2\right)\left(1+\frac{2v_\Delta^2}{v_\phi^2}\right),\label{mhp}\\
m_A^2 &=M_\Delta^2\left(1+\frac{4v_\Delta^2}{v_\phi^2}\right), \label{mA}\\
m_H^2&=\mathcal{M}_{11}^2\sin^2\alpha+\mathcal{M}_{22}^2\cos^2\alpha-\mathcal{M}_{12}^2\sin2\alpha,\label{mH}\\
m_h^2&=\mathcal{M}_{11}^2\cos^2\alpha+\mathcal{M}_{22}^2\sin^2\alpha+\mathcal{M}_{12}^2\sin2\alpha,
\end{align}
where $\mathcal{M}_{11}^2$, $\mathcal{M}_{22}^2$ and $\mathcal{M}_{12}^2$ are 
the elements of the mass matrix $\mathcal{M}_{ij}^2$ for the CP-even scalar states in the $(\phi,\delta)$ basis which are 
given by
\begin{align}
\mathcal{M}_{11}^2&=2v_\phi^2\lambda_1,\\
\mathcal{M}_{22}^2&=M_\Delta^2+2v_\Delta^2(\lambda_2+\lambda_3),\\
\mathcal{M}_{12}^2&=-\frac{2v_\Delta}{v_\phi}M_\Delta^2+v_\phi v_\Delta(\lambda_4+\lambda_5).
\end{align}
The six parameters $\mu$ and $\lambda_1$-$\lambda_5$ in the Higgs potential in Eq.~(\ref{pot_htm}) 
can be written in terms of the physical scalar masses, the mixing angle $\alpha$ and VEVs $v_\phi$ and $v_\Delta$ as
\begin{align}
\mu&=\frac{\sqrt{2}v_\Delta}{v_\phi^2}M_\Delta^2 =\frac{\sqrt{2}v_\Delta}{v_\phi^2+4v_\Delta^2}m_A^2, \label{mu}\\
\lambda_1 & = \frac{1}{2v_\phi^2}(m_h^2\cos^2\alpha+m_H^2\sin^2\alpha),\\
\lambda_2 & = \frac{1}{2v_\Delta^2}\left[2m_{H^{++}}^2+v_\phi^2\left(\frac{m_A^2}{v_\phi^2+4v_\Delta^2}-\frac{4m_{H^+}^2}{v_\phi^2+2v_\Delta^2}\right)
+m_H^2\cos^2\alpha+m_h^2\sin^2\alpha\right],\\
\lambda_3 & = \frac{v_\phi^2}{v_\Delta^2}\left(\frac{2m_{H^+}^2}{v_\phi^2+2v_\Delta^2}-\frac{m_{H^{++}}^2}{v_\phi^2}-\frac{m_A^2}{v_\phi^2+4v_\Delta^2}\right),\\
\lambda_4 & = \frac{4m_{H^+}^2}{v_\phi^2+2v_\Delta^2}-\frac{2m_A^2}{v_\phi^2+4v_\Delta^2}+\frac{m_h^2-m_H^2}{2v_\phi v_\Delta}\sin2\alpha,
\label{lam4}\\
\lambda_5 & = 4\left(\frac{m_A^2}{v_\phi^2+4v_\Delta^2}-\frac{m_{H^+}^2}{v_\phi^2+2v_\Delta^2}\right).
\end{align}

When the triplet VEV $v_\Delta$ is much less than the doublet VEV $v_\phi$, which is required by the rho parameter data, 
there appear relationships among the masses of the triplet-like Higgs bosons by neglecting $\mathcal{O}(v_\Delta^2/v_\phi^2)$ terms as 
\begin{align}
m_{H^{++}}^2-m_{H^{+}}^2&=m_{H^{+}}^2-m_{A}^2~~\left(=-\frac{\lambda_5}{4}v^2\right), \label{eq:mass_relation1}\\
m_A^2&=m_{H}^2~~(=M_\Delta^2). \label{eq:mass_relation2}
\end{align}
In the limit of $v_\Delta/v_\phi\to 0$, the four mass parameters of the triplet-like Higgs bosons are determined by two parameters. 
Eqs.~(\ref{eq:mass_relation1}) and (\ref{eq:mass_relation2}) can be regarded as the consequence of the global symmetries which are 
mentioned in just below Eq.~(\ref{pot_htm}).


\begin{figure}[t]
\begin{center}
\includegraphics[width=100mm]{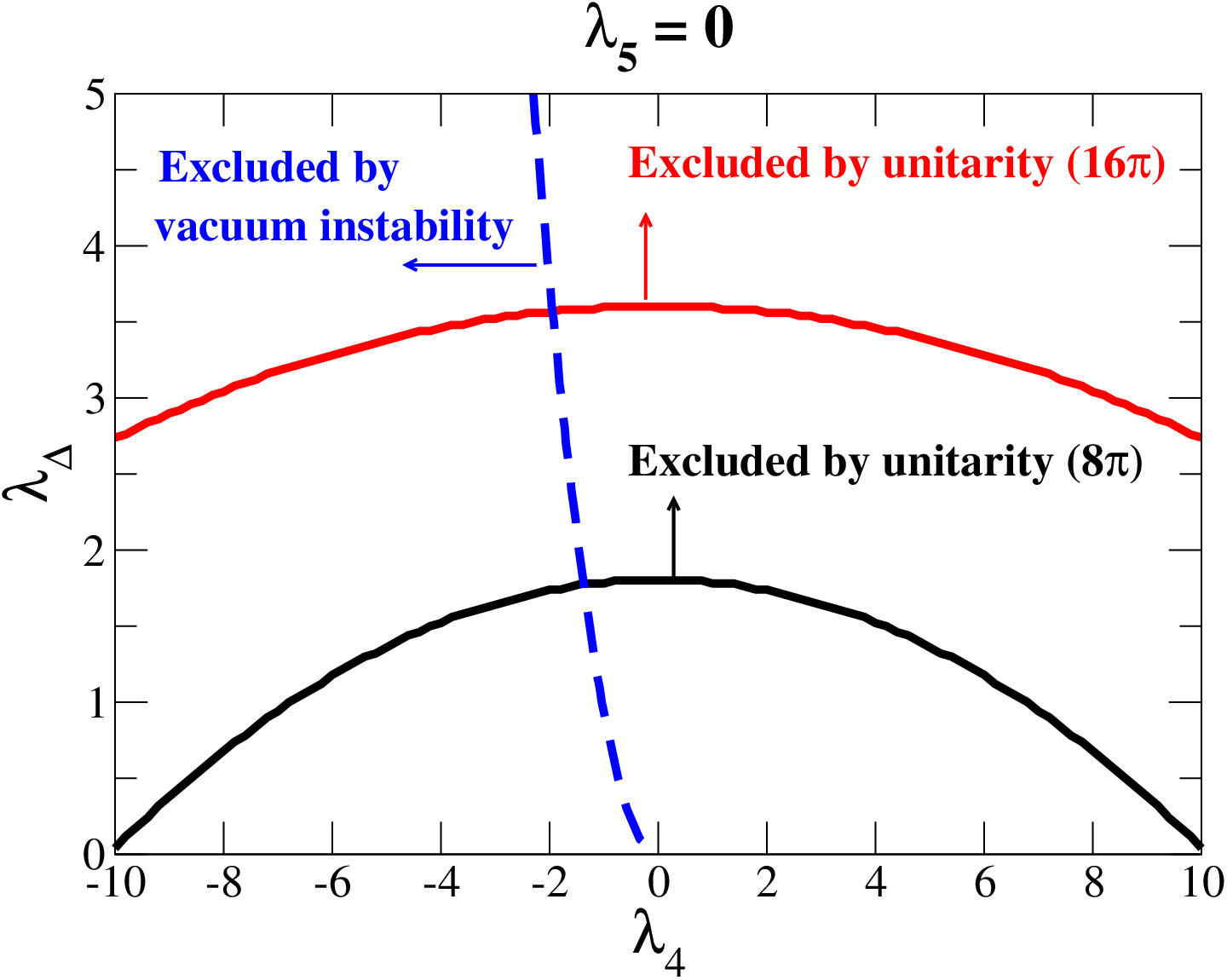}
\caption{
Constraints from the unitarity and vacuum stability bounds for $\lambda_1=m_h^2/(2v^2)\simeq 0.13$ and 
$\lambda_5 = 0$ in the $\lambda_4$-$\lambda_\Delta$ plane. }
\label{FIG:pu_vs1}
\end{center}
\end{figure}

\begin{figure}[t]
\begin{center}
\includegraphics[width=80mm]{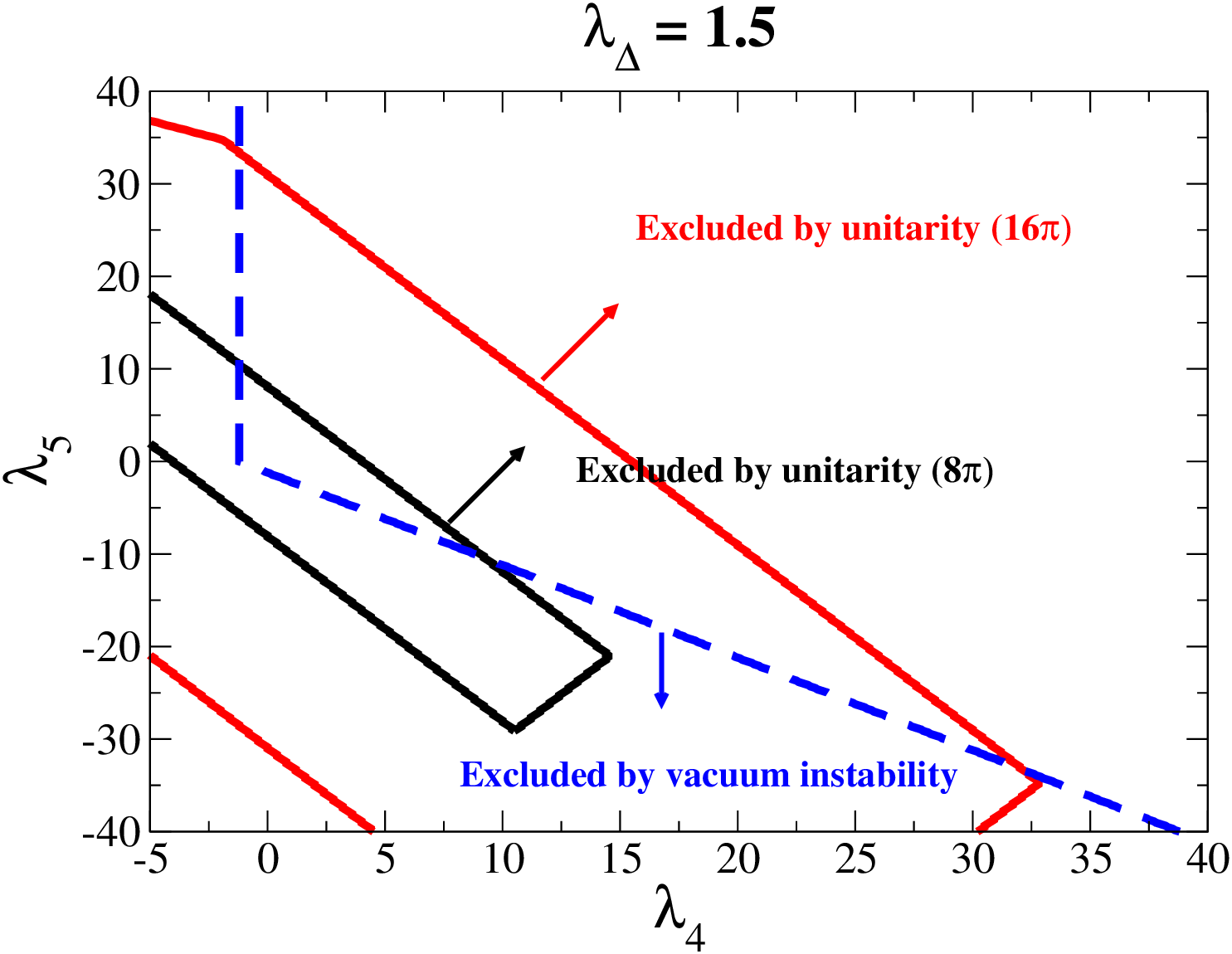}\hspace{3mm}
\includegraphics[width=80mm]{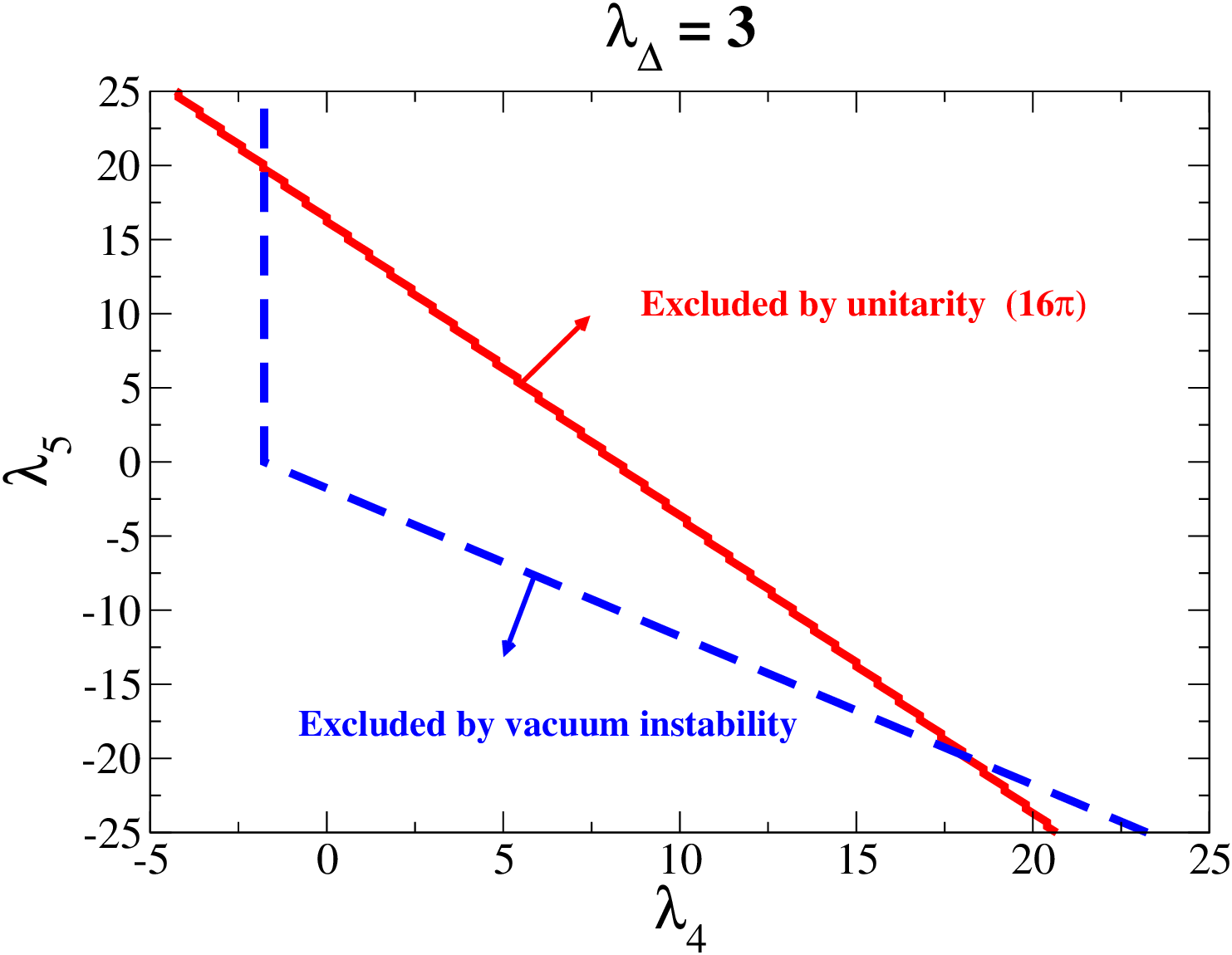}
\caption{Constraints from the unitarity and vacuum stability bounds for $\lambda_1=m_h^2/(2v^2)\simeq 0.13$
in the $\lambda_4$-$\lambda_5$ plane. 
We take $\lambda_\Delta=1.5$ for the left panel and $\lambda_\Delta=3$ for the right panel with 
$\lambda_\Delta=\lambda_2=\lambda_3$. 
}
\label{FIG:pu_vs2}
\end{center}
\end{figure}

From now on, we discuss the constraints from the unitarity and the vacuum stability. 
The condition for the vacuum stability bound has been derived in Ref.~\cite{Arhrib}, where 
we require that 
the Higgs potential is bounded from below in any direction of the large scalar fields region. 
The unitarity bound has been discussed in Ref.~\cite{Aoki-Kanemura} in the Gerogi-Machacek model~\cite{GM} 
which contains the HTM. The unitarity bound in the HTM has also been discussed in Ref.~\cite{Arhrib}. 

The necessary and sufficient condition for the requirement of the vacuum stability is given by~\cite{Arhrib}
\begin{align}
&\lambda_1 > 0,\quad 
\lambda_2+\text{MIN}\left[\lambda_3,~\frac{1}{2}\lambda_3\right] > 0,\notag\\
&\lambda_4+\text{MIN}[0,~\lambda_5] +2\text{MIN}[\sqrt{\lambda_1(\lambda_2+\lambda_3)},\sqrt{\lambda_1(\lambda_2+\lambda_3/2)}]> 0.
\label{vs_condition}
\end{align}
The definition of the dimension less scalar coupling constants used in Ref.~\cite{Arhrib} are different from that of ours. 
Conditions listed in Eq.~(\ref{vs_condition}) are correct in our notation. 
When we take $\lambda_\Delta\equiv \lambda_2=\lambda_3>0$, these inequalities can be written as the simple form;
\begin{align}
\lambda_1 > 0,\quad \lambda_\Delta > 0,\quad
2\sqrt{2\lambda_1\lambda_\Delta}+\lambda_4+\text{MIN}[0,~\lambda_5] > 0. 
\end{align}

In the unitarity bound, we require that the 
matrix of the $S$-wave amplitude for the elastic scatterings of 
two scalar boson states $\langle\varphi_3\varphi_4|a_0|\varphi_1\varphi_2\rangle$ 
are satisfied the following condition; 
\begin{align}
|\langle\varphi_3\varphi_4|a_0|\varphi_1\varphi_2\rangle|<1\quad 
\text{or}\quad |\text{Re}\langle\varphi_3\varphi_4|a_0|\varphi_1\varphi_2\rangle|<\frac{1}{2}, \label{uni1}
\end{align}
where $\varphi_i$ denote the NG bosons and the physical Higgs bosons. 
In the HTM, there are 35 possible scattering processes, i.e., 
15 neutral channels, 10 singly-charged channels, 7 doubly-charged channels, 2 triply-charged channels and one 
quadruply-charged channel. 
Thus, there are 35 corresponding eigenvalues, but some of them have the same expressions. 
In fact, 12 eigenvalues can be regarded as independent eigenvalues, these are 
\begin{align}
y_1 &= 2\lambda_1,\quad y_2  = 2(\lambda_2+\lambda_3),\quad y_3  = 2\lambda_2,\notag\\
y_4^\pm  &= \lambda_1+\lambda_2+2\lambda_3\pm\sqrt{\lambda_1^2-2\lambda_1(\lambda_2+2\lambda_3)
+\lambda_2^2+4\lambda_2\lambda_3+4\lambda_3^2+\lambda_5^2},\notag\\
y_5^\pm & = 3\lambda_1+4\lambda_2+3\lambda_3
\pm\sqrt{9\lambda_1^2-6\lambda_1(4\lambda_2+3\lambda_3)+16\lambda_2^2+24\lambda_2\lambda_3+9\lambda_3^2+6\lambda_4^2+2\lambda_5^2}, \notag\\
y_6 & = \lambda_4,\quad y_7 =\lambda_4+\lambda_5,\quad 
y_8 = \frac{1}{2}(2\lambda_4+3\lambda_5),\quad y_9 =\frac{1}{2}(2\lambda_4-\lambda_5),\quad 
y_{10} = 2\lambda_2-\lambda_3.
\end{align}
The unitarity constrains by the following condition: 
\begin{align}
|y_i|< \zeta,\quad i=1,\dots,10, \label{lqt}
\end{align}
where $\zeta$ is the upper limit for these eigenvalues. 
In Eq.~(\ref{uni1}), when we impose the former (latter) condition to the $S$-wave amplitude, 
this corresponds to $\zeta = 16\pi$ ($8\pi$). 
In our numerical analysis for the constraint from the unitarity bound, we take both the cases with $\zeta=8\pi$ and $\zeta=16\pi$. 
These eigenvalues can be rewritten as a simple form by using $\lambda_\Delta(>0)$  by 
\begin{align}
x_1 &= 3\lambda_1+7\lambda_\Delta+\sqrt{(3\lambda_1-7\lambda_\Delta)^2+\frac{3}{2}(2\lambda_4+\lambda_5)^2},\\
x_2 &= \frac{1}{2}(2\lambda_4+3\lambda_5),\\
x_3 &= \frac{1}{2}(2\lambda_4-\lambda_5).
\end{align}

In Fig.~\ref{FIG:pu_vs1}, the excluded regions by the unitarity bound and the vacuum stability condition are shown 
for $\lambda_5=0$ and $\lambda_1 = m_h^2/(2v^2)\simeq 0.13$ in the $\lambda_4$-$\lambda_\Delta$ plane. 
In this figure, the solid black (red) curve and the black (red) arrow indicate the excluded regions by the 
unitarity bound with $\zeta=8\pi$ ($16\pi$). 
The left-side regions from the blue dashed curve are excluded by the vacuum stability bound. 
It can be seen that in the case with larger $\lambda_\Delta$ values, 
the vacuum stability bound (unitarity bound) is relaxed (more severe) 
compared with the case with smaller $\lambda_\Delta$ values. 
When we take $\zeta=8\pi$ ($16\pi$), 
we obtain the constraint of $\lambda_4\gtrsim -1.4$ for $\lambda_\Delta\simeq 1.8$
($\lambda_4\gtrsim -2$ for $\lambda_\Delta\simeq 3.5$).  

In Fig.~\ref{FIG:pu_vs2}, the excluded regions by the unitarity bound and the vacuum stability condition are shown 
for $\lambda_1 = m_h^2/(2v^2)\simeq 0.13$ in the $\lambda_4$-$\lambda_5$ plane. 
We take $\lambda_\Delta=1.5$ $(3)$ in the left (right) panel. 
Excluded regions by the unitarity and vacuum stability bounds are 
shown by the same way as in the Fig.~\ref{FIG:pu_vs1}. 
In the case with $\lambda_\Delta=1.5$ (left), the allowed minimum value for $\lambda_4$ is about $-1.3$, 
in which $\lambda_5$ is constrained to be $0<\lambda_5<10~(33)$ for $\zeta=8\pi$ $(16\pi)$. 
In the case with $\lambda_\Delta=3$ (right), the allowed minimum value for $\lambda_4$ is about $-1.7$, 
in which $\lambda_5$ is constrained to be $0<\lambda_5<20$ for $\zeta=16\pi$. 
For $\lambda_\Delta=3$, there is no allowed regions by the unitarity bound with $\zeta=8\pi$, so that the black curve 
does not appear in the right plot. 
In both cases with $\lambda_\Delta=1.5$ and $\lambda_\Delta=3$, 
when $\lambda_4$ is taken to be a negative value, negative values for $\lambda_5$ are strongly constrained by the 
vacuum stability bound; i.e., 
the mass hierarchy of $m_{H^{++}}>m_{H^+}>m_A$ is disfavored in that case.

\section{Renormalization}

In this section, we define the on-shell renormalization scheme in order to calculate 
the one-loop corrected electroweak precision parameters and also the SM-like Higgs boson couplings: $hZZ$, $hWW$ and $hhh$. 
First, we discuss the renormalization of the electroweak sector to calculate the renormalized W boson mass, 
which can be used to constrain parameters such as the triplet-like Higgs boson masses in the HTM. 
Second, we consider the renormalization of parameters in the Higgs potential. 


\subsection{Renormalization of the electroweak parameters}

The renormalization prescription in models where the tree level rho parameter: $\rho_{\text{tree}}$ is predicted 
to be unity such as the SM 
is different from that in models without $\rho_{\text{tree}}=1$ such as the HTM. 
Therefore, we separately discuss the renormalization prescriptions in models with $\rho_{\text{tree}}=1$ and 
those with $\rho_{\text{tree}}\neq 1$ in order to clarify the difference between two prescriptions. 

\subsubsection{Models with $\rho_{\text{tree}}=1$}

We first discuss the renormalization of the electroweak precision parameters in models with $\rho_{\text{tree}}=1$ 
based on the on-shell renormalization prescription discussed in Ref.~\cite{Hollik_SM} in detail. 
In this class of models, the electroweak parameters are described by three independent input parameters. 
For instance, when we choose $m_W$, $m_Z$ and $\alpha_{\text{em}}$ as input parameters, 
all the other parameters are written in these parameters; 
\begin{align}
s_W^2 &=1-\frac{m_W^2}{m_Z^2},\label{swsq}\\
G_F &= \frac{\pi\alpha_{\text{em}}}{\sqrt{2} m_W^2s_W^2}.\label{GF}
\end{align}

In the renormalization calculation, 
we shift all the input parameters into the renormalized parameters and the counter-terms. 
Once we specify the input parameters, all the counter-terms are also described by the three 
counter-terms which are associated with the three input parameters. 
We shift all the parameters in the kinetic Lagrangian as follows 
\begin{align}
m_W^2 &\to m_W^2 +\delta m_W^2,\quad 
m_Z^2 \to m_Z^2 +\delta m_Z^2,\quad
\alpha_{\text{em}} \to \alpha_{\text{em}}  +\delta \alpha_{\text{em}} ,\notag\\
B_\mu &\to  B_\mu+\frac{1}{2}\delta Z_B,\quad
W_\mu^a \to W_\mu^a+\frac{1}{2}\delta Z_W.\label{shift}
\end{align}
The wave function renormalization for the photon and the Z boson can be obtained by the shift
\begin{align}
\left(
\begin{array}{c}
Z_\mu \\
A_\mu
\end{array}\right)
&\to
\left[1+
\frac{1}{2}\left(
\begin{array}{cc}
\delta Z_Z  & \delta Z_{Z\gamma } \\
\delta Z_{Z\gamma } & \delta Z_\gamma
\end{array}\right)
+\frac{1}{2s_Wc_W}
\left(
\begin{array}{cc}
0 & -\delta s_W^2 \\
\delta s_W^2 & 0
\end{array}\right)
\right]
\left(
\begin{array}{c}
Z_\mu \\
A_\mu 
\end{array}\right), \label{za_wave}
\end{align}
where the counter-terms $\delta Z_Z$, $\delta Z_\gamma$, $\delta Z_{Z\gamma}$ and $\delta s_W^2$ are expressed 
in terms of the counter-terms defined in Eq.~(\ref{shift}) as
\begin{align}
\left(\begin{array}{c}
\delta Z_Z\\
\delta Z_\gamma
\end{array}\right)
&=\left(
\begin{array}{cc}
c_W^2 & s_W^2\\
s_W^2 & c_W^2
\end{array}\right)
\left(\begin{array}{c}
\delta Z_W\\
\delta Z_B
\end{array}\right),  \label{delZAB}\\
\delta Z_{Z\gamma }&=c_Ws_W(\delta Z_W-\delta Z_B)=\frac{c_Ws_W}{c_W^2-s_W^2}(\delta Z_Z-\delta Z_\gamma), \label{Zzgam}\\
\frac{\delta s_W^2}{s_W^2} &= \frac{c_W^2}{s_W^2}\left(\frac{\delta m_Z^2}{m_Z^2}-\frac{\delta m_W^2}{m_W^2}\right).\label{delsw}
\end{align}
The renormalized two point functions for the gauge bosons can be expressed as
\begin{align}
\hat{\Pi}_{WW}[p^2]&=\Pi_{WW}^{\text{1PI}}(p^2)-\delta m_W^2+\delta Z_W(p^2-m_W^2),\\
\hat{\Pi}_{ZZ}[p^2]&=\Pi_{ZZ}^{\text{1PI}}(p^2)-\delta m_Z^2+\delta Z_{Z}(p^2-m_Z^2),\\
\hat{\Pi}_{\gamma\gamma}[p^2]&=\Pi_{\gamma\gamma}^{\text{1PI}}(p^2)+p^2\delta Z_{\gamma},\\
\hat{\Pi}_{Z\gamma }[p^2]&=\Pi_{Z\gamma }^{\text{1PI}}(p^2)-\delta Z_{Z\gamma }(p^2-\frac{1}{2}m_Z^2)-m_Z^2\frac{\delta s_W^2}{2s_Wc_W}, 
\end{align}
where $\Pi_{XY}^{\text{1PI}}$ ($XY=WW,ZZ,\gamma\gamma$ or $Z\gamma$) are the 1PI diagram contributions to the gauge boson two point functions. 
We here define derivatives of the renormalized two point functions and 1PI diagram contributions as 
$\hat{\Pi}_{XY}^{\prime}[m^2]  \equiv \frac{d}{dp^2}\hat{\Pi}_{XY}[p^2]\big|_{p^2=m^2}$ and 
$\Pi_{XY}^{\text{1PI}^\prime}(m^2)\equiv \frac{d}{dp^2}\Pi_{XY}^{\text{1PI}}(p^2)\big|_{p^2=m^2}$. 

In order to determine the counter-terms, we impose the following five renormalization conditions:
\begin{align}
&\text{Re}\hat{\Pi}_{WW}[m_W^2]=0,\quad \text{Re}\hat{\Pi}_{ZZ}[m_Z^2]=0,
\label{rc1}\\
&\hat{\Pi}^\prime_{\gamma\gamma}[0] =0,\quad 
\hat{\Pi}_{Z\gamma }[0]=0,\quad
\hat{\Gamma}_\mu^{\gamma ee }[q^2=0,~p_1\hspace{-3mm}/=p_2\hspace{-3mm}/=m_e]=ie\gamma_\mu \label{rc3}, 
\end{align}
where $\hat{\Gamma}_\mu^{\gamma ee }$ is the renormalized $\gamma ee$ vertex. 
By using these conditions, all the counter-terms in the electroweak sector can be determined as
\begin{align}
\delta m_W^2 &= \text{Re}\Pi_{WW}^{\text{1PI}}(m_W^2),\quad  \delta m_Z^2 = \text{Re}\Pi_{ZZ}^{\text{1PI}}(m_Z^2),\quad 
\frac{\delta \alpha_{\text{em}}}{\alpha_{\text{em}}} 
= \Pi_{\gamma\gamma}^{\text{1PI}^\prime}(0)-\frac{2s_W}{c_W}\frac{\Pi_{Z\gamma }^{\text{1PI}}(0)}{m_Z^2}, 
\label{count1}\\
\delta Z_\gamma &=  -\Pi_{\gamma\gamma}^{\text{1PI}^\prime}(0),\quad 
\delta Z_{Z\gamma } = -2\frac{\Pi_{Z\gamma }^{\text{1PI}}(0)}{m_Z^2}+\frac{\delta s_W^2}{s_Wc_W}, \label{zzz} \\
\delta Z_Z &= -\Pi_{\gamma\gamma}^{\text{1PI}^\prime}(0)-\frac{2(c_W^2-s_W^2)}{c_Ws_W}\frac{\Pi_{Z\gamma }^{\text{1PI}}(0)}{m_Z^2}
+\frac{c_W^2-s_W^2}{c_W^2}\frac{\delta s_W^2}{s_W^2},\\
\delta Z_W &= -\Pi_{\gamma\gamma}^{\text{1PI}^\prime}(0)
-\frac{2c_W}{s_W}\frac{\Pi_{Z\gamma }^{\text{1PI}}(0)}{m_Z^2}
+\frac{\delta s_W^2}{s_W^2}, 
\end{align}
The counter-term of $\delta s_W^2$ is also determined by using the relation given in Eq.~(\ref{delsw}) as
\begin{align}
\frac{\delta s_W^2}{s_W^2} = 
\frac{c_W^2}{s_W^2}\left[
\frac{\text{Re}\Pi_{ZZ}^{\text{1PI}}(m_Z^2)}{m_Z^2}-\frac{\text{Re}\Pi_{WW}^{\text{1PI}}(m_W^2)}{m_W^2}\right]. \label{delswsq}
\end{align}
Now we can calculate the one-loop level predictions for electroweak observables. 
The renormalized mass of the W boson $m_W^{\text{ren}}$ 
as well as the weak mixing angle which is defined in Eq.~(\ref{swsq}) $s_W^{\text{ren}}$ 
can be calculated thorough the $\Delta r$ parameter 
which summarizes the radiative corrections as
\begin{align}
m_W^2 &= \frac{\pi\alpha_{\text{em}}}{\sqrt{2} G_Fs_W^2}\frac{1}{1-\Delta r}. \label{delr}
\end{align}
$\Delta r$ can be determined from the muon decay process at the one-loop level under the on-shell renormalization conditions by 
\begin{align}
\Delta r &= \frac{\hat{\Pi}_{WW}(0)}{m_W^2} + \delta_{VB}\notag\\
& = \Pi_{\gamma\gamma}^{\text{1PI}^\prime}(0)
-\frac{c_W^2}{s_W^2}\left[
\frac{\text{Re}\Pi_{ZZ}^{\text{1PI}}(m_Z^2)}{m_Z^2}-\frac{\text{Re}\Pi_{WW}^{\text{1PI}}(m_W^2)}{m_W^2}
-\frac{2s_W}{c_W}\frac{\Pi_{Z\gamma }^{\text{1PI}}(0)}{m_Z^2}
\right]\notag\\
&+\frac{\Pi_{WW}^{\text{1PI}}(0)-\text{Re}\Pi_{WW}^{\text{1PI}}(m_W^2)}{m_W^2} 
 +\delta_{VB},  \label{delr22}
\end{align}
where $\delta_{VB}$ is the vertex and box diagram corrections to the muon decay process, which is given by 
\begin{align}
\delta_{VB}  = \frac{\alpha_{\text{em}}}{4\pi s_W^2}\left[6+\frac{10-10s_W^2-3m_W^2/(c_W^2m_Z^2)(1-2s_W^2)}{2\left(1-\frac{m_W^2}{m_Z^2}\right)}\ln \frac{m_W^2}{m_Z^2}\right].  
\end{align}
Eq.~(\ref{delr22}) can be rewritten in terms of the shift of the fine structure constant $\Delta \alpha_{\text{em}}$ from the electron 
mass scale to the Z boson mass scale, 
the radiative correction to the electroweak rho parameter $\Delta \rho$ and the reminding contribution $\Delta r_{\text{rem}}$ 
as
\begin{align}
\Delta r &= \Delta\alpha_{\text{em}}-\frac{c_W^2}{s_W^2}\Delta\rho+\Delta r_{\text{rem}}  \label{delr3}, 
\end{align}
where $\Delta \alpha_{\text{em}}$, $\Delta\rho$ and $\Delta r_{\text{rem}}$ can be expressed in terms of the gauge boson two point 1PI functions
\begin{align}
\Delta \alpha_{\text{em}} &= -\hat{\Pi}_{\gamma\gamma}^\prime(m_Z^2)=\Pi_{\gamma\gamma}'(0)-\Pi_{\gamma\gamma}^\prime(m_Z^2),\label{del_alpha}\\
\Delta \rho  &= \frac{\Pi_{ZZ}^{\text{1PI}}(0)}{m_Z^2}-\frac{\Pi_{WW}^{\text{1PI}}(0)}{m_W^2}-\frac{2s_W}{c_W}\frac{\Pi_{Z\gamma }^{\text{1PI}}(0)}{m_Z^2},\label{del_rho}\\
\Delta r_{\text{rem}}&=\frac{c_W^2}{s_W^2}\left[\frac{\Pi_{ZZ}^{\text{1PI}}(0)}{m_Z^2}-\frac{\text{Re}\Pi_{ZZ}^{\text{1PI}}(m_Z^2)}{m_Z^2}\right]+
\left(1-\frac{c_W^2}{s_W^2}\right)\left[\frac{\Pi_{WW}^{\text{1PI}}(0)}{m_W^2}-\frac{\text{Re}\Pi_{WW}^{\text{1PI}}(m_W^2)}{m_W^2}\right]\notag\\
&+\Pi_{\gamma\gamma}'(m_Z^2)+\delta_{VB}. \label{del_rem}
\end{align}
We note that 
light fermions can contribute to $\Delta\alpha$ by the logarithmic power of their masses, but 
heavy particles such as the top quark are suppressed by their inverse quadratic power, 
so that we can neglect these effects to $\Delta \alpha_{\text{em}}$. 
Mass splitting among the particles belonging to the same isospin multiplet can 
contribute to $\Delta\rho$ by the quadratic power; e.g., the $m_t^2$ dependence appears in $\Delta\rho$. 
This quadratic mass dependence can be regarded as the effect of the violation of the custodial symmetry. 
%
The renormalized W boson mass $m_W^{\text{ren}}$ and 
the weak mixing angle
$s_W^{\text{ren}}$ can be calculated through $\Delta r$ as 
\begin{align}
(m_W^{\text{ren}})^2 &= \frac{m_Z^2}{2}\left[1+\sqrt{1-\frac{4\pi \alpha_{\text{em}}}{\sqrt{2}G_Fm_Z^2(1-\Delta r)}}\right],~ 
(s_W^{\text{ren}})^2 = \frac{1}{2}\left[1-\sqrt{1-\frac{4\pi \alpha_{\text{em}}}{\sqrt{2}G_Fm_Z^2(1-\Delta r)}}\right].
\end{align}
Notice that $m_W$ written in the right-hand side of Eq.~(\ref{delr22}) is calculated by using the three experimental 
input values $\alpha_{\text{em}}$, $G_F$ and $m_Z$ via the 
tree level relation expressed in 
Eqs.~(\ref{swsq}) and (\ref{GF}). 

We here discuss the other definition of the weak mixing angle which is so-called the effective mixing angle denoted by 
$\sin\theta_\text{eff}^f$. 
This is defined by the 
ratio of the effective coupling constants $g_V^f$ and $g_A^f$ in the neutral current vertex at the Z boson resonance as follows~\cite{Heinemeyer}: 
\begin{align}
\sin^2\theta_{\text{eff}}^f = \frac{1}{4|Q_f|}\left[1-\frac{\text{Re}(g_V^f)}{\text{Re}(g_A^f)}\right],  \label{deleffsw}
\end{align}
where 
\begin{align}
g_V^f &= \left(\rho\frac{1-\Delta r}{1+\hat{\Pi}_{ZZ}'(m_Z^2)}\right)^{1/2}
\left[v_f+2Q_fs_Wc_W\frac{\hat{\Pi}_{Z\gamma }(m_Z^2)}{m_Z^2+\hat{\Pi}_{\gamma\gamma}(m_Z^2)}+\hat{\Lambda}_V^{Zff}(m_Z^2)\right],\label{gveff}\\
g_A^f &= \left(\rho\frac{1-\Delta r}{1+\hat{\Pi}_{ZZ}'(m_Z^2)}\right)^{1/2}
[a_f+\hat{\Lambda}_A^{Zff}(m_Z^2)].\label{gaeff}
\end{align}
In Eqs.~(\ref{gveff}) and (\ref{gaeff}), 
$\hat{\Lambda}_V^{Zff}$ ($\hat{\Lambda}_A^{Zff}$) is the one-loop contribution to the coefficient of the vector (axial vector) part of the renormalized $Zff$ vertex: 
\begin{align}
\hat{\Gamma}_{\mu}^{Zff}&=
 i\frac{e}{2s_Wc_W}\left[\gamma_\mu (v_f-\gamma_5 a_f)
+\gamma_\mu\hat{\Lambda}_V^{Zff}-\gamma_\mu\gamma_5\hat{\Lambda}_A^{Zff}\right], 
\end{align}
and 
\begin{align}
\hat{\Lambda}_V^{Zff}&=\Lambda_V^{Zff}
-v_f\Pi_{V,f}^{\text{1PI}}(m_f^2)-a_f\Pi_{A,f}^{\text{1PI}}(m_f^2)
-a_f\frac{c_W}{s_W}\frac{\Pi_{Z\gamma }^{\text{1PI}}(0)}{m_Z^2},\label{lamhatv}\\
\hat{\Lambda}_A^{Zff}&=\Lambda_A^{Zff}
-a_f\Pi_{V,f}^{\text{1PI}}(m_f^2)-v_f\Pi_{A,f}^{\text{1PI}}(m_f^2)
-a_f\frac{c_W}{s_W}\frac{\Pi_{Z\gamma }^{\text{1PI}}(0)}{m_Z^2}, \label{lamhata}\\
& \text{with }v_f =I_f-2Q_fs_W^2 ,\quad a_f=I_f,
\end{align}
where $\Lambda_{V}^{Zff}$ $(\Lambda_{A}^{Zff})$ and $\Pi_{V,f}^{\text{1PI}}$ ($\Pi_{A,f}^{\text{1PI}}$) 
are the vector (axial vector) part of the 1PI diagram contributions to the $Zff$ vertex and the fermion two point functions, 
respectively. 
By using Eqs.~(\ref{deleffsw})-(\ref{lamhata}), $\sin^2\theta_{\text{eff}}^f$ can be expressed as
\begin{align}
\sin^2\theta_{\text{eff}}^f=
\frac{1}{4|Q_f|}\Bigg\{1-\frac{v_f}{a_f}\Big[&1+\frac{2Q_fs_Wc_W}{v_f}\left(\frac{\Pi_{Z\gamma }^{\text{1PI}}(m_Z^2)}{m_Z^2}
-\frac{\delta s_W^2}{s_Wc_W}\right)\notag\\
&+\frac{\Lambda_V^{Zff}(m_Z^2)}{v_f}-\frac{\Lambda_A^{Zff}(m_Z^2)}{a_f}+\frac{v_f^2-a_f^2}{v_fa_f}\Pi_{A,f}^{\text{1PI}}(m_f^2)\Big]\Bigg\}. 
\label{sw_eff}
\end{align}

\subsubsection{Models without $\rho_{\text{tree}}=1$}

In this class of models, electroweak parameters are not described by the three input parameters such 
as $m_W$, $m_Z$ and $\alpha_{\text{em}}$ in the SM, but they are described by four input parameters, because 
the relation of $m_W/m_Z=\cos\theta_W$ does not hold. 
Therefore, we need one extra input parameter in addition to above three parameters. 
We here discuss two different renormalization schemes, where we call them as  $Scheme~I$ and $Scheme~II$. 

In Scheme~I, four input parameters are chosen from electroweak precision observables, i.e., 
$m_W$, $m_Z$, $\alpha_{\text{em}}$ and the effective mixing 
angle $\sin^2\theta_{\text{eff}}^e$ defined in Eq.~(\ref{sw_eff}) for $f=e$. 
We denote the weak mixing angle defined by the effective angle as $\hat{s}_W^2\equiv \sin^2\theta_{\text{eff}}^e$ 
($\hat{c}_W^2 = 1-\hat{s}_W^2$) to distinguish from the other definitions. 
This scheme has been first proposed in Ref.~\cite{Blank_Hollik}. 
In this scheme, the additional renormalization condition can be set by requiring the effective mixing angle 
is not changed from the tree level prediction. 
Namely, we impose 
\begin{align}
\frac{\text{Re}(g_V^e)}{\text{Re}(g_A^e)} = \frac{v_e}{a_e}. 
\end{align}
This reads
\begin{align}
\frac{\delta \hat{s}_W^2}{\hat{s}_W^2} &= \frac{\hat{c}_W}{\hat{s}_W}\frac{\Pi_{Z\gamma }^{\text{1PI}}(m_Z^2)}{m_Z^2}+\delta_{V}',~~\text{with }
\delta_{V}'=
-\frac{v_e}{2s_W^2}\left[\frac{\Lambda_V^{Zee}}{v_e}-\frac{\Lambda_A^{Zee}}{a_e}+\frac{v_e^2-a_e^2}{v_ea_e}\Pi_{A,e}^{\text{1PI}}(m_e^2)\right].  \label{delsw_s1}
\end{align}
Using the $\delta \hat{s}_W^2$ determined by Eq.~(\ref{delsw_s1}) instead of Eq.~(\ref{delswsq}), 
we obtain the formulae of $\Delta r$ as
\begin{align}
\Delta r^{\text{Scheme~I}} &=
\Delta\alpha +\Delta r_{\text{rem}}^{\text{Scheme~I}}, \label{delr33}
\end{align}
where $\Delta \alpha_{\text{em}}$ is defined in Eq.~(\ref{del_alpha}). 
There is no $\Delta\rho$ term in the expression of $\Delta r^{\text{Scheme~I}}$, because of the 
renormalization condition of $\hat{s}_W^2$. 
Thus, the $m_t^2$ dependence in $\Delta r^{\text{Scheme~I}}$ vanishes in Scheme~I. 
In Eq.~(\ref{delr33}), 
$\Delta r_{\text{rem}}^{\text{Scheme~I}}$ is 
\begin{align}
\Delta r^{\text{Scheme~I}}_{\text{rem}} &=
\frac{\Pi_{WW}^{\text{1PI}}(0)-\text{Re}\Pi_{WW}^{\text{1PI}}(m_W^2)}{m_W^2} +\Pi_{\gamma\gamma}^{\text{1PI}'}(m_Z^2)\notag\\
&+\frac{\hat{c}_W}{\hat{s}_W}\left[\frac{2\Pi_{Z\gamma }^{\text{1PI}}(0)-\text{Re}\Pi_{Z\gamma }^{\text{1PI}}(m_Z^2)}{m_Z^2}
\right]
+\delta_{VB}-\delta_V'.    
\end{align}

The one loop corrected W boson mass is calculated as
\begin{align}
(m_W^{\text{ren}})_{\text{Scheme~I}}^2 = 
\frac{\pi\alpha_{\text{em}}}{\sqrt{2}G_F\hat{s}_W^2(1-\Delta r^{\text{Scheme~I}})}. \label{mw_sch1}
\end{align}
The numerical evaluation of the renormalized W boson mass given by Eq.~(\ref{mw_sch1}) has been done in Ref.~\cite{Kanemura-Yagyu}. 
It has been found that the mass of $H^{\pm\pm}$ is of $\mathcal{O}(100-200)$ GeV, with $\Delta m$ to be a few hundred GeV and 
$v_\Delta$ of several GeV are preferred by the electroweak precision data. 

Next, we discuss the renormalization in Scheme~II in the HTM. 
In this scheme, three of four input parameters are chosen from the electroweak precision observables, i.e., 
$m_W$, $m_Z$ and $\alpha_{\text{em}}$ such as the SM. The other one is chosen from the mixing angle $\beta'$ between the CP-odd Higgs boson $A$ 
and the NG boson $G^0$ defined in Eqs.~(\ref{mixing1}) and (\ref{tan2a}). 
The counter-term of the mixing angle $\delta \beta'$ 
can be determined by the conditions: 
\begin{align}
\hat{\Pi}_{AG}[0]=\hat{\Pi}_{AG}[m_A^2] = 0, 
\end{align}
where $\hat{\Pi}_{AG}$ is the renormalized two point function of the $G^0$-$A$ mixing given in Eq.~(\ref{Gam_GA}). 
The other counter-terms are determined by the same renormalization conditions given in Eqs.~(\ref{rc1}) and (\ref{rc3}) 
as in the SM. 
In this scheme, the weak mixing angle is not the independent parameter, but it is determined by 
\begin{align}
\cos^2\bar{\theta}_W \equiv \bar{c}_W^2 = \frac{2m_W^2}{m_Z^2(1+c_{\beta'}^2)}. \label{swsq_2}
\end{align}
In order to distinguish the definition of the weak mixing angle in this scheme 
from the other definition, we introduced $\bar{c}_W^2$ $(\bar{s}_W^2 = 1-\bar{c}_W^2)$. 
The counter-term for the weak mixing angle is obtained by imposing 
Eq.~(\ref{count1}): 
\begin{align}
\delta \bar{s}_W^2 = -\delta \bar{c}_W^2 = 
\frac{2m_W^2}{m_Z^2(1+c_{\beta'}^2)}
\left[\frac{\text{Re}\Pi_{ZZ}^{\text{1PI}} (m_Z^2)}{m_Z^2}-\frac{\text{Re}\Pi_{WW}^{\text{1PI}} (m_W^2)}{m_W^2}-\frac{2 c_{\beta'}s_{\beta'}}{1+c_{\beta'}^2}\delta \beta'\right]. 
 \label{delsw_2}
\end{align}
We note that Eqs.~(\ref{swsq}) and (\ref{delsw}) can be reproduced by taking $s_{\beta'}\to 0$ (or $v_\Delta \to 0$) from 
Eqs.~(\ref{swsq_2}) and (\ref{delsw_2}). 
The expression for $\Delta r$ can be obtained in the same way as in models with $\rho_{\text{tree}}=1$: 
\begin{align}
\Delta r^{\text{Scheme~II}} &= 
\Delta\alpha-\frac{\bar{c}_W^2}{\bar{s}_W^2}\Delta\rho+\Delta r_{\text{rem}}^{\text{Scheme~II}}  \label{delr2}, 
\end{align}
where $\Delta\alpha$ and 
$\Delta r_{\text{rem}}^{\text{Scheme~II}}$ are given by the same formulae in Eqs.~(\ref{del_alpha}) and (\ref{del_rem}), 
but $s_W$ and $c_W$ should be replaced by $\bar{s}_W$ and $\bar{c}_W$. 
$\Delta\rho$ can be expressed as
\begin{align}
\Delta \rho &=\frac{\Pi_{ZZ}^{\text{1PI}}(0)}{m_Z^2}-\frac{\Pi_{WW}^{\text{1PI}}(0)}{m_W^2}
-\frac{2\bar{s}_W}{\bar{c}_W}\frac{\Pi_{Z\gamma }^{\text{1PI}}(0)}{m_Z^2}-\frac{2 c_{\beta'}s_{\beta'}}{1+c_{\beta'}^2}\delta \beta'. 
\label{del_rho2}
\end{align}
The quadratic mass dependences due to the custodial symmetry breaking 
such as $m_t^2$
appear in $\Delta r^{\text{Scheme~II}}$ through $\Delta \rho$. 
The one loop corrected W boson mass can be calculated as
\begin{align}
(m_W^{\text{ren}})_{\text{Scheme~II}}^2 = 
\frac{m_Z^2(1+c_{\beta'}^2)}{4}\left[1+\sqrt{1-\frac{8}{1+c_{\beta'}^2}\frac{\pi\alpha_{\text{em}}}{\sqrt{2}G_Fm_Z^2(1-\Delta r^{\text{Scheme~II}})}}\right].
\end{align}

In the following, we show numerical results for the renormalized 
$\Delta r$ and $m_W$ in 
both Scheme I and Scheme II in the HTM. 
The measured mass of the W boson is given by~\cite{PDG}
\begin{align}
m_W^{\text{exp}} = 80.385\pm0.015~\text{GeV}. 
\end{align}
We use following input values from the experiments: 
\begin{align}
&\alpha_{\text{em}}^{-1} = 137.035989,\quad 
\Delta\alpha_{\text{ferm}} = 0.06635,\quad
G_F = 1.16637\times 10^{-5} ~\text{GeV}^{-2},\quad
m_Z =91.1876 ~\text{GeV},\notag\\
&\alpha_s = 0.118,\quad m_t = 173.5~\text{GeV},\quad m_b = 4.7 ~\text{GeV},~m_h=126~\text{GeV}
\end{align}
where $\Delta\alpha_{\text{ferm}}$ is the light fermion contributions to $\Delta\alpha$~\cite{PDG}. 
The QCD corrections to the fermion-loop contributions to the gauge boson two point functions are taken into account 
according to Ref.~\cite{HHKM}. 
In Scheme I, we use $\hat{s}_W^2=0.23146$ as the forth low energy input. 
In Scheme II, we fix $\mu$ expressed in Eq.~(\ref{mu}) or $v_\Delta$ as an additional input parameter. 
Fermionic two-loop corrections to 
$\Delta r^{\text{Scheme~II}}$ is added 
as $\Delta r_{\text{ferm}}^{(\alpha^2)}=0.002856$~\cite{Weiglein}.  
We assume the mass relations among the triplet-like Higgs bosons expressed in Eqs.~(\ref{eq:mass_relation1}) and (\ref{eq:mass_relation2}), 
so that the four triplet mass parameters can be determined by fixing two input parameters. 
We choose the mass of the lightest triplet-like Higgs boson $m_{\text{lightest}}$ 
and the parameter $\xi$ or $\Delta m$ defined by
\begin{align}
\xi \equiv m_{H^{++}}^2-m_{H^+}^2=m_{H^{+}}^2-m_{A}^2,\quad 
\Delta m\equiv m_{H^+}-m_{\text{lightest}}, \text{ with }m_H=m_A \label{xi_delm}, 
\end{align}
instead of specifying two masses of the triplet-like Higgs bosons. 
By the definition given in Eq.~(\ref{xi_delm}), 
$\Delta m$ is positive, but the sign of $\xi$ depends on the mass hierarchy among the 
triplet-like Higgs bosons. 
When $H^{\pm\pm}$ ($A$ and $H$) is the lightest of all the triplet-like Higgs bosons, 
i.e., $m_A>m_{H^+}>m_{H^{++}}$ ($m_A<m_{H^+}<m_{H^{++}}$), then $\xi$ is negative (positive). 
We call the former case (latter case) as Case~I (Case~II). 
These two parameters $\xi$ and $\Delta m$ are related to each other by the equation: 
$\xi=\mp(2m_{\text{lightest}}+\Delta m)\Delta m$ (a negative sign for Case~I and a positive sign for Case~II). 
In addition to these input parameters written in the above, 
we input the mixing angle $\alpha$. 
But, we select $\lambda_4$ as an input parameter instead of $\alpha$. 
In that case, $\alpha$ is determined by 
\begin{align}
\alpha = \frac{1}{2}\arcsin\left[\frac{2v_\Delta v}{m_h^2-m_A^2}\left(\lambda_4+\frac{2m_A^2-4m_{H^+}^2}{v^2}\right)
+\mathcal{O}(v_\Delta^2/v^2)\right]. \label{alpha}
\end{align}

\begin{figure}[t]
\begin{center}
\includegraphics[width=80mm]{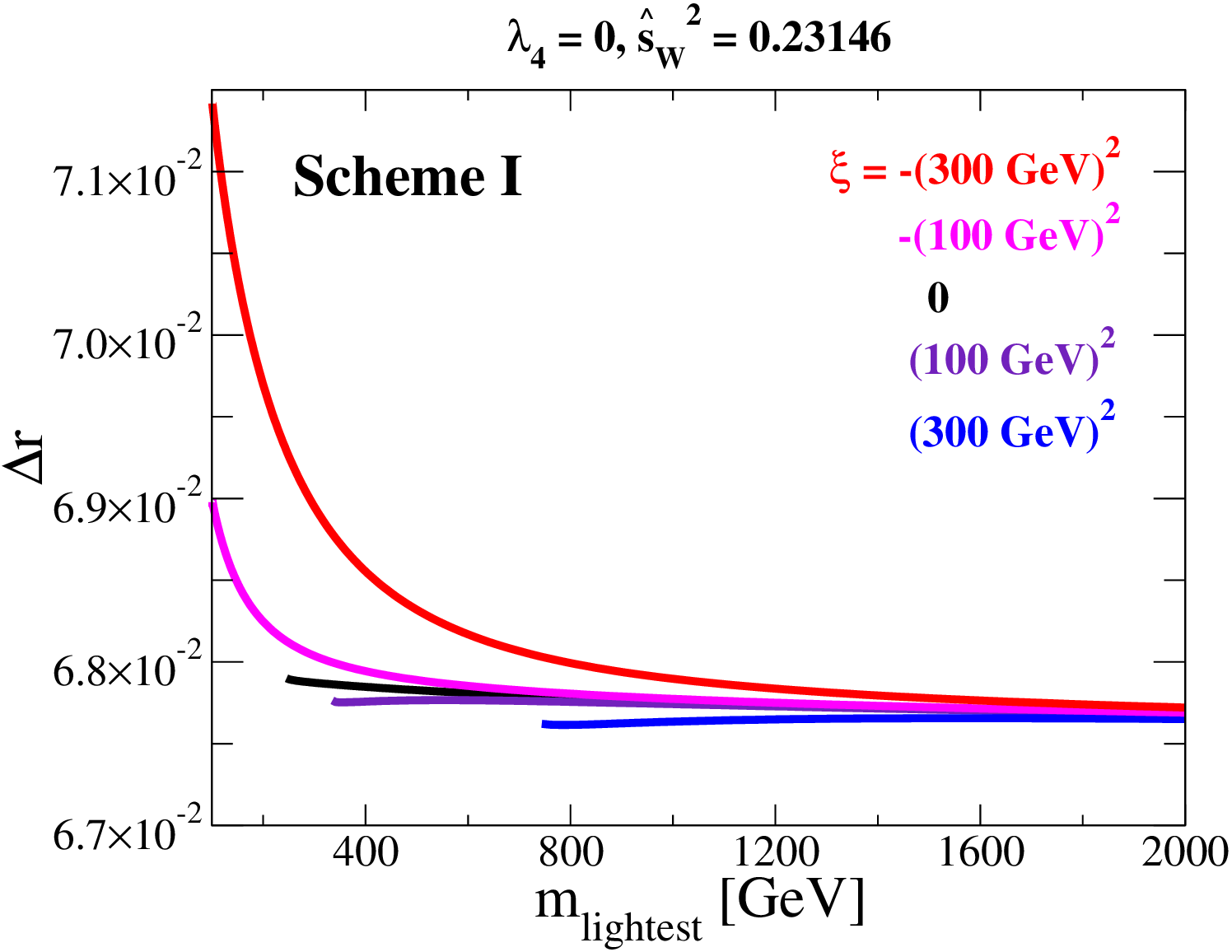}
\includegraphics[width=80mm]{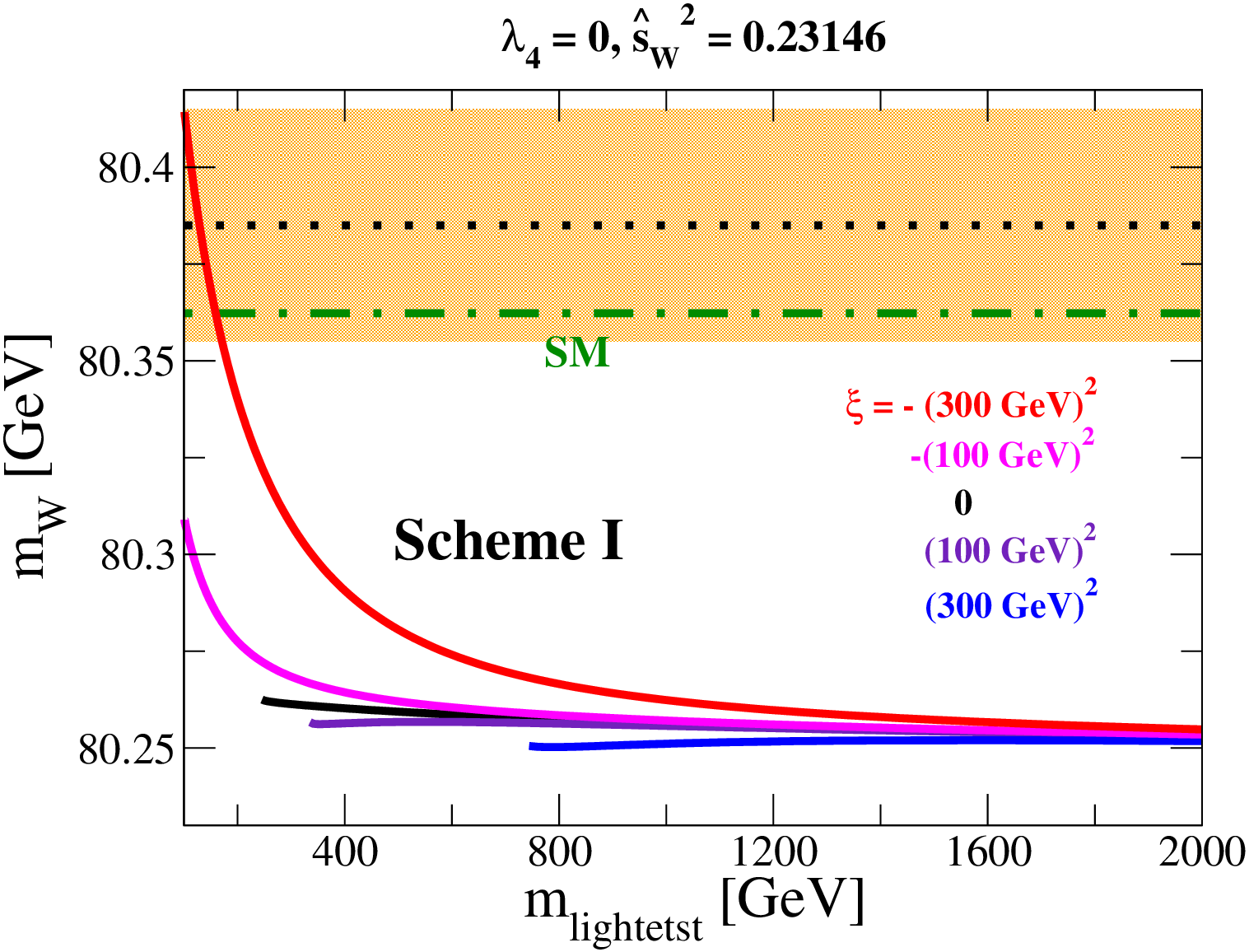}
\caption{Left (Right) plot shows 
$\Delta r$ (the renormalized W boson mass $m_W^{\text{ren}}$) 
calculated in Scheme~I as a function of $m_{\text{lightest}}$ for each fixed value of 
$\xi$.
In both the plots, we take $\hat{s}_W^2=0.23146$ and $\lambda_4=0$. 
In the right plot, the SM prediction is shown as the green dashed line, and 
the orange shaded regions are indicated the region within the 2$\sigma$ error bar of the measured W boson mass.  }
\label{FIG:delr}
\end{center}
\end{figure}

In Fig.~\ref{FIG:delr}, 
$\Delta r$ (left panel) 
and the renormalized W bosons mass $m_W^{\text{ren}}$ (right panel)
are shown as a function of the lightest triplet like Higgs boson mass 
$m_{\text{lightest}}$ in Scheme~I for each fixed value of $\xi$ in the case of 
$\lambda_4=0$ and $\hat{s}_W^2=0.23146$. 
In some regions of $m_{\text{lightest}}$, some curves are not displayed where the 
absolute value of the argument of $\arcsin$ in Eq.~(\ref{alpha}) is larger than 1. 
It is seen that 
the predictions for $\Delta r$ as well as $m_W^{\text{ren}}$ 
are asymptotically close to those in the case of $\xi=0$ in the large $m_{\text{lightest}}$ region. 
The asymptotic value of $m_W^{\text{ren}}$ is not coincident with the SM prediction, which is displayed as 
the green dashed curve. 
This is because the central value of $\hat{s}_W^2$
is different from that value predicted in the SM.
In other words, the triplet VEV is predicted by the four electroweak
precision data, and it is order 1 GeV by using the central value of
$\hat{s}_W^2$.
Thus, the discrepancy between the prediction of $m_W^{\text{ren}}$ 
in the HTM calculated in Scheme I and that in the SM is caused by such a non-zero
value of the triplet VEV. 
\begin{figure}[t]
\begin{center}
\includegraphics[width=80mm]{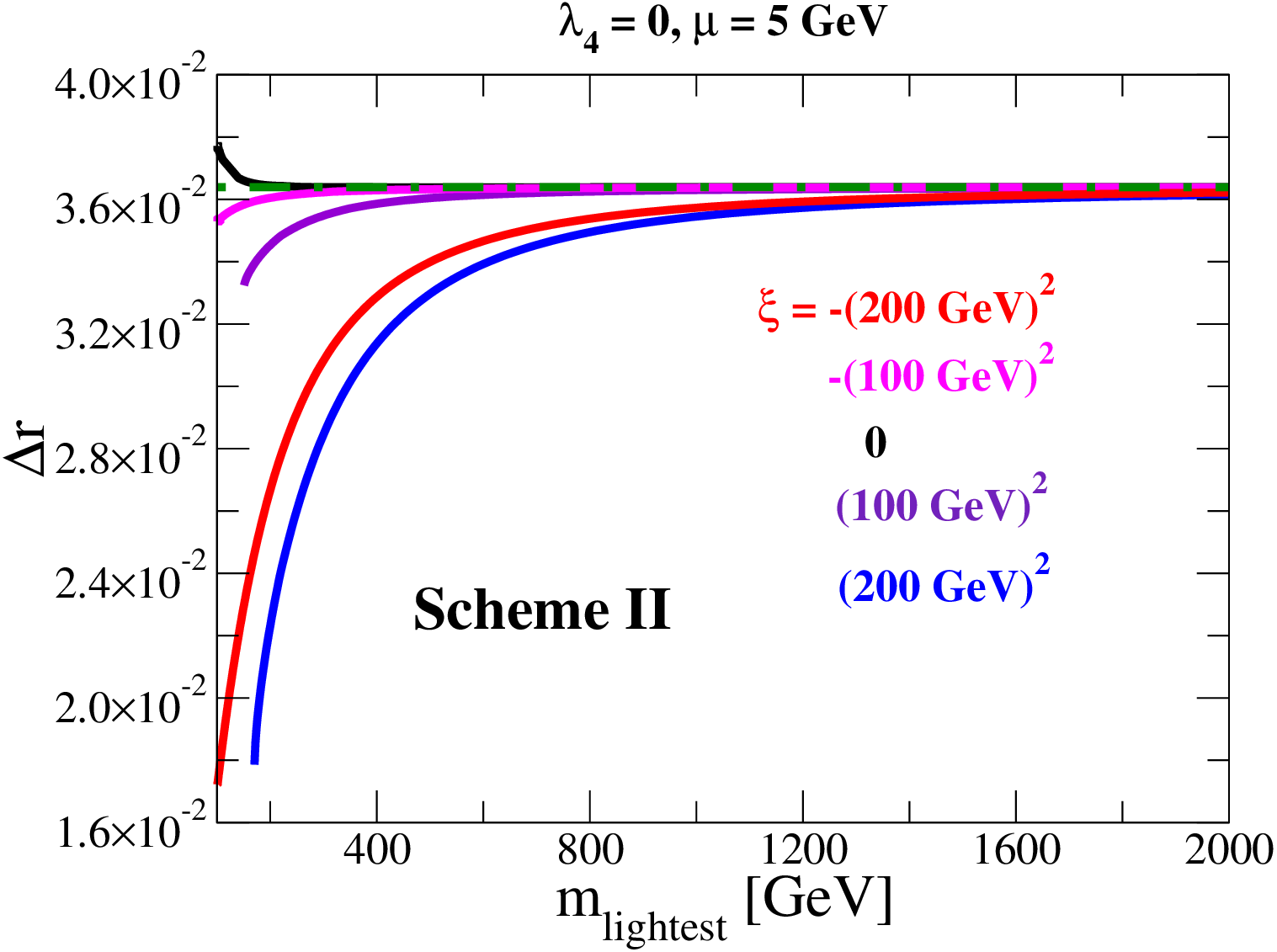}\hspace{3mm}
\includegraphics[width=80mm]{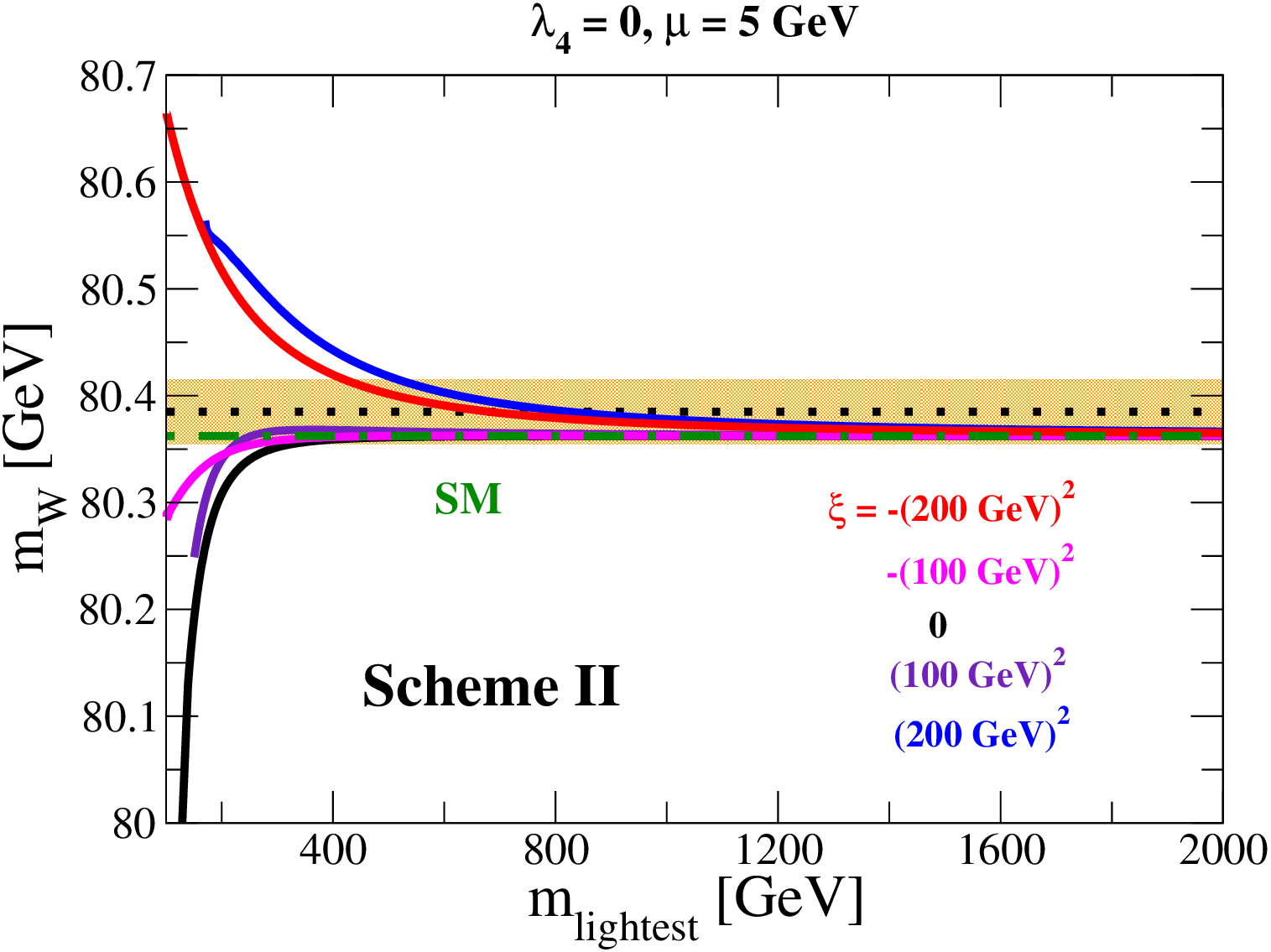}
\caption{Left (Right) plot shows 
$\Delta r$ (the renormalized W boson mass $m_W^{\text{ren}}$) 
calculated in Scheme~II are shown as a function of $m_{\text{lightest}}$ 
for each fixed value of $\xi$.
In both the plots, we take $\mu=5$ GeV and $\lambda_4=0$. 
The SM prediction is shown as the green dashed line
The orange shaded regions are indicated the region within the 2$\sigma$ error bar of the measured W boson mass. 
}
\label{FIG:mw}
\end{center}
\end{figure}

In Fig.~\ref{FIG:mw}, 
numerical values of $\Delta r$ (left panel) and the renormalized W bosons mass $m_W^{\text{ren}}$ (right panel)
are shown as a function of the lightest triplet like Higgs boson mass 
$m_{\text{lightest}}$ in Scheme~II for each fixed value of $\xi$.  
We take $\lambda_4=0$ and $\mu=5$ GeV. 
In some regions of $m_{\text{lightest}}$, some curves are not displayed where the 
absolute value of the argument of $\arcsin$ in Eq.~(\ref{alpha}) is larger than 1. 
It is seen that 
the predictions for $\Delta r$ as well as $m_W^{\text{ren}}$ 
are asymptotically close to those in the case of $\xi=0$ in the large $m_{\text{lightest}}$ limit which 
is coincident with the SM prediction.


\begin{figure}[t]
\begin{center}
\includegraphics[width=80mm]{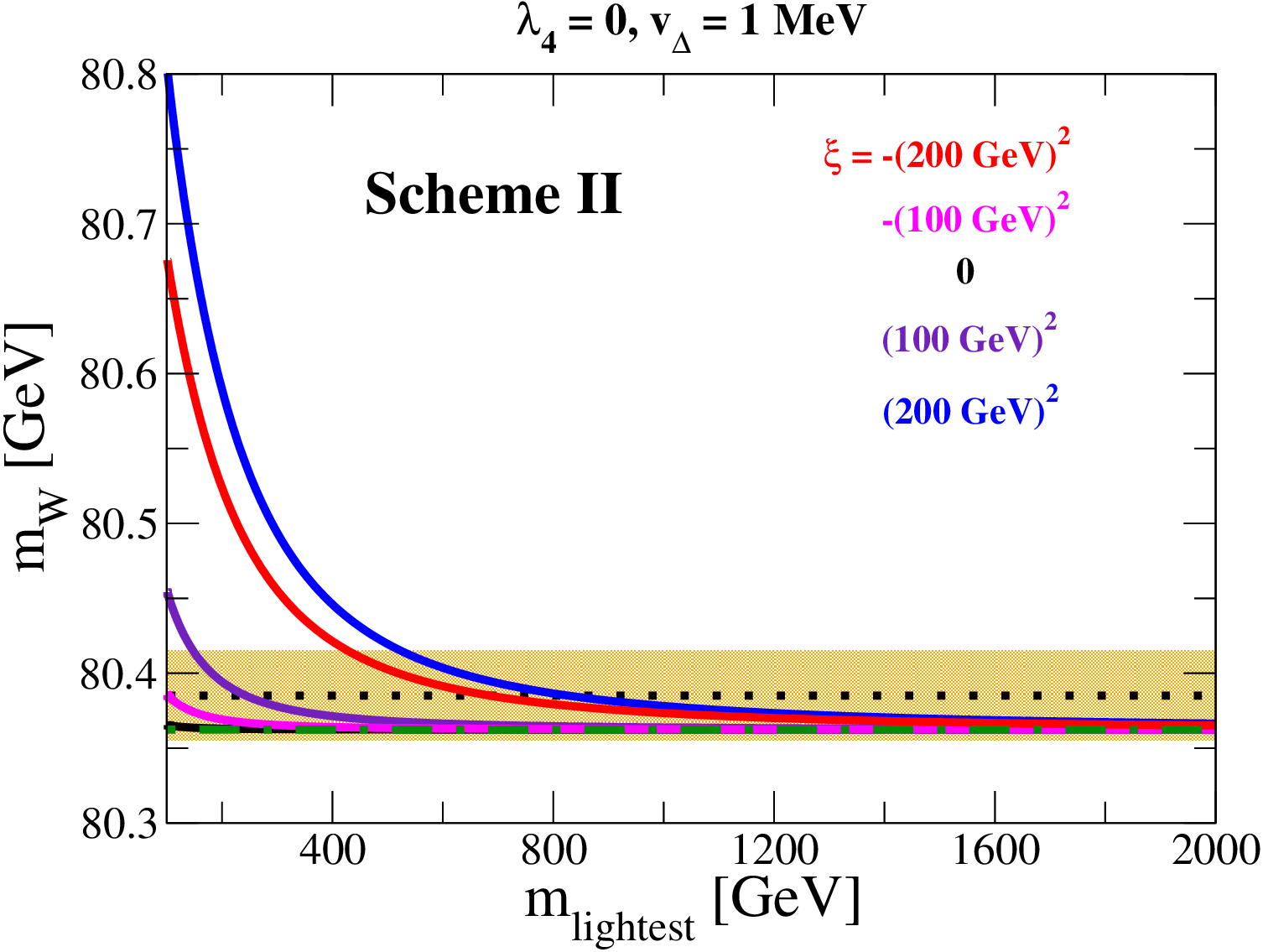}\hspace{3mm}
\includegraphics[width=80mm]{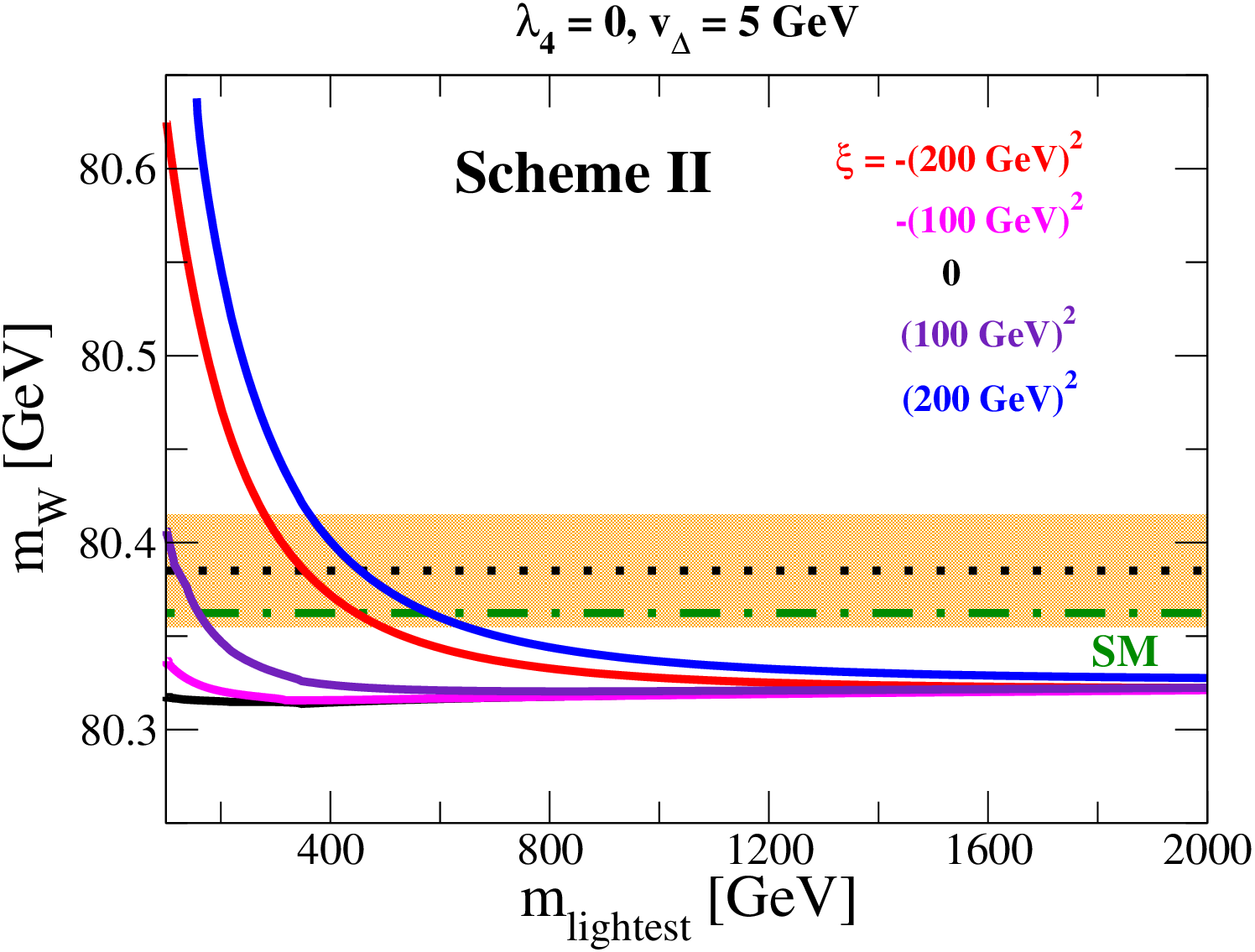}
\caption{Left (Right) figure shows the renormalized $m_W$ calculated in Scheme~II a function of $m_{\text{lightest}}$ 
in the case of $\lambda_4=0$ and $v_\Delta=1$ MeV (5 GeV) for each fixed value of $\xi$. 
The SM prediction is shown as the green dashed line. 
The orange shaded regions are indicated the region within the 2$\sigma$ error bar of the measured W boson mass. }
\label{FIG:mw2}
\end{center}
\end{figure}

In Fig.~\ref{FIG:mw2}, 
the renormalized W boson mass calculated in Scheme~II 
is shown as a function of $m_{\text{lightest}}$ 
in the case with $\lambda_4=0$ for each fixed value of $\xi$. 
In the left (right) plot, the triplet VEV $v_\Delta$ is taken to be 1 MeV and 5 GeV, respectively.
The orange shaded regions are indicated the region within the 2$\sigma$ error bar of the measured W boson mass. 
In the case with $v_\Delta=1$ MeV (left), the prediction of $m_W$ in the large $m_{\text{lightest}}$ is 
close to the SM prediction. 
However, in the case with $v_\Delta=5$ GeV (right), 
the asymptotic value of $m_W$ is not agree with the SM prediction. 

\begin{figure}[t]
\begin{center}
\includegraphics[width=50mm]{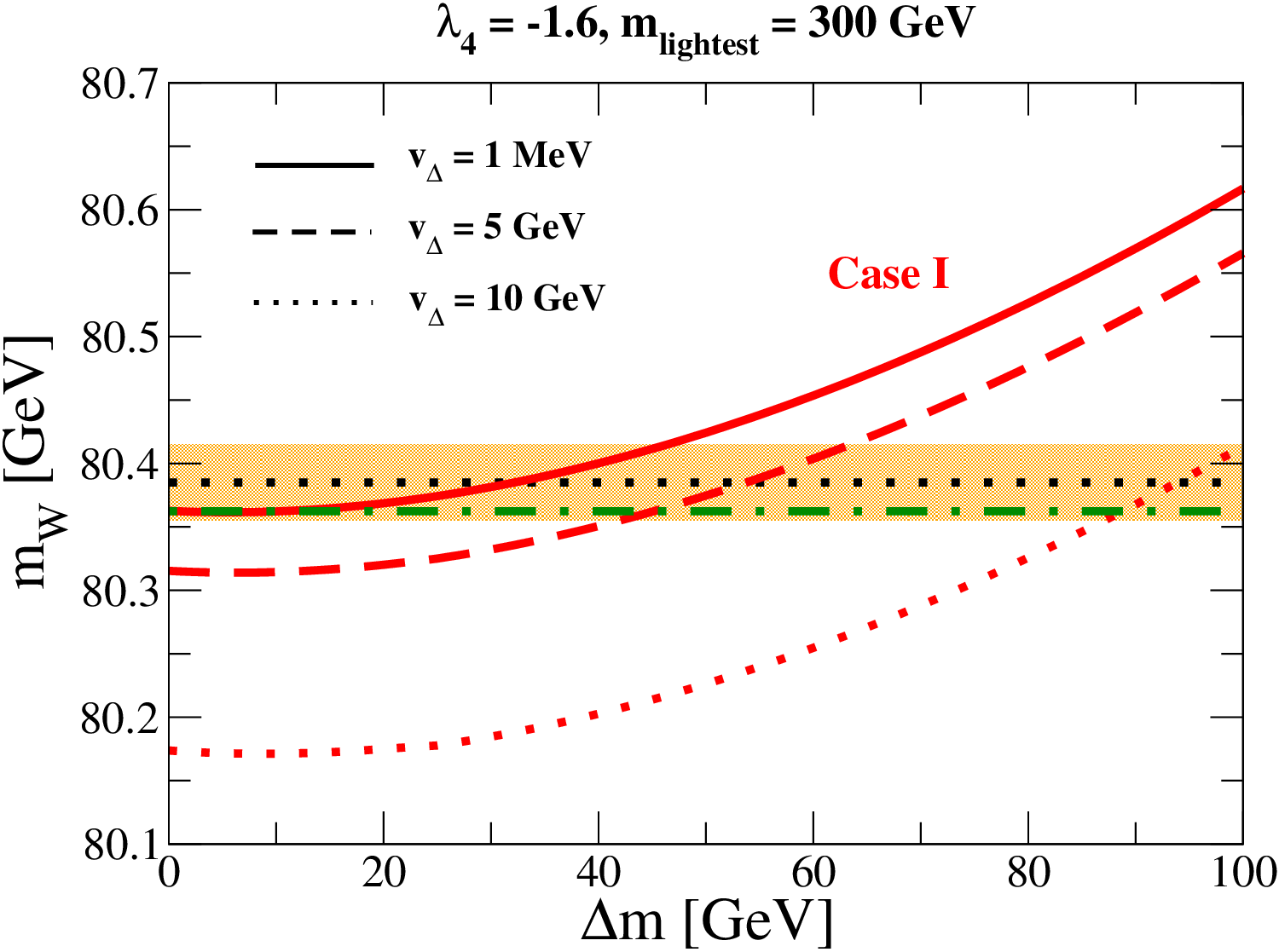}\hspace{3mm}
\includegraphics[width=50mm]{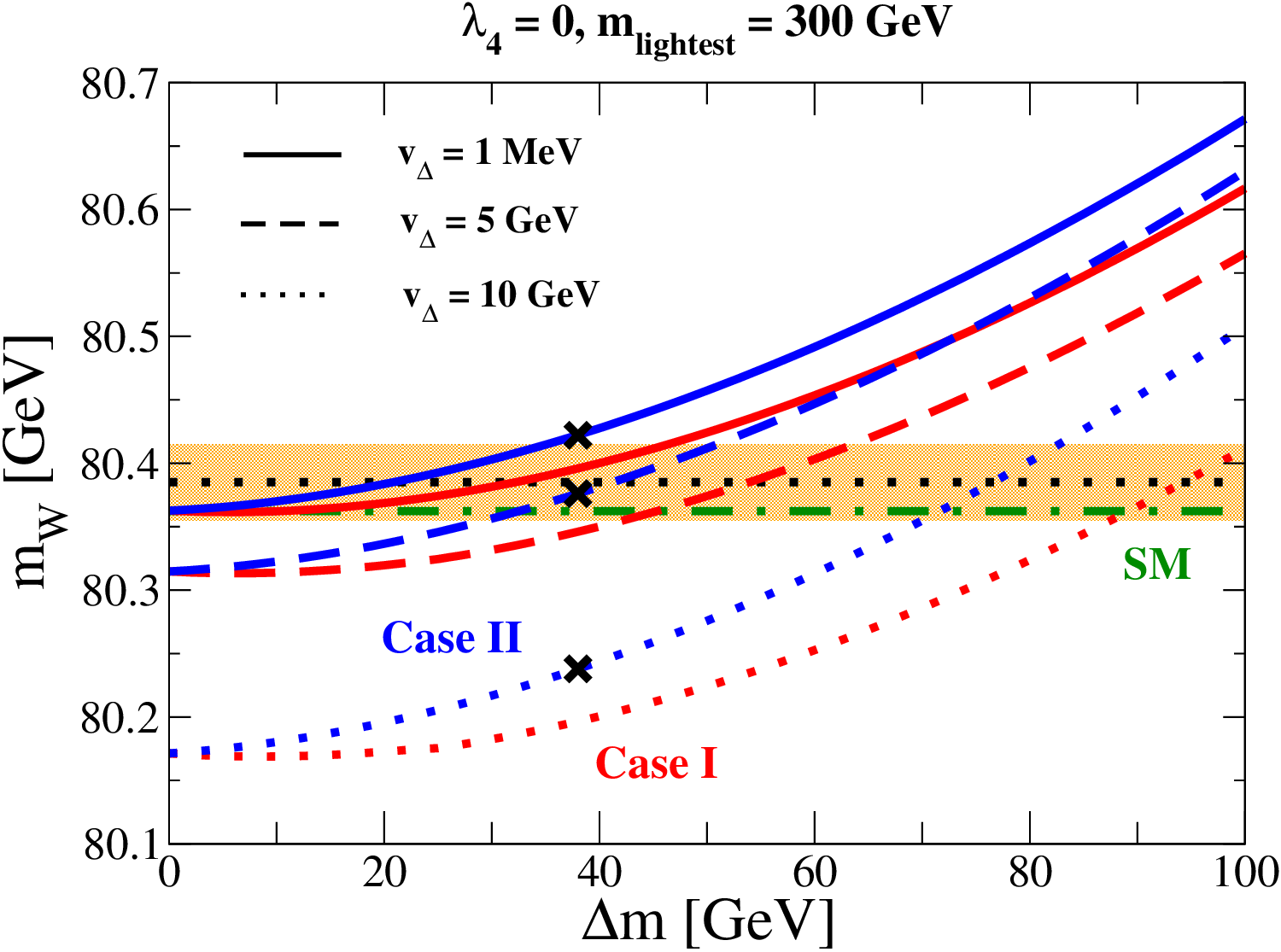}\hspace{3mm}
\includegraphics[width=50mm]{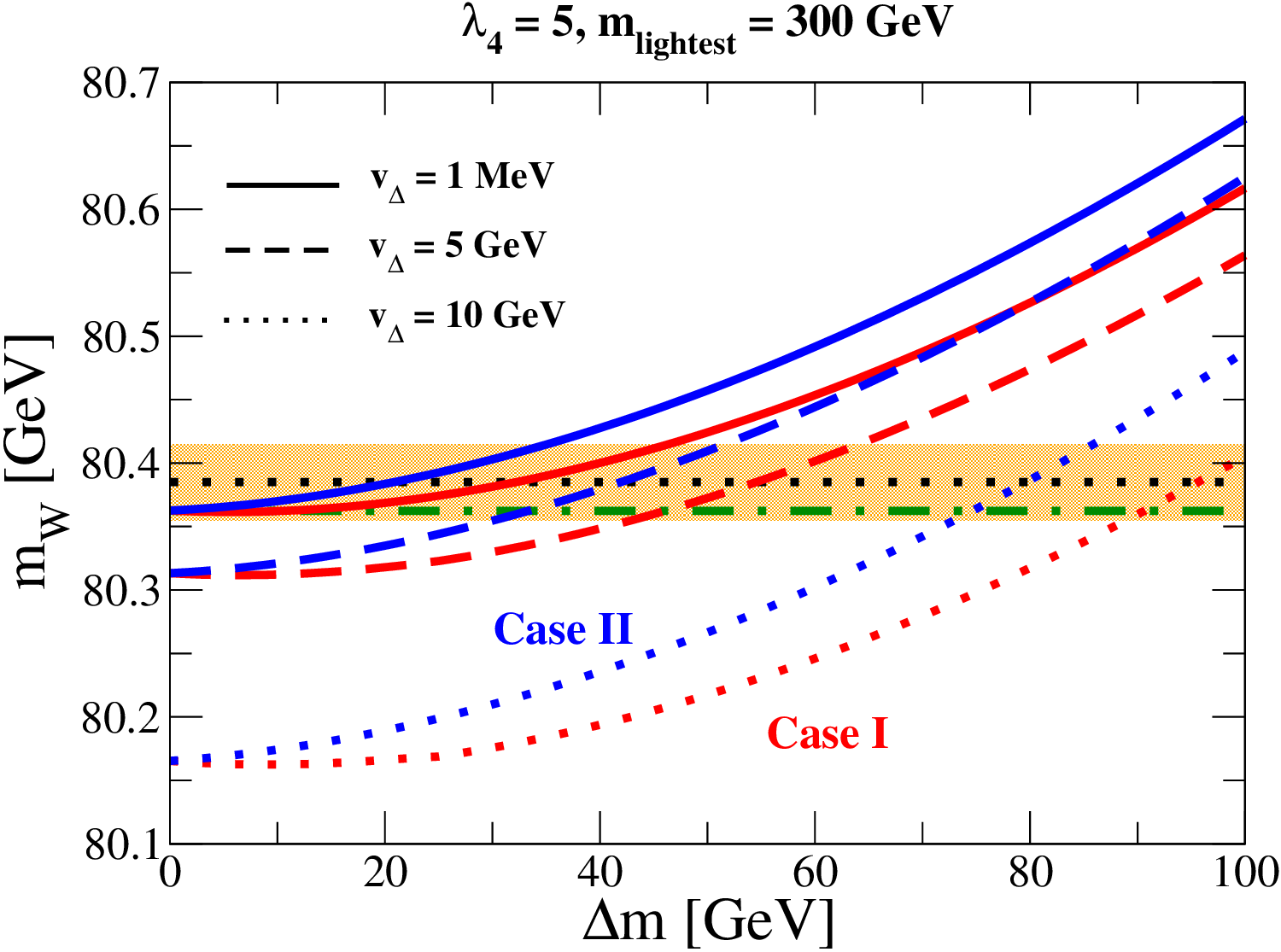}
\caption{The renormalized $m_W$ calculated in Scheme~II as a function of $\Delta m$
for both Case~I ($m_{H^{++}}<m_{H^+}<m_A$)
and Case~II ($m_{H^{++}}>m_{H^+}>m_A$) in the case of $m_{\text{lightest}}=300$ GeV. 
In the left, center and right panels, $\lambda_4$ are taken to be $-1.6$, 0 and 5, respectively. 
The SM prediction is shown as the green dashed line. 
The orange shaded regions are indicated the region within the 2$\sigma$ error bar of the measured W boson mass. 
The solid, dashed and dotted curves respectively show the cases with $v_\Delta=1$ MeV, 5 GeV and 10 GeV. 
The cross marked points are indicated the upper limit of $\Delta m$ from the theoretical bounds which is obtained 
by taking $\lambda_\Delta = 3$. }
\label{FIG:mw3}
\end{center}
\end{figure}

In Fig.~\ref{FIG:mw3}, the renormalized $m_W$ calculated in Scheme~II 
is shown as a function of $\Delta m$ in the case of $m_{\text{lightest}}=300$ GeV. 
In the left, center and right panels, we take $\lambda_4=-1.6$, 0 and 5, respectively. 
In all the figures, the solid, dashed and dotted curves respectively show the case with $v_\Delta=1$ MeV, 5 GeV and 10 GeV. 
In fact, the predictions in the cases with $v_\Delta=1$ MeV and $v_\Delta=1$ GeV are almost the same, so that 
the result in $v_\Delta=1$ MeV can be regarded as that in $v_\Delta \lesssim 1$ GeV. 
It can be seen that the dependence of $\lambda_4$ to $m_W$ is quite small.  
We find that in Case~I (Case~II) the mass difference $\Delta m $ is constrained by the electroweak precision data to be 
$0<\Delta m \lesssim 50$ GeV ($0<\Delta m \lesssim 30\text{ GeV}$) for $v_\Delta \lesssim 1$ GeV, 
$40\text{ GeV}\lesssim \Delta m\lesssim 60$ GeV ($30\text{ GeV}\lesssim \Delta m\lesssim 50$ GeV) for $v_\Delta=5$ GeV 
and $85\text{ GeV}\lesssim \Delta m\lesssim 100$ GeV ($70\text{ GeV}\lesssim \Delta m\lesssim 85$ GeV) 
for $v_\Delta=10$ GeV.

\subsection{Renormalization of the Higgs potential}
Next we discuss the renormalization of the parameters in the Higgs potential.
We here choose the set of
input parameters in the Higgs potential as~\cite{AKKY}
\begin{eqnarray}
v,~\alpha,~\beta,~\beta',~ m_h^2,~ m_H^2,~ m_A^2,~ m_{H^+}^2,~ m_{H^{++}}^2. 
\end{eqnarray}
We work on the mass eigenbasis for the Higgs fields.
Now we define the shift of the parameters as
\begin{align}
  T_\Phi &\to 0 + \delta T_\Phi,~T_\Delta \to 0 + \delta T_\Delta, \\
v &\to v + \delta v,~  \alpha \to \alpha + \delta \alpha,~\beta \to \beta + \delta \beta,~\beta' \to \beta' + \delta \beta'
\\
  m_\varphi^2 &\to m_\varphi^2 + \delta m_\varphi^2,
\end{align}
where $\varphi = h,H,A,H^+$ and $H^{++}$. 
In fact, $\beta$ and $\beta'$ are not independent from each other, so that the counter-terms 
$\delta\beta$ and $\delta\beta'$ are also not independent\footnote{
The counter-terms $\delta \beta$ and $\delta \beta'$ can be written in terms of 
$\delta v$ and $\delta v_\Delta$ as 
$\delta\beta =\frac{v_\Delta}{v}\sqrt{\frac{2}{1-2v_\Delta^2/v^2}}\left(\frac{\delta v_\Delta}{v_\Delta}-\frac{\delta v}{v}\right)$
 and 
$\delta\beta'=\frac{v_\Delta}{v}\frac{2}{(1+2v_\Delta^2/v^2)\sqrt{1-2v_\Delta^2/v^2}}\left(\frac{\delta v_\Delta}{v_\Delta}-\frac{\delta v}{v}\right)$. }. 
We start from the mass eigenstates for the Higgs fields at the tree
level. The wave function renormalization factors are defined as
\begin{align}
  H^{\pm\pm} &\to \left(1 + \frac{1}{2} \delta Z_{H^{++}}
                 \right)H^{\pm\pm},\notag\\
  \left(\begin{array}{c}
        G^\pm \\
        H^\pm \\
        \end{array}\right)
  &\to
  \left(\begin{array}{cc}
        1 + \frac{1}{2} \delta Z_{G^+} &
         \delta \beta + \delta C_{GH} \\
       - \delta \beta + \delta C_{HG} 
         &  1 + \frac{1}{2} \delta Z_{H^+}  \\
        \end{array}\right)
  \left(\begin{array}{c}
        G^\pm \\
        H^\pm \\
        \end{array}\right),\notag\\
  \left(\begin{array}{c}
        G^0 \\
        A \\
        \end{array}\right)
  &\to
  \left(\begin{array}{cc}
        1 + \frac{1}{2} \delta Z_{G^0} &
         \delta \beta' + \delta C_{GA} \\
       - \delta \beta' + \delta C_{AG} 
         &  1 + \frac{1}{2} \delta Z_{A}  \\
        \end{array}\right)
  \left(\begin{array}{c}
        G^0 \\
        A \\
        \end{array}\right), \notag\\
  \left(\begin{array}{c}
        h \\
        H \\
        \end{array}\right)
  &\to
  \left(\begin{array}{cc}
        1 + \frac{1}{2} \delta Z_{h} &
         \delta \alpha + \delta C_{hH} \\
       - \delta \alpha + \delta C_{Hh} 
         &  1 + \frac{1}{2} \delta Z_{H}  \\
        \end{array}\right)
  \left(\begin{array}{c}
        h \\
        H \\
        \end{array}\right). 
\end{align}
Hereafter, we set
$\delta C_{hH}= \delta C_{Hh}$,
$\delta C_{GH}= \delta C_{HG}$ and 
$\delta C_{GA}= \delta C_{AG}$ without loss of generality.
The counter-term of $\delta v$ can be determined by the renormalization of the electroweak 
parameters as
\begin{align}
\frac{\delta v}{v} = 
\frac{1}{2}\left(\frac{\delta m_W^2}{m_W^2}-\frac{\delta\alpha_{\text{em}}}{\alpha_{\text{em}}}
+\frac{\delta \bar{s}_W^2}{\bar{s}_W^2}\right), 
\end{align}
where counter-terms of $\delta m_W^2$ and $\delta\alpha_{\text{em}}$ are given in Eq.~(\ref{count1}) 
and that of $\delta \bar{s}_W^2$ is given in Eq.~(\ref{delsw_2}). 
At the one-loop level, one point functions of $h$ and $H$ are given by 
\begin{align}
\hat{T}_h &= \delta T_\Phi \cos\alpha+ \delta T_\Delta \sin\alpha +
 T_h^{\rm 1PI} , \\ 
\hat{T}_H &= -\delta T_\Phi \sin\alpha+ \delta T_\Delta \cos\alpha + T_H^{\rm 1PI} , 
\end{align}
where $T_h^{\rm 1PI}$ and $T_H^{\rm 1PI}$ are contributions of
one-particle-irreducible diagrams. 
The condition of vanishing the tadpoles at the one-loop level provides
\begin{align}
  \delta T_\Phi   &=    -T_h^{\rm 1PI} \cos\alpha +  T_H^{\rm
   1PI} \sin\alpha, \\ 
  \delta T_\Delta &=   -T_h^{\rm 1PI} \sin\alpha - T_H^{\rm 1PI} \cos\alpha . 
 \end{align}

Similar to the gauge boson two point functions, 
renormalized two point functions for scalar bosons can be expressed as $\hat{\Pi}_{\phi\phi}[p^2]$ 
and 1PI diagram contributions can also be denoted by $\Pi_{\phi\phi}^{\text{1PI}}(p^2)$. 
Derivatives for those functions can be defined as 
$\hat{\Pi}_{\phi\phi}^\prime[m^2]=\frac{d}{dp^2}\hat{\Pi}_{\phi\phi}(p^2)\big|_{p^2=m^2}$
and 
$\Pi_{\phi\phi}^{\text{1PI}^\prime}[m^2]=\frac{d}{dp^2}\Pi_{\phi\phi}^{\text{1PI}}(p^2)\big|_{p^2=m^2}$.

The renormalized scalar boson two point functions are given at the
one-loop level by
\begin{align}
 \hat{\Pi}_{H^{++}H^{--}}[p^2] &= (p^2 - m_{H^{++}}^2)\delta Z_{H^{++}}-\delta m_{H^{++}}^2
   +\frac{\sqrt{2}}{s_\beta}\frac{\delta T_\Delta}{v}+ \Pi^{\rm 1PI}_{H^{++}H^{--}}(p^2),  \\
 \hat{\Pi}_{H^+H^-}[p^2] &= (p^2 - m_{H^+}^2)\delta Z_{H^+}-\delta m_{H^+}^2
    + \frac{s_\beta^2}{c_\beta}
    \frac{\delta T_\Phi}{v}
    + \frac{\sqrt{2}c_\beta^2}{s_\beta}
    \frac{\delta T_\Delta}{v}
 + \Pi^{\rm 1PI}_{H^+H^-}(p^2),  \\
 \hat{\Pi}_{G^+G^-}[p^2] &=  p^2\delta Z_{G^+}
  + \frac{c_\beta \delta T_\Phi  + \sqrt{2}s_\beta  \delta T_\Delta }{v}
+ \Pi^{\rm 1PI}_{G^+G^-}(p^2),  \\  
 \hat{\Pi}_{AA}[p^2] &= (p^2 - m_{A}^2)\delta Z_A- \delta m_A^2
    + \frac{s_{\beta'}^2}{c_\beta}
    \frac{\delta T_\Phi}{v}
    + \frac{\sqrt{2}c_{\beta'}^2}{s_\beta}
    \frac{\delta T_\Delta}{v} + \Pi^{\rm 1PI}_{AA}(p^2),\\ 
 \hat{\Pi}_{GG}[p^2] &= p^2\delta Z_{G^0}
+\frac{c_{\beta'}^2}{c_\beta}\frac{\delta T_\Phi}{v}+\frac{s_{\beta'}^2}{s_\beta}\frac{\delta T_\Delta}{v}
  + \Pi^{\rm1PI}_{GG}(p^2),  \\  
 \hat{\Pi}_{HH}[p^2] &=  (p^2 - m_H^2)\delta Z_H-\delta m_H^2 +\frac{s_\alpha^2}{c_\beta}\frac{\delta T_\Phi }{v}
 +\frac{\sqrt{2}c_\alpha^2}{s_\beta} \frac{\delta T_\Delta }{v}
  +\Pi^{\rm1PI}_{HH}(p^2),  \\
 \hat{\Pi}_{hh}[p^2] &= (p^2 - m_h^2)\delta Z_h-\delta m_h^2 +\frac{c_\alpha^2}{c_\beta}\frac{\delta T_\Phi }{v}
 +\frac{\sqrt{2}s_\alpha^2}{s_\beta}\frac{\delta T_\Delta }{v} 
  +\Pi^{\rm1PI}_{hh}(p^2). 
\end{align}
The renormalized two point functions for the scalar boson mixing are given by
\begin{align}
 \hat{\Pi}_{H^+G^-}[p^2] &= \delta C_{HG} (2 p^2 - m_{H^+}^2)
  - \frac{s_\beta \delta T_\Phi - \sqrt{2} c_\beta\delta
  T_\Delta}{v}
  + m_{H^+}^2 \delta \beta  + \Pi^{\rm 1PI}_{G^+H^-}(p^2),\\
 \hat{\Pi}_{AG}[p^2] &=  \delta C_{AG} (2 p^2 - m_A^2)
-\frac{s_{\beta'}^2}{\sqrt{2}s_\beta}\frac{\delta T_\Phi}{v}
+\frac{2c_{\beta'}^2}{c_\beta}\frac{\delta T_\Delta}{v}
+ m_A^2 \delta \beta' + \Pi^{\rm 1PI}_{AG}(p^2),  \label{Gam_GA}\\
 \hat{\Pi}_{Hh}[p^2] &= \frac{c_\alpha s_\alpha}{c_\beta s_\beta}
\frac{\sqrt{2}c_\beta\delta T_\Delta-s_\beta\delta T_\Phi}{v}
  -\delta\alpha (m_h^2 - m_H^2) \notag\\
&+\delta C_{Hh} (2 p^2 - m_h^2 - m_H^2)
+\Pi^{\rm 1PI}_{Hh}(p^2),   
\end{align}
The counter-terms of the doubly-charged Higgs boson mass $\delta m_{H^{++}}^2$ 
and its wave function renormalization 
factor $\delta Z_{H^{++}}$ are determined by the following renormalization conditions: 
\begin{align}
\hat{\Pi}_{H^{++}H^{--}}[m_{H^{++}}^2] = 0,\quad \hat{\Pi}_{H^{++}H^{--}}^\prime[m_{H^{++}}^2]  = 0,  
\end{align}
which yield
\begin{align}
\delta m_{H^{++}}^2 &= \frac{\sqrt{2}\delta T_\Delta}{vs_\beta} +
 \Pi_{H^{++}H^{--}}^{\rm 1PI}(m_{H^{++}}^2),\quad  
\delta Z_{H^{++}} = - \Pi_{H^{++}H^{--}}^{\rm 1PI^\prime}(m_{H^{++}}^2).  \label{Hpp_count}
\end{align}
The five parameters related to the CP-odd scalar states 
($\delta m_{A}^2$, $\delta Z_{G^0}$, $\delta Z_{A}$, $\delta C_{GA}$ and $\delta\beta'$)
are determined by imposing the following five renormalization conditions
\begin{align}
&  \hat{\Pi}_{AA}[m_A^2] = 0, \quad  \hat{\Pi}_{AA}^\prime[m_A^2] = 0,\\
&\hat{\Pi}_{GG}^\prime[0] = 0, \quad 
\hat{\Pi}_{AG}[0] = 0,  \quad
\hat{\Pi}_{AG}[m_A^2]=0. 
\end{align}
by which we obtain 
\begin{align}
 \delta m_A^2 &=
   \frac{s_{\beta'}^2}{c_\beta}
    \frac{\delta T_\Phi}{v}
    + \frac{\sqrt{2}c_{\beta'}^2}{s_\beta}
    \frac{\delta T_\Delta}{v} + \Pi^{\rm 1PI}_{AA}(m_A^2),\\
 \delta Z_{G^0} &= - \Pi_{GG}^{\rm 1PI^\prime}(0), \quad \delta Z_{A} = - \Pi_{AA}^{\rm 1PI^\prime}(m_A^2),\\ 
\delta C_{AG} &= \frac{1}{2m_A^2}\left[\Pi_{AG}^{\text{1PI}}(0)-\Pi_{AG}^{\text{1PI}}(m_A^2)\right],\\
\delta \beta' &= -\frac{1}{2m_A^2}\left[\Pi_{AG}^{\text{1PI}}(0)
+\Pi_{AG}^{\text{1PI}}(m_A^2)
-\frac{\sqrt{2}s_{\beta'}^2}{s_\beta}\frac{\delta T_\Phi}{v}
-\frac{4c_{\beta'}^2}{c_\beta}\frac{\delta T_\Delta}{v}\right].  
\end{align}
The four counter-terms related to the singly-charged Higgs boson 
($\delta m_{H^+}^2$, $\delta Z_{G^+}$, $\delta Z_{H^+}$ and $\delta C_{GH}$)
are determined by imposing the following four renormalization conditions
\begin{align}
&  \hat{\Pi}_{H^+H^-}[m_{H^+}^2] = 0, \quad   \hat{\Pi}_{G^+G^-}^\prime[0] = 0, \\ 
&  \hat{\Pi}_{H^+H^-}^\prime[m_{H^+}^2]= 0, \quad  \hat{\Pi}_{HG}[0] = 0,  
\end{align}
by which we obtain
\begin{align}
 \delta m_{H^+}^2 &= \frac{s_\beta^2}{c_\beta}
    \frac{\delta T_\Phi}{v}
    + \frac{\sqrt{2}c_\beta^2}{s_\beta}
    \frac{\delta T_\Delta}{v} + \Pi^{\rm1PI}_{H^+H^-}(m_{H^+}^2),
    \label{eq:delmchsq}\\ 
\delta Z_{G^+} &= - \Pi^{\rm 1PI^\prime}_{G^+G^-}(0), \quad
\delta Z_{H^+} = - \Pi^{\rm 1PI^\prime}_{H^+H^-}(m_{H^+}^2), \\
\delta C_{HG} &=  \delta\beta+\frac{1}{m_{H^+}^2} \left[
\Pi_{H^+G^-}^{\rm 1PI}(0)+
\frac{-s_\beta \delta T_\Phi+\sqrt{2}c_\beta \delta T_\Delta}{v}\right],
\label{delC_GH}
\end{align}
where $\delta \beta$ is determined through $\delta \beta'$ as
\begin{align}
\delta \beta = \frac{1+s_\beta^2}{\sqrt{2}}\delta \beta'.
\end{align}
Finally, the six parameters related to the CP-even Higgs states 
($\delta \alpha$, $\delta m_h^2$, $\delta m_H^2$,
$\delta Z_{h}$, $\delta Z_H$ and $\delta C_{Hh}$)
are determined by imposing the following six renormalization conditions
\begin{align}
&  \hat{\Pi}_{hh}[m_h^2] = 0,\quad \hat{\Pi}_{hh}^\prime[m_h^2] = 0, \\
&  \hat{\Pi}_{HH}[m_H^2] = 0,\quad \hat{\Pi}_{HH}^\prime[m_H^2] = 0, \\ 
&  \hat{\Pi}_{Hh}[m_h^2] = 0,\quad   \hat{\Pi}_{Hh}[m_H^2] = 0,
\end{align}
by which we obtain
\begin{align}
 \delta m_h^2 &= \frac{\delta T_\Phi}{v} \frac{c^2_\alpha}{c_\beta}
               +\frac{\sqrt{2}\delta T_\Delta}{v} \frac{s^2_\alpha}{s_\beta}
               + \Pi_{hh}^{\rm 1PI}(m_h^2),\quad 
 \delta m_H^2 = \frac{\delta T_\Phi}{v} \frac{s^2_\alpha}{c_\beta}
               +\frac{\sqrt{2}\delta T_\Delta}{v} \frac{c^2_\alpha}{s_\beta}
               + \Pi_{HH}^{\rm 1PI}(m_H^2), \\
\delta Z_h &= -  \Pi_{hh}^{\rm 1PI^\prime}(m_h^2), \quad
\delta Z_H = - \Pi_{HH}^{\rm 1PI^\prime}(m_H^2), \\ 
\delta \alpha &= \frac{1}{2(m_h^2-m_H^2)}
\left[\Pi_{Hh}^{\rm 1PI}(m_h^2) + \Pi_{Hh}^{\rm 1PI}(m_H^2)
- \frac{2s_\alpha c_\alpha}{ s_\beta c_\beta} \left(\frac{\delta T_\Phi}{v}s_\beta
               - \frac{\sqrt{2}\delta T_\Delta}{v}c_\beta\right)\right] ,\\
\delta C_{Hh} &= 
\frac{1}{2(m_h^2-m_H^2)}
\left[ \Pi_{Hh}^{\rm 1PI}(m_H^2)-\Pi_{Hh}^{\rm 1PI}(m_h^2)
\right]. 
\end{align}

\begin{figure}[t]
\begin{center}
\includegraphics[width=80mm]{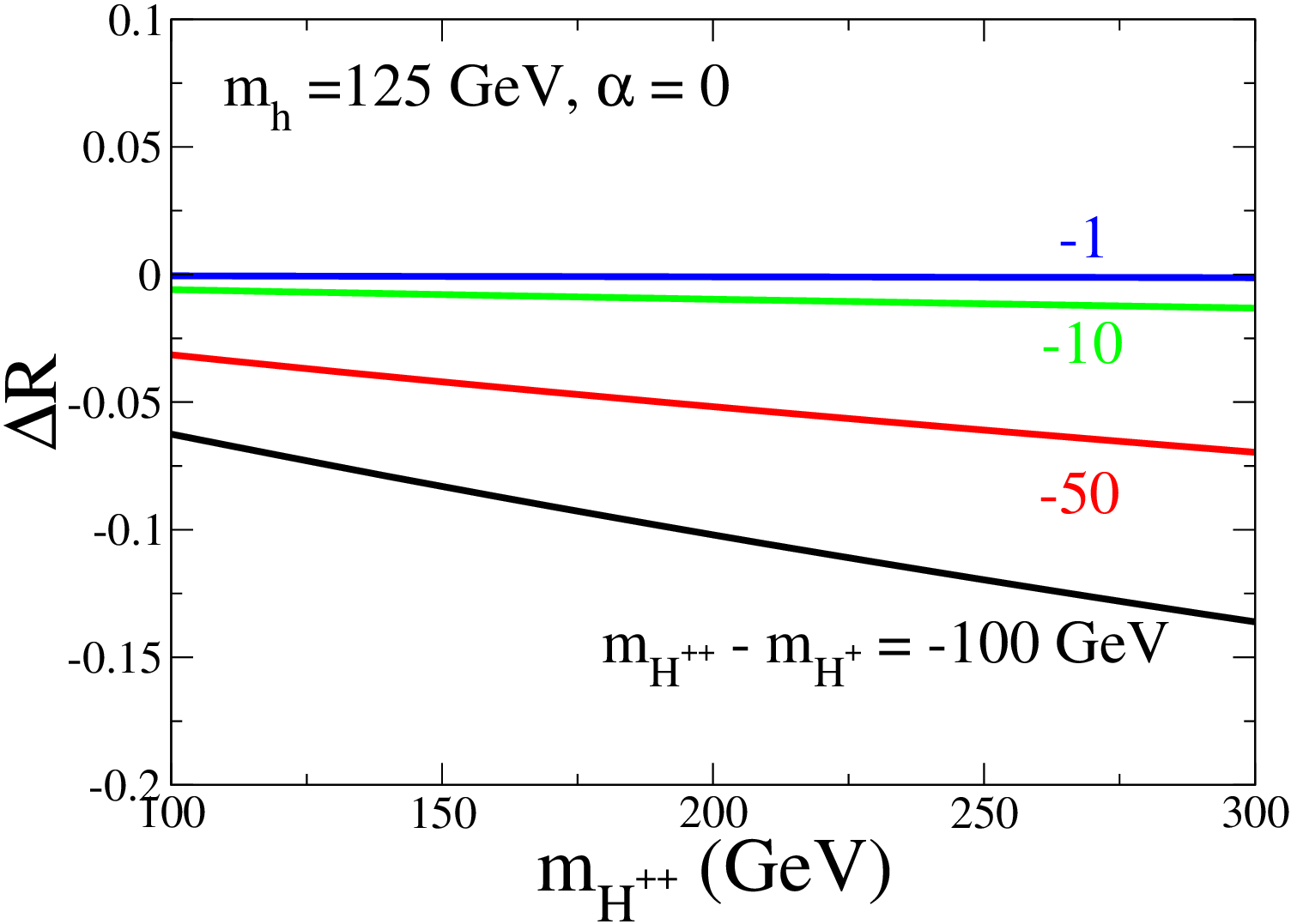}\hspace{3mm}
\includegraphics[width=80mm]{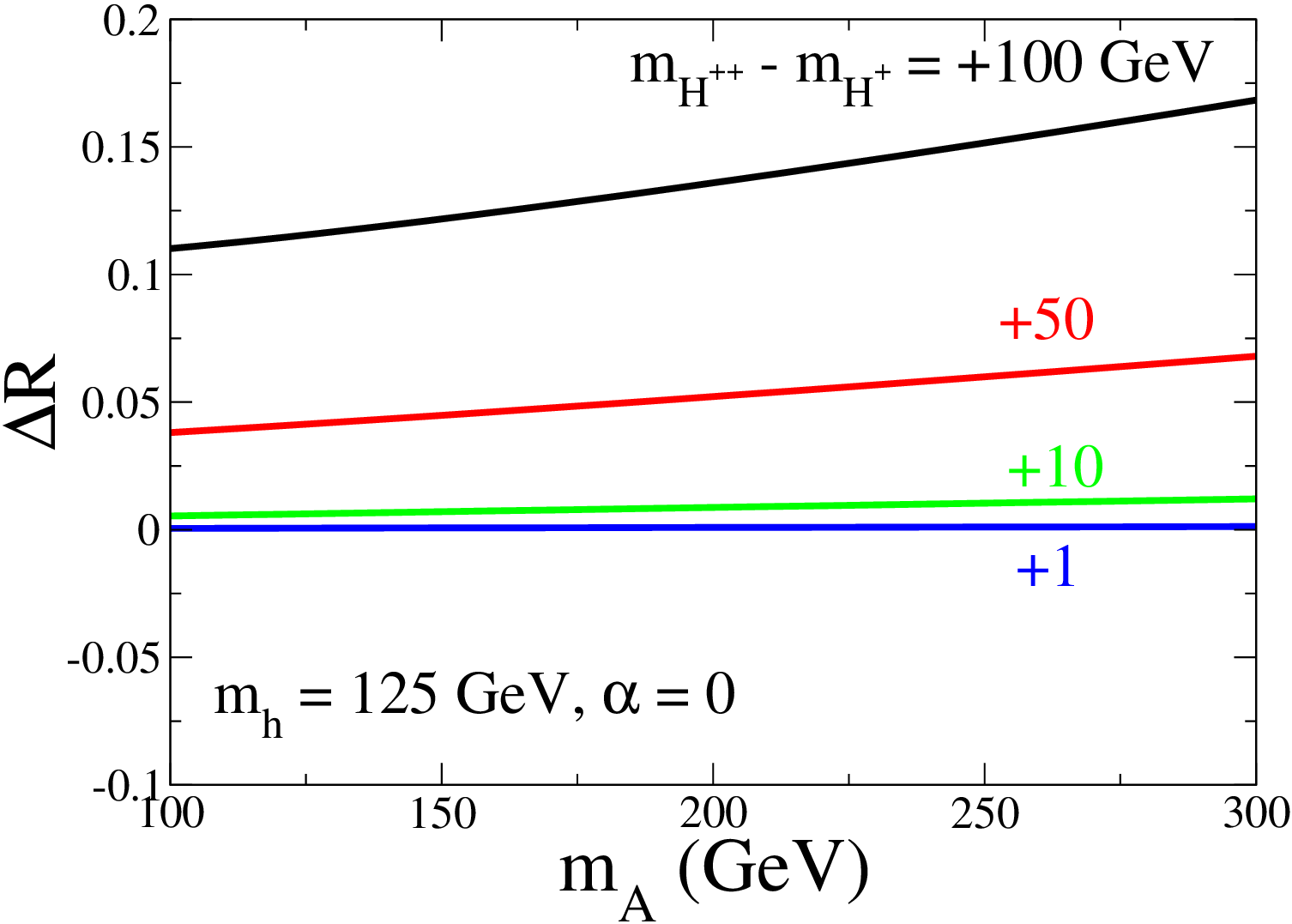}
\caption{$\Delta R$ is shown as a function of the lightest triplet-like Higgs boson mass in the case of 
$\alpha=0$ and $m_h=125$ GeV for each fixed value of $m_{H^{++}}-m_{H^+}$~\cite{AKKY}. 
The left (right) panel shows the results in Case~I (Case~II). }
\label{FIG:DeltaR}
\end{center}
\end{figure}

In the limit of $v_\Delta/v \to 0$, the mass of $A$ is no more independent parameter, which is determined by the 
masses of $H^{\pm\pm}$ and $H^\pm$. 
In this limit, the CP-odd Higgs boson mass is expressed at the tree level as
\begin{align}
(m_A^2)_{\text{tree}}\equiv \lim_{v_\Delta/v\to 0}m_A^2=
2m_{H^{+}}^2-m_{H^{++}}^2. 
\end{align}
The renormalized pole mass of $A$, which has been discussed in Ref.~\cite{AKKY},  
can be defined by the following equations
\begin{align}
\lim_{v_\Delta/v\to 0}\hat{\Pi}_{AA}[p^2=(m_A^2)_{\text{pole}}]=0.\label{mApole}
\end{align}
From Eq.~(\ref{mApole}), we obtain
\begin{align}
(m_A^2)_{\text{pole}}&\simeq(m_A^2)_{\text{tree}}+\frac{1}{1+\delta Z_A}
\left[\delta m_A^2-\frac{\delta T_\Delta}{v_\Delta}-\Pi_{AA}^{\text{1PI}}[(m_A^2)_{\text{tree}}]\right],
\end{align}
where we use $\Pi_{AA}[(m_A^2)_{\text{pole}}]\simeq\Pi_{AA}[(m_A^2)_{\text{tree}}]$. 
The counter-term $\delta m_A^2$ is not independent parameter in the limit of $v_\Delta/v\to0$, 
but they can be given by $\delta m_{A}^2=2\delta m_{H^+}^2-\delta m_{H^{++}}^2$. 
By using Eqs.~(\ref{Hpp_count}) and (\ref{eq:delmchsq}), we obtain
\begin{align}
(m_A^2)_{\text{pole}}&\simeq(m_A^2)_{\text{tree}}+
\left[2\Pi_{H^+H^-}^{\text{1PI}}(m_{H^+}^2)-\Pi_{H^{++}H^{--}}^{\text{1PI}}(m_{H^{++}}^2)-\Pi_{AA}^{\text{1PI}}[(m_A^2)_{\text{tree}}]\right].
\end{align}
Above the equation indicates that the tree level mass relations among the triplet-like Higgs bosons which are 
written in Eqs.~(\ref{eq:mass_relation1}) and (\ref{eq:mass_relation2}) can be deviated by the effects of the radiative correction. 
The magnitude of this deviation can be parameterized as
\begin{align}
\Delta R= \frac{m_{H^{++}}^2-m_{H^+}^2}{m_{H^{+}}^2-(m_A^2)_{\text{pole}}}-1. 
\end{align}
In Ref.~{\cite{AKKY}}, $\Delta R$ has been evaluated numerically in the case of $\alpha=0$ and $v_\Delta/v\to 0$ as shown in 
Fig.~\ref{FIG:DeltaR}. 

\section{Higgs couplings at the one-loop level}

In this section, we discuss the SM-like Higgs boson $h$ couplings with the gauge bosons 
($\gamma\gamma$, $W^+W^-$ and $ZZ$) and the Higgs selfcoupling $hhh$ at the one-loop level in the favored 
parameter regions by the unitarity bound, the vacuum stability bound and by the 
measured W boson mass discussed in previous sections.

\subsection{Higgs to the diphoton decay}

First, we discuss the decay of the Higgs to diphoton channel: $h\to\gamma\gamma$, which is important in the Higgs boson search at the LHC. 
The current experimental value of the signal strength for the diphoton mode is $1.8\pm0.5$ at the ATLAS~\cite{Higgs_ATLAS} and
$1.6\pm0.4$ at the CMS~\cite{Higgs_CMS}. 
This process is induced at the one-loop level, because the Higgs boson does not couple to the photon at the tree level. 
Thus, the decay of $h\to\gamma\gamma$ is sensitive to effects of new charged particle which can couple to the Higgs boson. 
In the HTM, the doubly-charged Higgs boson $H^{\pm\pm}$ and the singly-charged Higgs boson $H^\pm$ 
can contribute to the Higgs to diphoton decay.  
In particular, 
the contribution from the $H^{\pm\pm}$ loop to the $h\to \gamma\gamma$
is quite important compared to that from $H^{\pm}$, because $H^{\pm\pm}$ 
contribution is roughly 4 times larger than that from $H^{\pm}$ contribution at the amplitude level. 

The decay rate of $h\to \gamma\gamma$ is calculated at the one-loop level by
\begin{align}
\Gamma(h\to \gamma\gamma)&=\frac{G_F\alpha_{\text{em}}^2m_h^3}{128\sqrt{2}\pi^3}
\Bigg|-2\frac{c_\alpha}{c_\beta}\sum_fN_f^cQ_f^2\tau_f[1+(1-\tau_f)f(\tau_f)]\notag\\
&+(c_\beta c_\alpha +\sqrt{2}s_\alpha s_\beta)[2+3\tau_W+3\tau_W(2-\tau_W)f(\tau_W)]\notag\\
&-Q_{H^{++}}^2\frac{2v\lambda_{H^{++}H^{--}h}}{m_h^2}[1-\tau_{H^{++}}f(\tau_{H^{++}})]
-Q_{H^{+}}^2\frac{2v\lambda_{H^{+}H^{-}h}}{m_h^2}[1-\tau_{H^{+}}f(\tau_{H^{+}})]\Bigg|^2, \label{hgg}
\end{align}
where the function $f(x)$ is given by 
\begin{align}
f(x)=\left\{
\begin{array}{c}
[\arcsin(1/\sqrt{x})]^2, \quad \text{if }x\geq 1,\\
-\frac{1}{4}[\ln \frac{1+\sqrt{1-x}}{1-\sqrt{1-x}}-i\pi]^2, \quad \text{if }x< 1
\end{array}\right.. 
\end{align}
In Eq.~({\ref{hgg}}),  $Q_F$ is the electric charge of the field $F$, 
$N_f^c$ is the color factor and $\tau_F= 4m_F^2/m_h^2$.  
In the HTM, the coupling constants $\lambda_{ H^{++}H^{--}h}$ and $\lambda_{H^+H^-h}$ are given without 
any approximation in Eqs.~(\ref{Eq:hHppHmm}) and (\ref{Eq:hHpHm}), respectively. 
These couplings can be expressed quite a simple form by neglecting the terms proportional to $v_\Delta$: 
\begin{align}
\lambda_{H^{++}H^{--}h} &\simeq -v\lambda_4,\quad
\lambda_{H^{+}H^{-}h} \simeq -\frac{v}{2}(2\lambda_4+\lambda_5).   \label{couplings}
\end{align}
It is well known that 
the W boson loop contribution to the $h\to \gamma\gamma$ decay rate is dominant 
compared to the top quark loop contribution in the SM, so that when a new physics effect to the amplitude of 
the $h\to \gamma\gamma$ process has the same sign of the W-loop contribution, then 
the decay rate is enhanced compared with the SM prediction. 
In the HTM, when the sign of the coupling $\lambda_{H^{++}H^{--}h}$ is positive (negative), then the $H^{\pm\pm}$ loop 
contribution has the same (opposite) sign of the W loop contribution, which 
can be achieved by taking $\lambda_4$ to be a negative (positive) value~\cite{hgg_HTM}. 
From Eq.~(\ref{couplings}), it can be seen that the $\lambda_5$ coupling only affects to the singly-charged Higgs 
boson coupling with $h$: $\lambda_{H^+H^-h}$, so that the $h\to \gamma\gamma$ decay rate is not sensitive to the 
magnitude of $\lambda_5$. 
In other words, the mass difference among the triplet-like Higgs boson is not so important in the $h\to \gamma\gamma$ decay process 
as long as we keep a fixed value of $m_{H^{++}}$. 

\begin{figure}[t]
\begin{center}
\includegraphics[width=80mm]{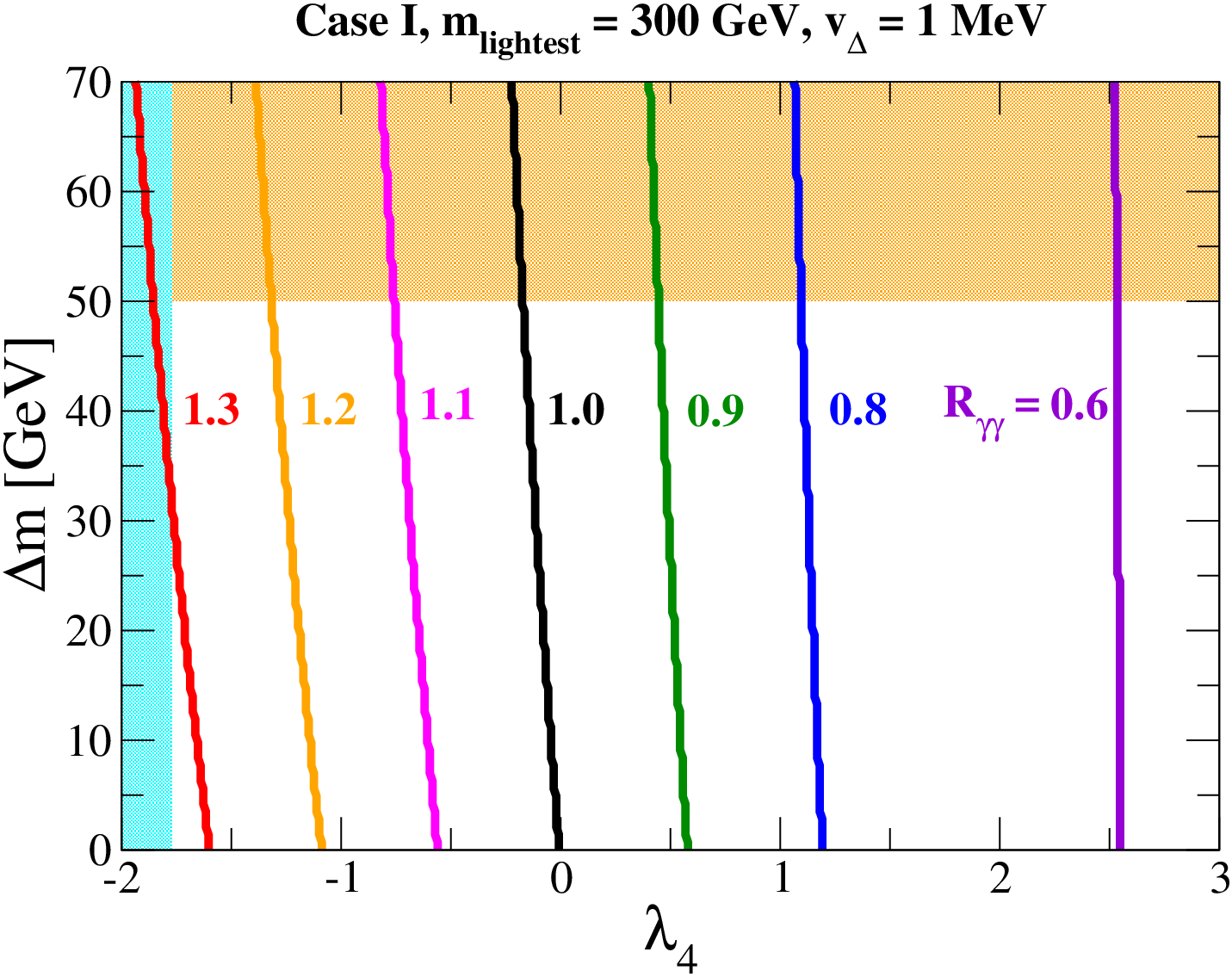}\hspace{3mm}
\includegraphics[width=80mm]{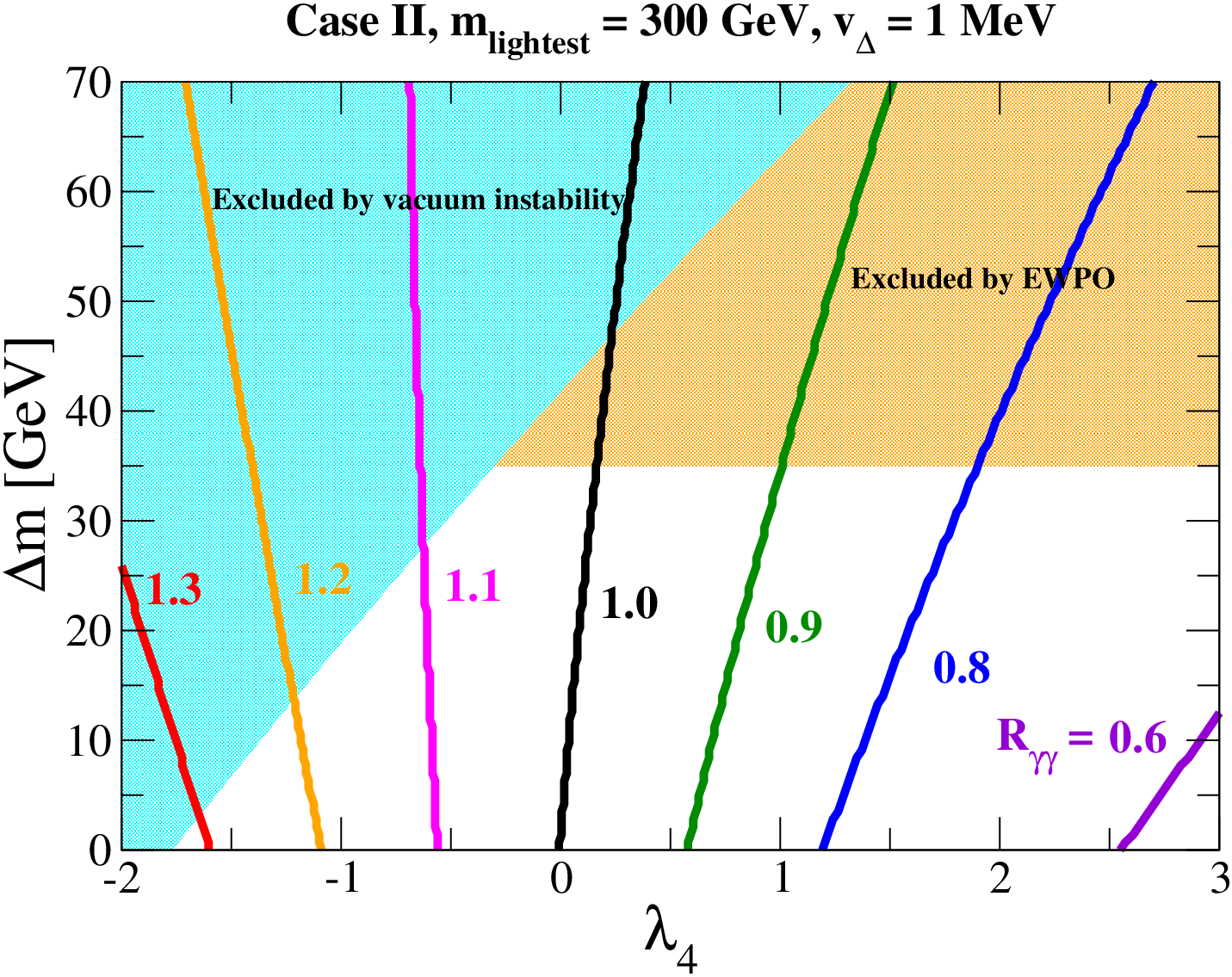}
\caption{Contour plots of $R_{\gamma\gamma}$ for $v_\Delta=1$ MeV and $m_{\text{lightest}}=300$ GeV 
in the $\lambda_4$-$\Delta m$ plane. 
The left panel (right panel) shows the result in Case~I (Case~II).  
The blue and orange shaded regions are excluded by the vacuum stability bound and the measured $m_W$ data, respectively. 
}
\label{FIG:hgg}
\end{center}
\end{figure}

In this subsection, we study the deviation of the $h\to\gamma\gamma$ event rate in the HTM 
from that in the SM taking into account 
the constraint from the perturbative unitarity, the vacuum stability and the electroweak precision data. 
We also investigate the correlation between the $h\to\gamma\gamma$ decay rate and the $hhh$, $hWW$ and $hZZ$ couplings at the 
one-loop level. 
To compare the Higgs to diphoton event rate from the SM prediction, we define 
\begin{align}
R_{\gamma\gamma}\equiv 
\frac{\sigma(gg\to h)_{\text{HTM}}\times 
\text{BR}(h\to \gamma\gamma)_{\text{HTM}}}{\sigma(gg\to h)_{\text{SM}}\times \text{BR}(h\to \gamma\gamma)_{\text{SM}}}, 
\end{align}
where $\sigma(gg\to h)_{\text{model}}$ is the cross section of the gluon fusion process, and 
$\text{BR}(h\to \gamma\gamma)_{\text{model}}$ is the branching fraction of the $h\to \gamma\gamma$ mode in a model. 
In fact, the ratio of the cross section $\sigma(gg\to h)_{\text{HTM}}/\sigma(gg\to h)_{\text{SM}}$ can be 
replaced by the factor $c_\alpha^2/c_\beta^2$.

In Fig.~\ref{FIG:hgg}, we show the contour plots of $R_{\gamma\gamma}$ for $v_\Delta=1$ MeV and $m_{\text{lightest}}=300$ GeV
in the $\lambda_4$-$\Delta m$ plane. 
The left panel (right panel) shows the result in Case~I (Case~II).  
The blue and orange shaded regions are those excluded by the vacuum stability bound (assuming $\lambda_\Delta=3$) 
and the measured $m_W$ data, respectively. 
In both cases, $R_{\gamma\gamma}$ can be greater (smaller) than 1 for negative (positive) values of $\lambda_4$. 
In Case I, no large $\Delta m$ dependence appears, while in Case II $R_{\gamma\gamma}$ slightly depends on $\Delta m$ 
due to the larger values of $m_{H^{++}}$ which affect $R_{\gamma\gamma}$ via $\Delta m$. 
Under the constraint of the vacuum stability and the electroweak precision observable $m_W$, larger 
$\Delta m$ can be allowed in Case I than in Case II. 
We find that predicted values of $R_{\gamma\gamma}$ are about 1.3 (about 0.6) 
in this case when $\lambda_4$ is about $-1.7$ (about $3$) in both Case~I and Case~II.

\subsection{Renormalized $hVV$ coupling}

The most general form factors of the $hVV$ coupling ($V=W^\pm$ or $Z$) can be written as 
\begin{align}
M_{hVV}^{\mu\nu}=M_1^{hVV}g^{\mu\nu}+M_2^{hVV}\frac{p_1^\mu p_2^\nu}{m_V^2}
+iM_3^{hVV}\epsilon^{\mu\nu\rho\sigma}\frac{p_{1\rho} p_{2\sigma}}{m_V^2}, \label{form_factor}
\end{align}
where $m_V$ is the mass of the gauge boson $V$, $p_1$ and $p_2$ are the incoming momenta of $V$. 
The renormalized form factors are given by 
\begin{align}
M_i^{hVV}=M_{i,\text{tree}}^{hVV}+\delta M_i^{hVV}+M_{i,\text{1PI}}^{hVV},\quad (i=1-3).
\end{align}
The tree-level contributions for these form factors are 
\begin{align}
M_{1,\text{tree}}^{hZZ}&=\frac{2m_Z^2}{v^2+2v_\Delta^2}(v_\phi c_\alpha+4v_\Delta s_\alpha),\quad
M_{1,\text{tree}}^{hWW}=\frac{2m_W^2}{v^2}(v_\phi c_\alpha+2v_\Delta s_\alpha),\notag\\
M_{2,\text{tree}}^{hVV(\text{tree})}&=M_{3,\text{tree}}^{hVV}=0.
\end{align}
The counter-term contributions are 
\begin{align}
\delta M_1^{hZZ}&=\frac{2m_Z^2c_\alpha}{v^2+2v_\Delta^2}
\Bigg\{\frac{v_\phi \delta m_Z^2}{m_Z^2} +\frac{v_\phi}{2}(\delta Z_h +2\delta Z_Z)+4v_\Delta \delta C_{Hh}\notag\\
&\quad\quad\quad\quad\quad+\frac{1}{v_\phi(v^2+2v_\Delta^2)}[2v_\Delta(2v_\Delta^2-3v^2)\delta v_\Delta
-v(v^2-6v_\Delta^2)\delta v]\Bigg\}\notag\\
&+\frac{2m_Z^2s_\alpha}{v^2+2v_\Delta^2}
\left[\frac{4v_\Delta \delta m_Z^2}{m_Z^2} +2v_\Delta(\delta Z_h +2\delta Z_Z)-v_\phi \delta C_{Hh}
+\frac{4(v_\phi^2\delta v_\Delta -2vv_\Delta\delta v)}{v^2+2v_\Delta^2}\right]
\notag\\
\delta M_1^{hWW}&=\frac{v_\phi m_W^2c_\alpha}{v^2}\Bigg\{\frac{2\delta m_W^2}{m_W^2}+2\delta Z_W+\delta Z_h
+\frac{4v_\Delta}{v_\phi}\delta C_{Hh}
-\frac{2(v^2-4v_\Delta^2)}{v_\phi^2}\frac{\delta v}{v}-\frac{4v_\Delta^2}{v_\phi^2}\frac{\delta v_\Delta}{v_\Delta}\Bigg\}\notag\\
&+\frac{2m_W^2 v_\Delta s_\alpha}{v^2}\left(\frac{2\delta m_W^2}{m_W^2}+2\delta Z_W+\delta Z_h
-\frac{v_\phi}{v_\Delta}\delta C_{Hh}-\frac{4\delta v}{v}+\frac{2\delta v_\Delta}{v_\Delta}\right)
,\notag\\
\delta M_{2}^{hVV}&=\delta M_{3}^{hVV}=0.
\end{align}

\begin{figure}[t]
\begin{center}
\includegraphics[width=75mm]{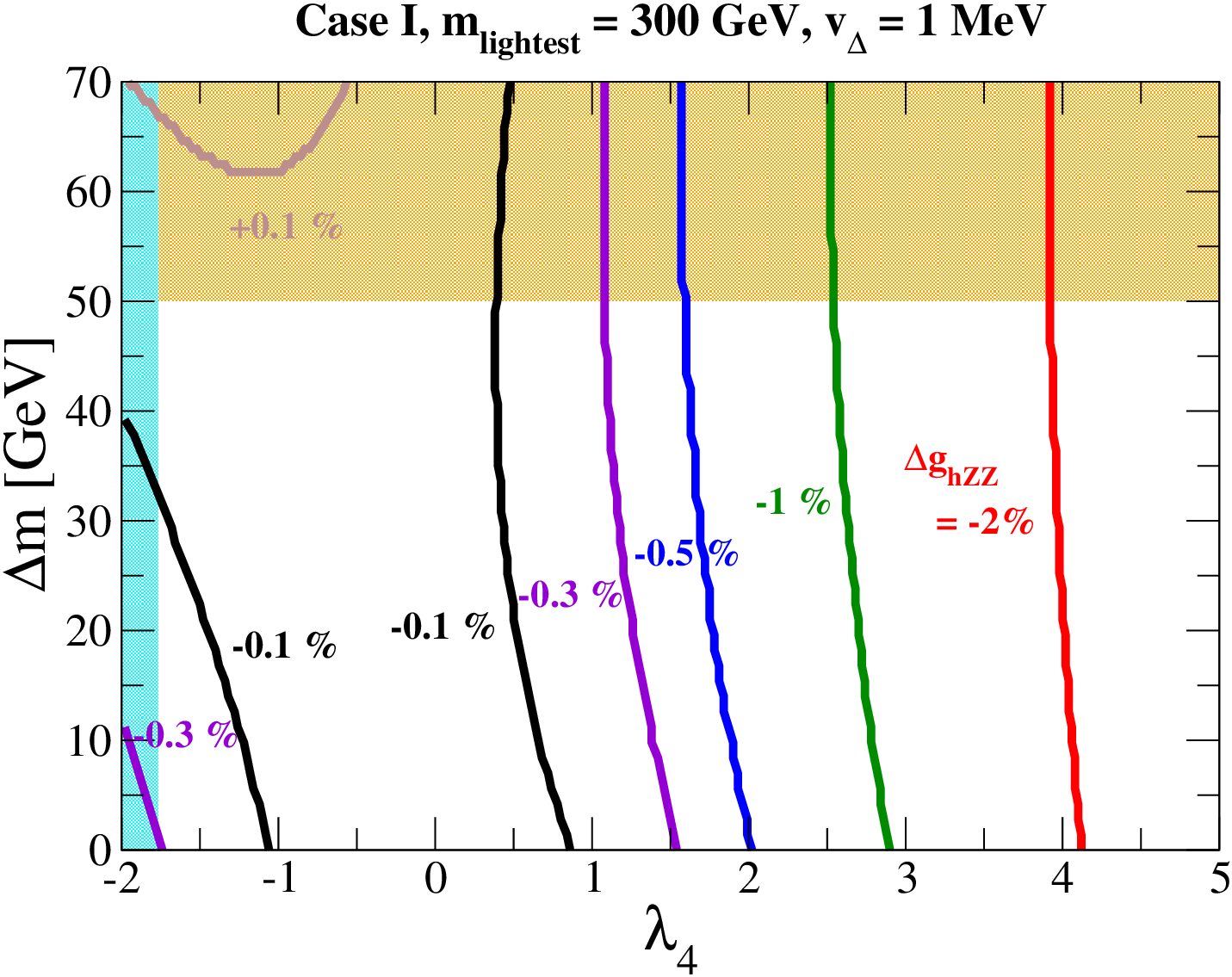}\hspace{5mm}
\includegraphics[width=75mm]{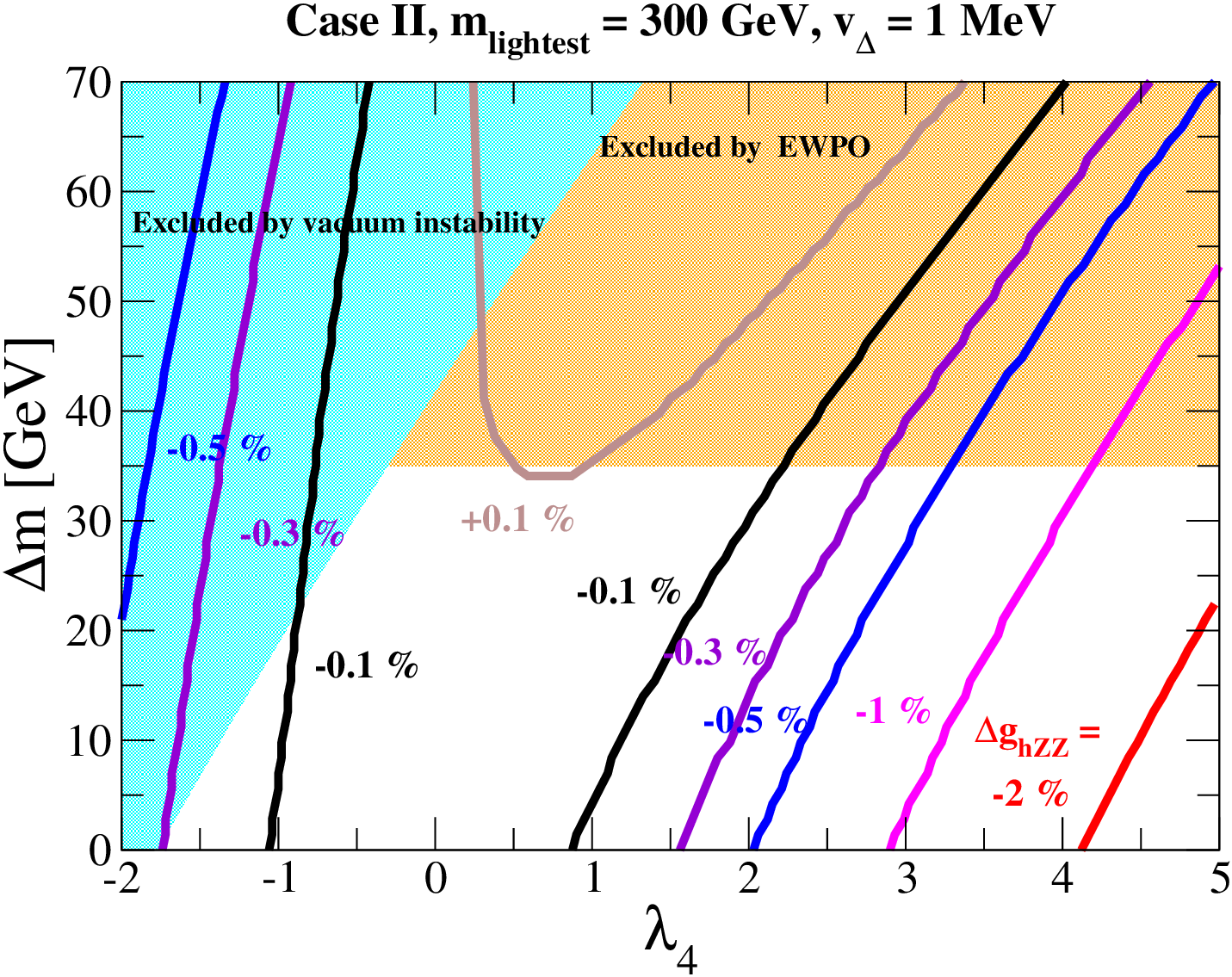}
\caption{Contour plots of $\Delta g_{hZZ}$ in Eq.~(\ref{hgVV}) for $m_{\text{lightest}}=300$ GeV and $v_\Delta$=1 MeV in the 
$\lambda_4$-$\Delta m$ plane. 
The left panel (right panel) shows the result in Case~I (Case~II).  
The blue and orange shaded regions are excluded by the vacuum stability bound and the measured $m_W$ data, respectively. }
\label{FIG:hZZ}
\end{center}
\end{figure}

\begin{figure}[t]
\begin{center}
\includegraphics[width=75mm]{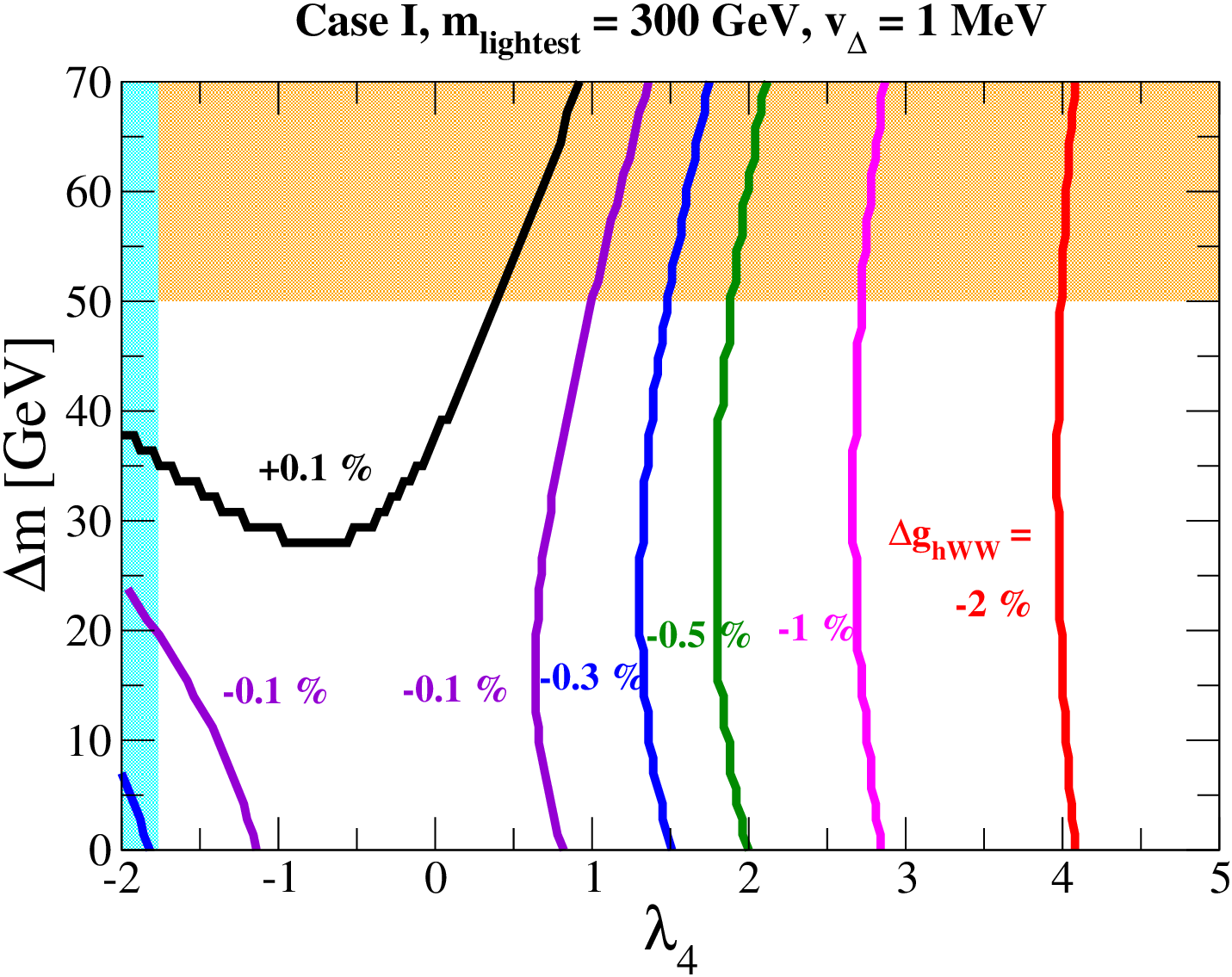}\hspace{5mm}
\includegraphics[width=75mm]{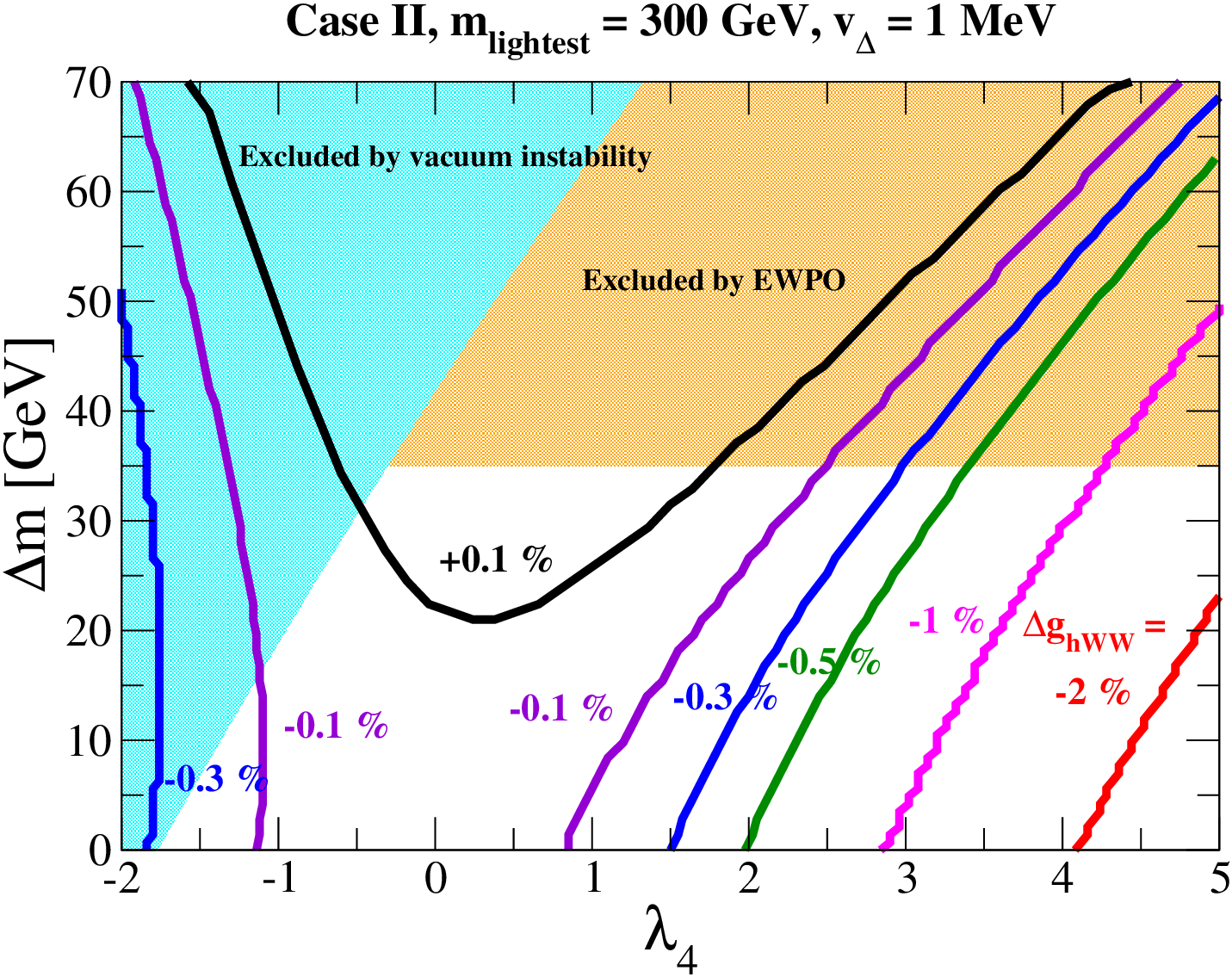}
\caption{Contour plots of $\Delta g_{hWW}$ defined in Eq.~(\ref{hgVV}) for $m_{\text{lightest}}=300$ GeV and $v_\Delta$=1 MeV in the 
$\lambda_4$-$\Delta m$ plane. 
The left panel (right panel) shows the result in Case~I (Case~II).  
The blue and orange shaded regions are excluded by the vacuum stability bound and the measured $m_W$ data, respectively. }
\label{FIG:hWW}
\end{center}
\end{figure}

We define the following quantity to study the deviation of the $hVV$ coupling from the SM prediction: 
\begin{align}
\Delta g_{hVV} \equiv \frac{\text{Re}M_1^{hVV}-\text{Re}M_1^{hVV}(\text{SM})}{\text{Re}M_1^{hVV}(\text{SM})},  \label{hgVV}
\end{align}
where $M_1^{hVV}=M_1^{hVV}(m_V^2,(m_h-m_V)^2,m_h^2)$ and $M_1^{hVV}(\text{SM})$ is the corresponding SM prediction. 

In Fig.~\ref{FIG:hZZ}, 
we show the contour plots for
$\Delta g_{hZZ}$ for $m_{\text{lightest}}=300$ GeV and $v_\Delta=1$ MeV 
in the $\lambda_4$-$\Delta m$ plane. 
The left (right) plot shows the result in Case~I (Case~II).  
The blue and orange shaded regions are excluded by the vacuum stability bound and the measured $m_W$ data, respectively. 
The magnitude of the negative corrections is larger for positive larger values of $\lambda_4$ 
for smaller values of $\Delta m $. 
For the cases with large $\Delta m$ such as about 30 GeV, 
the region with positive corrections appears. 
This is the striking feature of the HTM. 
On the contrary, in multi Higgs doublet models the correction is always negative~\cite{KOSY}. 
Under the constraint of the vacuum stability bound and the electroweak precision observable $m_W$, larger 
$\Delta m$ can be allowed in Case I than in Case II. 
We find that $\Delta g_{hZZ}$ is predicted to be at most a few $\%$. 
We can expect that such a deviation will be testable at the ILC~\cite{Peskin}. 

In Fig.~\ref{FIG:hWW}, the similar contour plots are shown for $\Delta g_{hWW}$ with the same parameter sets in the same plane. 
The behavior of $\Delta g_{hWW}$ in this plane is similar to that of $\Delta g_{hZZ}$. 
However, the correction can be positive for smaller values of $\Delta m$. 
We also show the same constraints from the vacuum stability bound and from the electroweak precision observable $m_W$. 
Magnitudes of maximum value of the correction are almost the same those of $\Delta g_{hZZ}$, especially for $\lambda_4>0$. 

\subsection{Renormalized $hhh$ counpling}

\begin{figure}[t]
\begin{center}
\includegraphics[width=80mm]{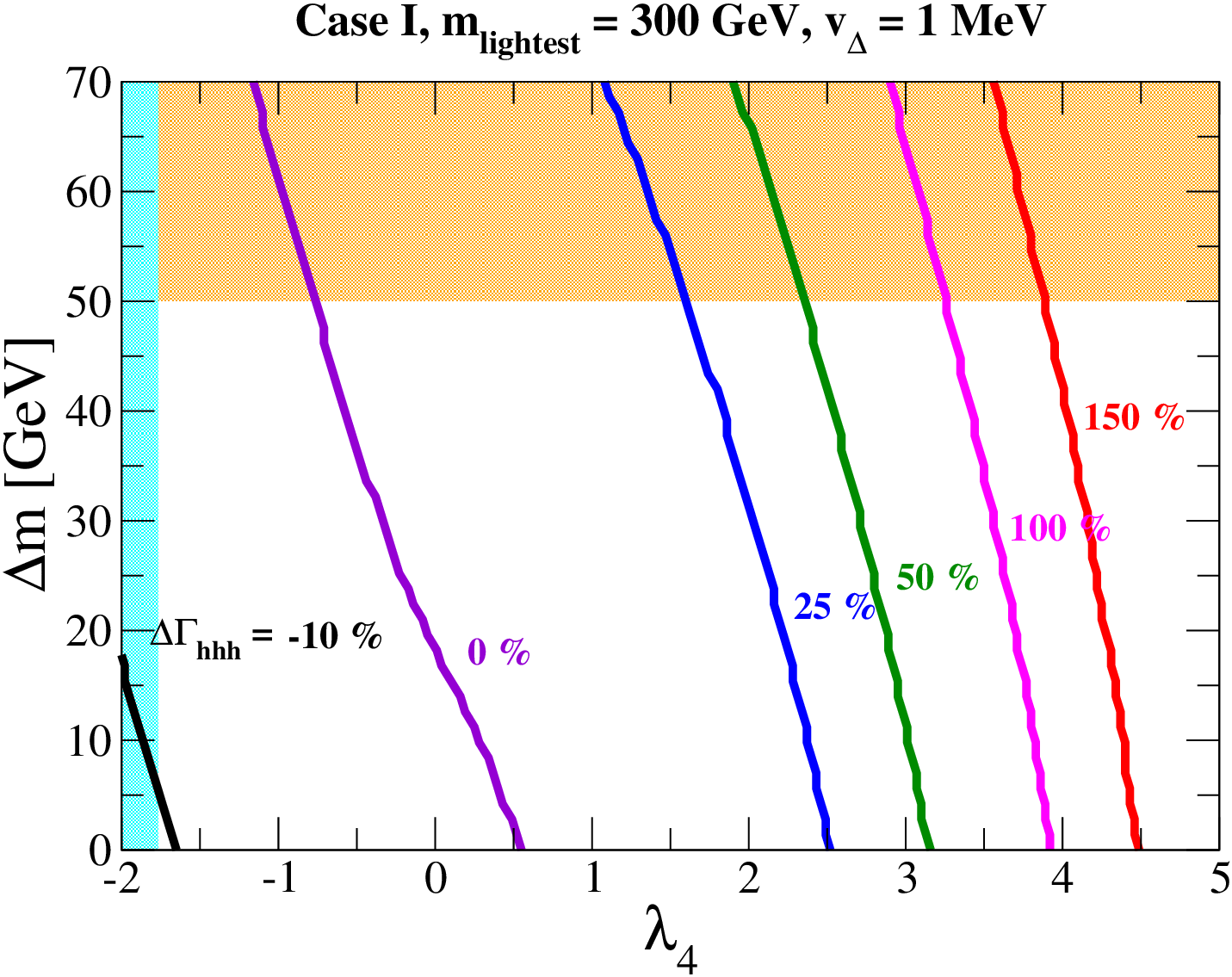}\hspace{3mm}
\includegraphics[width=80mm]{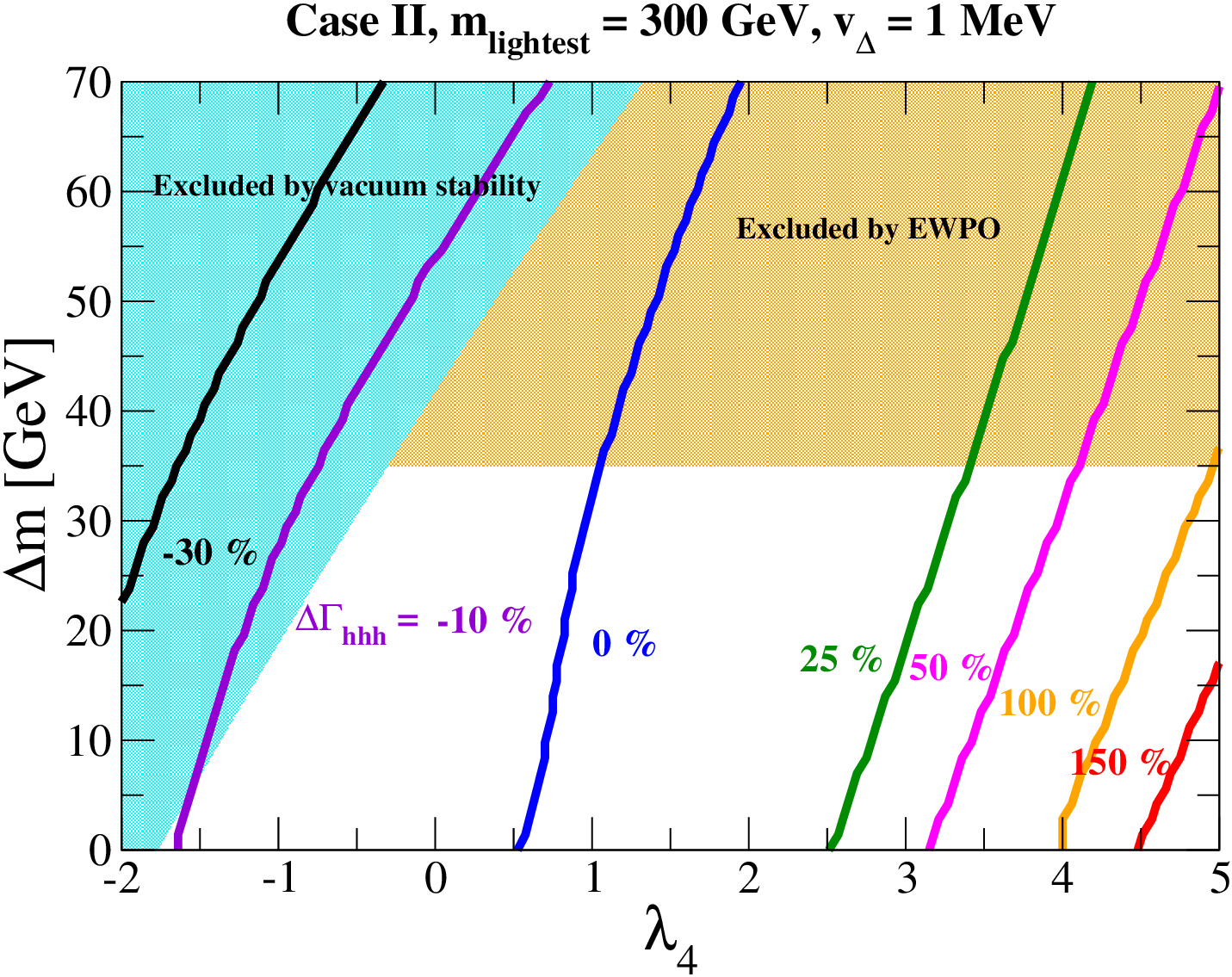}
\caption{Contour plots of
$\Delta\Gamma_{hhh}$ defined 
in Eq.~(\ref{delhhh}) for $m_{\text{lightest}}=300$ GeV and $v_\Delta=1$ MeV. 
The left panel (right panel) shows the result in Case~I (Case~II).  
The blue and orange shaded regions are excluded by the vacuum stability bound and the measured $m_W$ data, respectively. }
\label{FIG:hhh}
\end{center}
\end{figure}

Finally, we show the numerical results for the deviation of the Higgs trilinear coupling $hhh$ from the SM prediction.
The renormalized $hhh$ coupling can be expressed as a function of the external incoming momenta $p_1$ and $p_2$ 
and the outgoing momentum $q=p_1+p_2$ as
\begin{align}
\Gamma_{hhh}(p_1^2,p_2^2,q^2)=\Gamma_{hhh}^{\text{tree}}+\delta \Gamma_{hhh}+\Gamma_{hhh}^{\text{1PI}}(p_1^2,p_2^2,q^2), 
\end{align}
where the first, second and last terms are corresponding to the tree level, the counter-term, and 
the 1PI diagram contributions, respectively. 
The tree level contribution $\Gamma_{hhh}^{\text{tree}}$ is calculated as
\begin{align}
\Gamma_{hhh}^{\text{tree}}=
-6\left[\left(\frac{c_\alpha^3}{v_\phi}+\frac{s_\alpha^3}{v_\Delta}\right)\frac{m_h^2}{2}
+\frac{s_\alpha^2}{v^2+2v_\Delta^2}\left(v_\phi c_\alpha-\frac{v_\phi^2}{2v_\Delta}s_\alpha\right)m_A^2\right].
\end{align}
The counter-term contribution $\delta\Gamma_{hhh}$ is evaluated by
\begin{align}
&\delta\Gamma_{hhh}=
-3\left(\frac{c_\alpha^3}{v_\phi}+\frac{s_\alpha^3}{v_\Delta}\right)\delta m_h^2
-\frac{6s_\alpha^2}{v^2+2v_\Delta^2}\left(v_\phi c_\alpha-\frac{v_\phi^2s_\alpha}{2v_\Delta}\right)\delta m_A^2
-3(m_h^2-m_H^2)\left(\frac{s_\alpha^2c_\alpha}{v_\Delta}-\frac{c_\alpha^2s_\alpha}{v_\phi}\right)\delta \alpha
\notag\\
&-\frac{9}{2}\left[m_h^2\left(\frac{c_\alpha^3}{v_\phi}+\frac{s_\alpha^3}{v_\Delta}\right)
+\frac{m_A^2v_\phi s_\alpha^2}{v_\Delta(v^2+2v_\Delta^2)}(2v_\Delta  c_\alpha-v_\phi s_\alpha)\right]\delta Z_h\notag\\
&-3s_\alpha\Bigg\{(2m_h^2+m_H^2)\left(\frac{s_\alpha c_\alpha}{v_\Delta}-\frac{c_\alpha^2}{v_\phi}\right)
+\frac{m_A^2v_\phi }
{v_\Delta (v^2+2v_\Delta^2)}\left[2v_\Delta(2c_\alpha^2-s_\alpha^2)-3s_\alpha c_\alpha v_\phi\right]\Bigg\}\delta C_{Hh}
\notag\\
&-\frac{3}{v_\phi}\Bigg\{\frac{m_h^2}{v_\phi^2}\left[\frac{v_\Delta}{2}(3c_\alpha+c_{3\alpha})
-\frac{v_\phi^3s_\alpha^3}{v_\Delta^2}\right]
+\frac{4m_A^2s_\alpha^2}{(v^2+2v_\Delta^2)^2}\left[c_\alpha v_\Delta(2v_\Delta^2-3v^2)+v_\phi s_\alpha \left(\frac{v^4}{4v_\Delta^2}+2v^2-v_\Delta^2\right)\right]
\Bigg\}\delta v_\Delta
\notag\\
&+\frac{3}{v_\phi}\Bigg\{\frac{m_h^2v}{v_\phi^2}c_\alpha^3
+\frac{2m_A^2vs_\alpha^2}{(v^2+2v_\Delta^2)^2}
\left[(v^2-6v_\Delta^2)c_\alpha+4v_\Delta v_\phi s_\alpha \right]\Bigg\}\delta v. \label{hhha}
\end{align}
Contributions of the 1PI diagram to the $hhh$ coupling is listed in Appendix~B. 

In the limit of $v_\Delta/v \to 0$, these expressions are reduced to the same expressions in the SM as\footnote{
As long as we take $\lambda_4$ to be $\mathcal{O}(1)$ or less 
and the triplet-like Higgs boson masses to be $\mathcal{O}(100)$ GeV or more, the magnitude of the mixing angle $\alpha$ is 
as large as that of $v_\Delta$ (see Eq.~(\ref{alpha})). }
\begin{align}
\Gamma_{hhh}^{\text{tree}}&\to
\frac{-3m_h^2}{v}, \\
\delta\Gamma_{hhh}&\to 
-\frac{3\delta m_h^2}{v}
-\frac{9}{2}\frac{m_h^2}{v}\delta Z_h
+\frac{3m_h^2}{v^2}\delta v.
\end{align}
In this limit, the top quark loop and the gauge boson loop contributions 
to the 1PI diagram is the same as the SM. 
However, the scalar boson loop contributions can be different from the SM case, because
the triplet-like Higgs boson loop contributions can be remained even in this limit. 
Approximately, the triplet-like Higgs boson loop contributions can be expressed as
\begin{align}
\Gamma_{hhh}
&\simeq -\frac{3m_h^2}{v}\left[
1-\frac{v}{48\pi^2m_h^2}\left(\frac{\lambda_{H^{++}H^{--}h}^3}{m_{H^{++}}^2}
+\frac{\lambda_{H^{+}H^{-}h}^3}{m_{H^+}^2}
+\frac{4\lambda_{AAh}^3}{m_A^2}
+\frac{4\lambda_{HHh}^3}{m_H^2}
\right)+ \cdots\right]\notag\\
&\simeq 
-\frac{3m_h^2}{v}\left\{1
+\frac{v^4}{48\pi^2m_h^2}\left[\frac{\lambda_4^3}{m_{H^{++}}^2}
+\frac{(\lambda_4+\frac{\lambda_5}{2})^3}{m_{H^+}^2}
+\frac{(\lambda_4+\lambda_5)^3}{2m_A^2}
+\frac{(\lambda_4+\lambda_5)^3}{2m_H^2}
   \right]+  \cdots\right\}, \label{hhh_HTM}
\end{align}
where dotted terms mean the same correction given in the SM. 
Therefore,  
we find that the triplet-like Higgs boson loop contribution to the $hhh$ vertex 
gives a positive (negative) correction compared to the SM prediction when 
$\lambda_4$ is taken to be a positive (negative) value and $\lambda_5\simeq 0$. 

To illustrate the deviation of the $hhh$ coupling from the SM prediction, 
we define the following quantity: 
\begin{align}
\Delta \Gamma_{hhh}\equiv \frac{\text{Re}\Gamma_{hhh}-\text{Re}\Gamma_{hhh}^{\text{SM}}}{\text{Re}\Gamma_{hhh}^{\text{SM}}}, \label{delhhh}
\end{align}
where $\Gamma_{hhh}=\Gamma_{hhh}(m_h^2,m_h^2,4m_h^2)$ and $\Gamma_{hhh}^{\text{SM}}$ is the corresponding prediction in the SM.  

In Fig.~\ref{FIG:hhh}, 
contour plot for 
the deviation of the one-loop corrected $hhh$ coupling from the SM prediction $\Delta\Gamma_{hhh}$ defined 
in Eq.~(\ref{delhhh}) is shown for $m_{\text{lightest}}=300$ GeV and $v_\Delta=1$ MeV
in the $\lambda_4$-$\Delta m$ plane. 
The left (right) plot shows the result in Case~I (Case~II).  
The blue and orange shaded regions are excluded by the vacuum stability bound and the measured $m_W$ data, respectively. 
In both cases, positive (negative) values of $\Delta \Gamma_{hhh}$ are predicted in the case with a positive (negative) $\lambda_4$
whose magnitudes can be greater than about +150\% ($-10\%$). 
Such a deviation in $\Delta\Gamma_{hhh}$ is expected to be measured at the ILC with a center of mass energy 
to be 1 TeV and integrated luminosity being 2 ab$^{-1}$~\cite{Fujii}.
We note that there is a relations among the one-loop corrected Higgs boson couplings $h\gamma\gamma$, $hVV$ and $hhh$. 
In particular, a strong correlation can be found in deviations in $R_{\gamma\gamma}$ and $\Gamma_{hhh}$. 
If $R_{\gamma\gamma}<1$ ($R_{\gamma\gamma}>1$) which is predicted for $\lambda_4 >0$ ($\lambda_4 <0$), 
$\Delta \Gamma_{hhh}$ takes sufficiently large positive (negative) values. 
In conclusion, by measuring these coupling constants accurately, 
we can discriminate the HTM of the other models, even when additional particles are not directly discovered.

\section{Conclusions}

We have calculated a full set of one-loop corrections to 
the Higgs boson coupling constants as well as 
the electroweak parameters.  
Renormalization calculations are performed in the on-shell scheme, 
where 
two different schemes has been discussed in the renormalization of the one-loop calculation of electroweak parameters. 
We have computed the decay rate of the SM-like Higgs boson $h$ into diphoton. 
Renormalized Higgs couplings with the weak gauge bosons $hZZ$, $hWW$ and 
the trilinear coupling $hhh$ has also been calculated at the one-loop level. 
Magnitudes of the deviations in these quantities from the SM predictions 
have been evaluated in the parameter regions where the unitarity and vacuum 
stability bounds are satisfied and the predicted W boson mass is consistent with the data.
There are strong correlations among deviations in the Higgs boson couplings 
$R_{\gamma\gamma}$, $\Delta g_{hVV}$ and $\Delta\Gamma_{hhh}$. 
For example, when $R_{\gamma\gamma}$ is predicted by about 1.3 (0.6), 
$\Delta g_{hVV}$ and $\Delta\Gamma_{hhh}$ are respectively predicted as about $-0.1\%$ and $-10\%$ 
($-2\%$ and $+150\%$). 
By measuring these deviations in Higgs boson couplings accurately, 
the HTM can be distinguished from the other models. 
These deviations in the Higgs boson couplings may be detected at future colliders such as 
the LHC with 3000 fb$^{-1}$ and at the ILC. 
\\\\
\noindent
$Acknowledgments$

The paper was supported in part by Grant-in-Aid for Scientific Research, Nos. 22244031 [S.K.], 22740137 [M.A.],  
23104006 [S.K.] and 24340046 [M.A. and S.K.].
K.Y. was supported in part by the National Science Council of R.O.C. under Grant No. NSC-101-2811-M-008-014.

\begin{appendix}

\section{Tree level scalar boson couplings}

In this Appendix, 
we list the tree level scalar boson couplings. 
The Higgs-gauge-gauge type, the Higgs-ghost-anti ghost type and the 
Higgs-Higgs-gauge-gauge type vertices and those corresponding coefficients are respectively listed 
in Table~\ref{fr1}, Table~\ref{fr2} and Table~\ref{fr3}. 
In Table~\ref{fr4}, the Higgs-Higgs-gauge type vertices and those corresponding coefficients are listed, 
where $p_1$ and $p_2$ are the incoming momenta of the first and second particles in the vertices column.

\begin{table}[t]
\begin{center}
{\renewcommand\arraystretch{1.5}
\begin{tabular}{|c|c||c|c|}\hline
Vertices& Coefficient &Vertices& Coefficient  \\\hline\hline
$h W_\mu^+ W_\nu^-$  & $gm_W(c_\beta c_\alpha+\sqrt{2}s_\beta s_\alpha)g_{\mu\nu}$ &$h Z_\mu Z_\nu $ & 
$\frac{g_Zm_Z}{2}(c_{\beta'}c_\alpha+2s_{\beta'} s_\alpha)g_{\mu\nu}$ \\\hline
$H W_\mu^+ W_\nu^-$  & $gm_W(-c_\beta s_\alpha+\sqrt{2}s_\beta c_\alpha)g_{\mu\nu}$ &$H Z_\mu Z_\nu $ & 
$\frac{g_Zm_Z}{2}(-c_{\beta'}s_\alpha+2s_{\beta'}c_\alpha)g_{\mu\nu}$ \\\hline
$H^\pm W_\mu^\mp Z_\nu$  & $-g_Zm_W s_\beta c_\beta g_{\mu\nu} $& $G^\pm W_\mu^\mp Z_\nu$ & $-g_Zm_W\left(s_W^2
+s_\beta^2\right)g_{\mu\nu}$ \\\hline
$H^{\pm\pm} W_\mu^\mp W_\nu^\mp$  & $gm_Ws_\beta  g_{\mu\nu}$ & $G^\pm W_\mu^\mp A_\nu $ & $em_Wg_{\mu\nu}$ \\\hline
\end{tabular}}
\caption{The Higgs-gauge-gauge type vertices and those corresponding coefficients.}
\label{fr1}
\end{center}
\end{table}

\begin{table}[t]
\begin{center}
{\renewcommand\arraystretch{1.5}
\begin{tabular}{|c|c||c|c|}\hline
Vertices& Coefficient &Vertices& Coefficient  \\\hline\hline
$h \bar{c}^\pm c^\mp$  & $-\frac{gm_W}{2}(c_\beta c_\alpha+\sqrt{2}s_\beta s_\alpha)$ &$h \bar{c}_Z c_Z$ & 
$-\frac{g_Zm_Z}{2}(c_{\beta'}c_\alpha+2s_{\beta'} s_\alpha)$ \\\hline
$H \bar{c}^\pm c^\mp$  & $-\frac{gm_W}{2}(-c_\beta s_\alpha+\sqrt{2}s_\beta c_\alpha)$ &$H \bar{c}_Z c_Z$ & 
$-\frac{g_Zm_Z}{2}(-c_{\beta'}s_\alpha+2s_{\beta'} c_\alpha)$ \\\hline
$G^\pm \bar{c}_Z c^\mp$  & $i\frac{g_Zm_W}{2}(1+s_\beta^2)$ &$G^\pm \bar{c}^\mp  c_Z $ &$i\frac{g_Zm_W}{2}(-c_{2W}c_\beta^2+2s_W^2s_\beta^2)$ \\\hline
$H^\pm \bar{c}_Z c^\mp$  & $i\frac{g_Zm_W}{2}c_\beta s_\beta$ &$H^\pm \bar{c}^\mp  c_Z $ &
$i\frac{g_Zm_W}{2}c_\beta s_\beta(c_{2W}+2s_W^2)$ \\\hline
\end{tabular}}
\caption{The Higgs-ghost-anti ghost type vertices and those corresponding coefficients.}
\label{fr2}
\end{center}
\end{table}

\begin{table}[t]
\begin{center}
{\renewcommand\arraystretch{1.5}
\begin{tabular}{|c|c||c|c|}\hline
Vertices& Coefficient&Vertices& Coefficient   \\\hline\hline
$H^{++}H^{--}W_\mu^+ W_\nu^-$  & $g^2g_{\mu\nu}$  &
$H^{++}H^{--}Z_\mu Z_\nu$  & $   g_Z^2(c_W^2-s_W^2)^2g_{\mu\nu}$    \\\hline
$H^{+}H^{-}W_\mu^+ W_\nu^-$  & $\frac{g^2}{4}(5+3c_{2\beta})g_{\mu\nu}$&
$H^{+}H^{-}Z_\mu Z_\nu$   & $\frac{g_Z^2}{8}\left(2+c_{4W}-4c_{2W}c_{\beta}^2+c_{2\beta}\right)g_{\mu\nu}$   \\\hline
$G^{+}G^{-}W_\mu^+ W_\nu^-$  & $\frac{g^2}{4}(5-3c_{2\beta})g_{\mu\nu}$&  
$G^{+}G^{-}Z_\mu Z_\nu$   & $\frac{g_Z^2}{8}\left(2+c_{4W}-4c_{2W}s_{\beta}^2-c_{2\beta}\right)g_{\mu\nu}$\\\hline
$HHW_\mu^+ W_\nu^-$  & $\frac{g^2}{8}(3+c_{2\alpha})g_{\mu\nu}$&
$H^\pm G^\mp Z_\mu Z_\nu$  & $\frac{g_Z^2}{8}s_{2\beta}(1-2c_{2W})g_{\mu\nu}$    \\\hline
$hhW_\mu^+ W_\nu^-$  & $\frac{g^2}{8}(3-c_{2\alpha})g_{\mu\nu}$ &
$AAZ_\mu Z_\nu$   & $\frac{g_Z^2}{16}\left(5+3c_{2\beta'}\right)g_{\mu\nu}$    \\\hline
$AAW_\mu^+ W_\nu^-$  & $\frac{g^2}{8}(3+c_{2\beta'})g_{\mu\nu}$ & 
$G^0G^0Z_\mu Z_\nu$  & $\frac{g_Z^2}{16}\left(5-3c_{2\beta'}\right)g_{\mu\nu}$   \\\hline
$G^0G^0W_\mu^+ W_\nu^-$  & $\frac{g^2}{8}(3-c_{2\beta'})g_{\mu\nu}$&
$AG^0Z_\mu Z_\nu$  & $\frac{3g_Z^2}{8}s_{2\beta'}g_{\mu\nu}$   \\\hline
$H^\pm G^\mp W_\mu^+ W_\nu^-$  & $\frac{3}{4}g^2 s_{2\beta}g_{\mu\nu}$& 
$HHZ_\mu Z_\nu$  & $\frac{g_Z^2}{16}\left(5+3c_{2\alpha}\right)g_{\mu\nu}$   \\\hline
$A G^0 W_\mu^+ W_\nu^-$  &$\frac{1}{4}g^2 s_{2\beta'}g_{\mu\nu}$ & 
$hhZ_\mu Z_\nu$  & $\frac{g_Z^2}{16}\left(5-3c_{2\alpha}\right)g_{\mu\nu}$   \\\hline
$Hh W_\mu^+ W_\nu^-$  &$\frac{1}{4}g^2 s_{2\alpha}g_{\mu\nu}$ & 
$HhZ_\mu Z_\nu$  & $\frac{3g_Z^2}{8}s_{2\alpha}g_{\mu\nu}$    \\\hline
$H^{++}H^{--}A_\mu Z_\nu$  & $4eg_Z(\hat{c}_W^2-\hat{s}_W^2)g_{\mu\nu}$&
$H^{++}H^{--}A_\mu A_\nu$  &  $4e^2g_{\mu\nu}$\\\hline
$H^+H^-A_\mu Z_\nu$  & $eg_Z(\hat{c}_W^2-\hat{s}_W^2- c^2_{\beta})g_{\mu\nu}$&
$H^{+}H^{-}A_\mu A_\nu$ &$e^2g_{\mu\nu}$  \\\hline
$G^+G^-A_\mu Z_\nu$  & $eg_Z(\hat{c}_W^2-\hat{s}_W^2-s^2_{\beta})g_{\mu\nu}$&
$G^{+}G^{-}A_\mu A_\nu$ &$e^2g_{\mu\nu}$\\\hline
$G^\pm hW_\mu^\mp Z_\nu$  & $\frac{gg_Z}{2}[-c_\alpha c_\beta s_W^2 +\sqrt{2}(c_W^2-2)s_\alpha s_\beta]g_{\mu\nu}$&
$G^\pm hW_\mu^\mp A_\nu$ & $\frac{ge}{2}(c_\alpha c_\beta +\sqrt{2}s_\alpha s_\beta)g_{\mu\nu}$\\\hline
$H^\pm hW_\mu^\mp Z_\nu$  & $\frac{gg_Z}{2}[c_\alpha s_\beta s_W^2+\sqrt{2}(c_W^2-2)s_\alpha c_\beta]g_{\mu\nu}$&
$H^\pm hW_\mu^\mp A_\nu$ & $\frac{ge}{2}(-c_\alpha s_\beta +\sqrt{2}s_\alpha c_\beta)g_{\mu\nu}$\\\hline
$H^{\pm\pm} hW_\mu^\mp W_\nu^{\mp}$  & $\frac{g^2}{\sqrt{2}}s_\alpha g_{\mu\nu}$& & \\\hline
\end{tabular}}
\caption{The Higgs-Higgs-gauge-gauge type vertices and those corresponding coefficients.}
\label{fr3}
\end{center}
\end{table}

\begin{table}[t]
\begin{center}
{\renewcommand\arraystretch{1.5}
\begin{tabular}{|c|c||c|c|}\hline
Vertices& Coefficient &Vertices& Coefficient  \\\hline\hline
$H^{\pm\pm}H^\mp W_\mu^\mp$  & $\pm gc_\beta(p_1-p_2)_\mu$&$H^{++}H^{--}A_\mu$&$2e(p_1-p_2)_\mu$\\\hline
$H^{\pm\pm}G^\mp W_\mu^\mp$  & $\pm gs_\beta(p_1-p_2)_\mu$&$H^{+}H^{-}A_\mu$&$e(p_1-p_2)_\mu$\\\hline
$H^\pm AW_\mu^\mp$  & $-i\frac{g}{2}(s_{\beta'} s_\beta + \sqrt{2}c_{\beta'} c_\beta)(p_1-p_2)_\mu$&$G^{+}G^{-}A_\mu$&$e(p_1-p_2)_\mu$\\\hline
$H^\pm HW_\mu^\mp$  & $\pm\frac{g}{2}(s_\alpha s_\beta + \sqrt{2}c_\alpha c_\beta)(p_1-p_2)_\mu$&$H^{++}H^{--}Z_\mu$ &$g_Z(\hat{c}_W^2-\hat{s}_W^2)(p_1-p_2)_\mu$\\\hline
$H^\pm hW_\mu^\mp$  & $\pm\frac{g}{2}(-c_\alpha s_\beta+\sqrt{2}s_\alpha c_\beta)(p_1-p_2)_\mu$&$H^{+}H^{-}Z_\mu$&$\frac{g_Z}{2}(\hat{c}_W^2-\hat{s}_W^2-c_{\beta}^2)(p_1-p_2)_\mu$\\\hline
$H^\pm G^0W_\mu^\mp$  & $-i\frac{g}{2}(-c_{\beta'} s_\beta + \sqrt{2} s_{\beta'} c_{\beta})(p_1-p_2)_\mu$&$G^{+}G^{-}Z_\mu$&$\frac{g_Z}{2}(\hat{c}_W^2-\hat{s}_W^2-s_{\beta}^2)(p_1-p_2)_\mu$\\\hline
$G^\pm AW_\mu^\mp$  & $-i\frac{g}{2}(-s_{\beta'} c_\beta+\sqrt{2} c_{\beta'} s_\beta)(p_1-p_2)_\mu=0$&$H^{\pm}G^{\mp}Z_\mu$&$\mp\frac{g_Z}{4}s_{2\beta}(p_1-p_2)_\mu$\\\hline
$G^\pm HW_\mu^\mp$  & $\pm\frac{g}{2}(-s_\alpha c_{\beta}+\sqrt{2}c_\alpha s_\beta)(p_1-p_2)_\mu$&$AHZ_\mu$&$-i\frac{g_Z}{2}(2c_\alpha c_{\beta'}+s_\alpha s_{\beta'})(p_1-p_2)_\mu$
\\\hline
$G^\pm hW_\mu^\mp$  & $\pm\frac{g}{2}(c_\alpha c_\beta+\sqrt{2} s_\alpha s_\beta)(p_1-p_2)_\mu$&$AhZ_\mu$&$-i\frac{g_Z}{2}(-c_\alpha s_{\beta'}+2s_\alpha c_{\beta'})(p_1-p_2)_\mu$\\\hline
$G^\pm G^0W_\mu^\mp$  & $-i\frac{g}{2}(c_{\beta'} c_\beta+\sqrt{2} s_{\beta'} s_\beta)(p_1-p_2)_\mu$&$G^0HZ_\mu$&$-i\frac{g_Z}{2}(-c_{\beta'} s_{\alpha}+2c_\alpha s_{\beta'})(p_1-p_2)_\mu$\\\hline
&&$G^0hZ_\mu$&$-i\frac{g_Z}{2}(c_\alpha c_{\beta'}+2s_\alpha s_{\beta'})(p_1-p_2)_\mu$\\\hline
\end{tabular}}
\caption{The Higgs-Higgs-gauge type vertices and those corresponding coefficients.}
\label{fr4}
\end{center}
\end{table}

Coefficients for the scalar three-point vertices defined as
\begin{align}
\mathcal{L}=\lambda_{\phi_1\phi_2\phi_3}\phi_1\phi_2\phi_3+\cdots,
\end{align}
can be written in terms of the physical parameters as follows;
\begin{align}
\lambda_{H^{++}H^{--}H}&=\frac{2v_\phi}{v^2+2v_\Delta^2}\left[2m_{H^+}^2\left(1+\frac{2v_\Delta^2}{v^2}\right)-m_A^2\right]s_\alpha\notag\\
&-\frac{1}{v_\Delta}\left[2m_{H^{++}}^2-4m_{H^+}^2\frac{v_\phi^2}{v^2}
+m_A^2\left(1-\frac{4v_\Delta^2}{v^2+2v_\Delta^2}\right)+m_H^2\right]c_\alpha,\\
\lambda_{H^{++}H^{--}h}&=-\frac{2v_\phi}{v^2+2v_\Delta^2}\left[2m_{H^+}^2\left(1+\frac{2v_\Delta^2}{v^2}\right)-m_A^2\right]c_\alpha\notag\\
&-\frac{1}{v_\Delta}\left[2m_{H^{++}}^2-4m_{H^+}^2\frac{v_\phi^2}{v^2}
+m_A^2\left(1-\frac{4v_\Delta^2}{v^2+2v_\Delta^2}\right)+m_h^2\right]s_\alpha \label{Eq:hHppHmm},\\
\lambda_{H^{+}H^{-}H}&=\frac{1}{v_\phi}
\left[2m_{H^+}^2\frac{v_\phi^2}{v^2}+m_H^2\frac{2v_\Delta^2}{v^2}\right]s_\alpha
-\frac{1}{v_\Delta}\left[4m_{H^+}^2\frac{v_\Delta^2}{v^2}-m_A^2\frac{v^2}{v^2+2v_\Delta^2}+m_H^2\frac{v_\phi^2}{v^2}
\right]c_\alpha,\\
\lambda_{H^{+}H^{-}h}&=-\frac{1}{v_\phi}
\left(2m_{H^+}^2\frac{v_\phi^2}{v^2}+m_h^2\frac{2v_\Delta^2}{v^2}\right)c_\alpha
-\frac{1}{v_\Delta}\left(4m_{H^+}^2\frac{v_\Delta^2}{v^2}-m_A^2\frac{v^2}{v^2+2v_\Delta^2}+m_h^2\frac{v_\phi^2}{v^2}
\right)s_\alpha,\label{Eq:hHpHm}
\end{align}

\begin{align}
\lambda_{H^+G^- A}&=i\frac{\sqrt{2}(m_A^2-m_{H^+}^2)}{\sqrt{v^2+2v_\Delta^2}},\\
\lambda_{H^+G^- G^0}&=-i\sqrt{\frac{2}{v^2+2v_\Delta^2}}\frac{v_\Delta v_\phi}{v^2}m_{H^+}^2,\\
\lambda_{H^{+}G^{-}H}&=\frac{\sqrt{2}(m_{H^+}^2-m_H^2)}{v^2}(v_\phi c_\alpha+v_\Delta s_\alpha),\\
\lambda_{H^{+}G^{-}h}&=-\frac{\sqrt{2}(m_{H^+}^2-m_h^2)}{v^2}(v_\Delta c_\alpha -v_\phi s_\alpha),\\
\lambda_{G^{+}G^{-}H}&=\frac{m_H^2}{v^2}(-2v_\Delta c_\alpha+v_\phi s_\alpha),\\
\lambda_{G^{+}G^{-}h}&=-\frac{m_h^2}{v^2}(v_\phi c_\alpha +2v_\Delta s_\alpha),\\
\lambda_{AAH}&=\frac{v^2}{v_\phi}\frac{1}{v^2+2v_\Delta^2}
\left[m_A^2\left(1-\frac{2v_\Delta^2}{v^2}\right)+m_H^2\frac{2v_\Delta^2}{v^2}\right]s_\alpha\notag\\
&+\frac{1}{2v_\Delta}\left[m_A^2\left(1-\frac{8v_\Delta^2}{v^2+2v_\Delta^2}\right)
-m_H^2\left(1-\frac{4v_\Delta^2}{v^2+2v_\Delta^2}\right)\right]c_\alpha,\\
\lambda_{AAh}&=-\frac{v^2}{v_\phi}\frac{1}{v^2+2v_\Delta^2}
\left[m_A^2\frac{v_\phi^2}{v^2}+m_h^2\frac{2v_\Delta^2}{v^2}\right]c_\alpha\notag\\
&+\frac{1}{2v_\Delta}\left[m_A^2\left(1-\frac{8v_\Delta^2}{v^2+2v_\Delta^2}\right)
-m_h^2\left(1-\frac{4v_\Delta^2}{v^2+2v_\Delta^2}\right)\right]s_\alpha,\\
\lambda_{AG^0H}&=\frac{2(m_A^2-m_H^2)}{v^2+2v_\Delta^2}(v_\phi c_\alpha+v_\Delta s_\alpha),\\
\lambda_{AG^0h}&=-\frac{2(m_A^2-m_h^2)}{v^2+2v_\Delta^2}(v_\Delta c_\alpha-v_\phi s_\alpha),\\
\lambda_{G^0G^0H}&=\frac{m_H^2}{2(v^2+2v_\Delta^2)}(-4v_\Delta c_\alpha+v_\phi s_\alpha),\\
\lambda_{G^0G^0h}&=-\frac{m_h^2}{2(v^2+2v_\Delta^2)}(v_\phi c_\alpha+4v_\Delta s_\alpha),\\
\lambda_{HHH}&=-\frac{1}{8v_\Delta}\left(3c_\alpha+c_{3\alpha}-\frac{4v_\Delta}{v_\phi}s_\alpha^3\right)m_H^2
+\frac{v_\phi c_\alpha^2}{2v_\Delta(v^2+2v_\Delta^2)}(v_\phi c_\alpha+2v_\Delta  s_\alpha)m_A^2,\\
\lambda_{HHh}&=-\frac{1}{2v_\Delta}\left(c_\alpha+\frac{v_\Delta}{v_\phi}s_\alpha\right)\left(m_H^2+\frac{m_h^2}{2}\right)s_{2\alpha}
+\frac{v_\phi c_\alpha}{4(v^2+2v_\Delta^2)}\left(2-6c_{2\alpha}+\frac{3v_\phi}{v_\Delta}s_{2\alpha}\right)m_A^2,\\
\lambda_{Hhh}&=-\frac{s_{2\alpha}}{2v_\Delta}\left(s_\alpha-\frac{v_\Delta}{v_\phi}c_\alpha\right)\left(m_h^2+\frac{m_H^2}{2}\right)
-\frac{v_\phi s_\alpha}{4(v^2+2v_\Delta^2)}\left(2+6c_{2\alpha}-\frac{3v_\phi}{v_\Delta}s_{2\alpha}\right)m_A^2,\\
\lambda_{hhh}&=-\left(\frac{c_\alpha^3}{v_\phi}+\frac{s_\alpha^3}{v_\Delta}\right)\frac{m_h^2}{2}
-\frac{s_\alpha^2}{v^2+2v_\Delta^2}\left(v_\phi c_\alpha-\frac{v_\phi^2}{2v_\Delta}s_\alpha\right)m_A^2.
\end{align}
Coefficients for the scalar four-point vertices defined as
\begin{align}
\mathcal{L}=\lambda_{\phi_1\phi_2\phi_3\phi_4}\phi_1\phi_2\phi_3\phi_4+\cdots,
\end{align}
can be written in terms of the physical parameters as follows;
\begin{align}
\lambda_{H^{++}H^{--}AA}&=
\frac{1}{v_\Delta^2(v^2+2v_\Delta^2)}\Bigg[
-v_\phi^2 m_{H^{++}}^2+2(v^2-4v_\Delta^2)m_{H^+}^2+\frac{-v^4+4v^2v_\Delta^2+4v_\Delta^4}{2(v^2+2v_\Delta^2)}m_A^2\notag\\
&+\frac{1}{2}(2v_\Delta^2-v^2)(c_\alpha^2 m_H^2+s_\alpha^2 m_h^2)+\frac{v_\Delta^3}{v_\phi}s_{2\alpha} (m_H^2-m_h^2)
\Bigg],\\
\lambda_{H^{++}H^{--}AG^0}&=
\frac{1}{v_\Delta(v^2+2v_\Delta^2)}
\Bigg[-4v_\phi m_{H^{++}}^2+\frac{8v_\phi}{v^2}(v^2-v_\Delta^2)m_{H^+}^2\notag\\
&-\frac{2v^2v_\phi}{v^2+2v_\Delta^2}m_A^2-2c_\alpha (v_\phi c_\alpha+v_\Delta s_\alpha)m_H^2
+(-2v_\phi s_\alpha^2+v_\Delta s_{2\alpha})m_h^2
\Bigg],\\
\lambda_{H^{++}H^{--}HH}&=
-\frac{1}{v_\Delta^2}\Bigg\{
c_\alpha^2 m_{H^{++}}^2+\left[\frac{v_\Delta^2}{v^2}(3+c_{2\alpha})-(1+c_{2\alpha})\right]m_{H^+}^2
+\frac{1}{4}\frac{v^2(1+c_{2\alpha})-4v_\Delta^2}{v^2+2v_\Delta^2}m_A^2\notag\\
&+\frac{c_\alpha}{2} \left(c_\alpha^3-\frac{v_\Delta}{v_\phi}s_\alpha^3\right)m_H^2
+\frac{c_\alpha s_\alpha^2}{2}\left(c_\alpha+\frac{v_\Delta}{v_\phi}s_\alpha\right)m_h^2\Bigg\},\\
\lambda_{H^{++}H^{--}Hh}&=-\frac{s_{2\alpha}}{v_\Delta^2}
\Bigg[
m_{H^{++}}^2
+2\left(\frac{v_\Delta^2}{v^2}-1\right)m_{H^+}^2
+\frac{v^2}{2(v^2+2v_\Delta^2)}m_A^2\notag\\
&+\frac{1}{2}\left(c_\alpha^2+\frac{v_\Delta}{2v_\phi}s_{2\alpha}\right)m_H^2
-\frac{1}{2}\left(\frac{v_\Delta}{2v_\phi}s_{2\alpha}-s_\alpha^2\right)m_h^2
\Bigg],\\
\lambda_{H^{++}H^{--}hh}&=\frac{1}{v_\Delta^2}\Bigg\{
-s_\alpha^2 m_{H^{++}}^2+2\left[\left(1-\frac{v_\Delta^2}{v^2}\right)s_\alpha^2-\frac{v_\Delta^2}{v^2}\right]m_{H^+}^2\notag\\
&\hspace{-15mm}-\frac{v^2}{2(v^2+2v_\Delta^2)}\left(s_\alpha^2-\frac{2v_\Delta^2}{v^2}\right)m_A^2
+\frac{c_\alpha^2 s_\alpha}{2}\left(\frac{v_\Delta}{v_\phi}c_\alpha-s_\alpha\right)m_H^2
-\frac{s_\alpha}{2}\left(s_\alpha^3+\frac{v_\Delta}{v_\phi}c_\alpha^3\right)m_h^2\Bigg\},\\
\lambda_{H^+H^-AA}&=\frac{v^6-6v^4v_\Delta^2 +16v_\Delta^6}{2v^2v_\Delta^2(v^2+2v_\Delta^2)^2}m_A^2
-\frac{v^6-6v^4v_\Delta^2+12v^2v_\Delta^4}{4v^2v_\Delta^2(v^4-4v_\Delta^4)}(m_H^2+m_h^2)\notag\\
&\hspace{-15mm}-\frac{1}{4v^2v_\Delta^2(v^4-4v_\Delta^4)}[(v^6-6v^4v_\Delta^2+12v^2v_\Delta^4-16v_\Delta^6)c_{2\alpha}-6v_\Delta^3 v_\phi^3s_{2\alpha}](m_H^2-m_h^2),\\
\lambda_{H^+H^-AG^0}&=\frac{1}{v^2v_\Delta v_\phi (v^2+2v_\Delta^2)}
\Bigg\{
\frac{2(v^6-4v^4v_\Delta^2+8v_\Delta^6)}{v^2+2v_\Delta^2}m_A^2\notag\\
&+\left(-v^4+4v^2v_\Delta^2 -2v_\Delta^4
\right)(m_H^2+m_h^2)\notag\\
&+\left[-(v^4-4v^2v_\Delta^2+6v_\Delta^4)c_{2\alpha}-v_\Delta v_\phi (v^2-4v_\Delta^2)s_{2\alpha}\right](m_H^2-m_h^2)
\Bigg\},
\end{align}

\begin{align}
\lambda_{H^{+}H^{-}HH}&=
\frac{1}{2v^4}\left[-v^2-2v_\Delta^2+(v^2-6v_\Delta^2)c_{2\alpha}+4v_\Delta v_\phi s_{2\alpha}\right]m_{H^+}^2\notag\\
&+\frac{c_\alpha}{2v^2v_\Delta^2(v^2+2v_\Delta^2)}\left[(v^4-4v^2v_\Delta^2+8v_\Delta^4)c_\alpha-8v_\Delta^3v_\phi s_\alpha\right]m_A^2
\notag\\
&+\frac{1}{16}\left(-\frac{8c_\alpha^4}{v_\Delta^2}+\frac{9+4c_{2\alpha}+3c_{4\alpha}}{v^2}+\frac{8c_\alpha s_\alpha^3}{v_\Delta v_\phi}-\frac{8s_\alpha^4}{v_\phi^2}+\frac{4v_\Delta s_{4\alpha}}{v^2v_\phi}\right)m_H^2\notag\\
&+\frac{s_{2\alpha}}{8v^2v_\Delta v_\phi^3}\left[-v^4+2v^2v_\Delta^2+(v^4-6v^2v_\Delta^2+8v_\Delta^4)c_{2\alpha}-\frac{v_\phi}{v_\Delta}(v^4-4v^2v_\Delta^2+6v_\Delta^4)s_{2\alpha}\right]m_h^2,\\
\lambda_{H^{+}H^{-}Hh}&=
\frac{1}{v^4}\left[-4v_\Delta v_\phi c_{2\alpha}+(v^2-6v_\Delta^2)s_{2\alpha}\right]m_{H^+}^2\notag\\
&+\frac{1}{2v^2v_\Delta^2(v^2+2v_\Delta^2)}
\left[8v_\Delta^3v_\phi c_{2\alpha}+(v^4-4v^2v_\Delta^2+8v_\Delta^4)s_{2\alpha}\right]m_A^2\notag\\
&-\frac{s_{2\alpha}}{4v^2v_\Delta^2v_\phi^2}\left(v^4-4v^2v_\Delta^2+2v_\Delta^4\right)(m_H^2+m_h^2)\notag\\
&-\frac{s_{2\alpha}}{4v^2v_\Delta^2v_\phi^2}\left[(v^4-4v^2v_\Delta^2+6v_\Delta^4)c_{2\alpha}
+\frac{v_\Delta}{v_\phi}(v^4-6v^2v_\Delta^2+8v_\Delta^4)s_{2\alpha}
\right](m_H^2-m_h^2),\\
\lambda_{H^+H^-hh}&=-\frac{1}{2v^2}\left[1+\frac{2v_\Delta^2}{v^2}+\left(1-\frac{6v_\Delta^2}{v^2}\right)c_{2\alpha}+\frac{4v_\Delta v_\phi}{v^2} s_{2\alpha}\right]m_{H^+}^2\notag\\
&+\frac{v^2s_\alpha}{2v_\Delta^2(v^2+2v_\Delta^2)}\left[\frac{8v_\Delta^3v_\phi}{v^4} c_\alpha
+\left(1-\frac{4v_\Delta^2}{v^2}+\frac{8v_\Delta^4}{v^4}\right)s_\alpha\right]m_A^2\notag\\
&+\frac{v^2s_{2\alpha}}{8v_\Delta v_\phi^3}\left[\frac{v_\phi^2}{v^2}+ \left(1-\frac{6v_\Delta^2}{v^2}+\frac{8 v_\Delta^4}{v^4}\right)c_{2\alpha}-\frac{v_\phi}{v_\Delta}\left(1-\frac{4v_\Delta^2}{v^2}+\frac{6v_\Delta^4}{v^4}\right)s_{2\alpha}\right]m_H^2\notag\\
&+\frac{1}{16}\left[-\frac{8c_\alpha^4}{v_\phi^2}-\frac{8s_\alpha}{v_\Delta^2}\left(\frac{v_\Delta}{v_\phi}c_\alpha^3
+s_\alpha^3\right)+\frac{1}{v^2}\left(9-4c_{2\alpha}+3c_{4\alpha}+\frac{4v_\Delta}{v_\phi}s_{4\alpha}\right)\right]m_h^2,\\
\lambda_{H^+G^-hh}&=
\frac{1}{\sqrt{2}v^2}\Bigg\{
\frac{1}{v^2}
\left[v_\Delta v_\phi(1-3c_{2\alpha})
+(v^2-4v_\Delta^2)s_{2\alpha}
\right]m_{H^+}^2\notag\\
&+\frac{(v^2-4v_\Delta^2)s_\alpha}{v_\Delta v_\phi (v^2+2v_\Delta^2)}
\left[v^2 s_\alpha-2v_\Delta (v_\phi c_\alpha +v_\Delta s_\alpha)\right]m_A^2\notag\\
&+\frac{s_{2\alpha}}{4v_\Delta v_\phi}
\left[2v_\Delta v_\phi c_{2\alpha}-(v^2-3v_\Delta^2)s_{2\alpha}\right]m_H^2\notag\\
&+\frac{1}{8v_\Delta v_\phi}
\left[-3v^2+9v_\Delta^2+4(v^2-v_\Delta^2)c_{2\alpha}-(v^2-3v_\Delta^2)c_{4\alpha}-2v_\Delta v_\phi s_{4\alpha}\right]m_h^2
\Bigg\},\\
\lambda_{G^+G^-AA}&=\frac{1}{v^2+2v_\Delta^2}\Bigg\{
-2m_{H^+}^2 +\frac{2v_\phi^2}{v^2}m_A^2-\frac{1}{2}(m_H^2+m_h^2)\notag\\
&+\frac{1}{4v^2}\left[-2(v^2-4v_\Delta^2)c_{2\alpha}+\frac{1}{v_\phi v_\Delta}(v^4-4v^2v_\Delta^2 +12 v_\Delta^4)s_{2\alpha}\right](m_H^2-m_h^2)\Bigg\},
\end{align}

\begin{align}
\lambda_{G^+G^-AG^0}&=\frac{1}{v^2(v^2+2v_\Delta^2)}
\Bigg\{
-4v_\Delta v_\phi (m_{H^+}^2-m_A^2)-v_\Delta v_\phi (m_H^2+m_h^2)\notag\\
&+[-3v_\Delta v_\phi c_{2\alpha}+(v^2-4v_\Delta^2)s_{2\alpha}](m_H^2-m_h^2)\Bigg\},\\
\lambda_{G^+G^-Hh}&=\frac{1}{v^2}\Bigg\{
\frac{1}{v^2}\left[4v_\Delta v_\phi c_{2\alpha}-2(v^2-3v_\Delta^2)s_{2\alpha}\right]m_{H^+}^2\notag\\
&+\frac{1}{v^2+2v_\Delta^2}\left[-4v_\Delta v_\phi c_{2\alpha}+2v_\phi^2 s_{2\alpha}\right]m_A^2\notag\\
&-\frac{s_{2\alpha}}{4}(m_H^2+m_h^2)-\frac{s_{2\alpha}}{4}\left[3c_{2\alpha}-\frac{v^2-4v_\Delta^2}{v_\Delta v_\phi}s_{2\alpha}\right](m_H^2-m_h^2)\Bigg\},\\
\lambda_{G^+G^-hh}&=
\frac{1}{v^2}\Bigg\{
\frac{1}{v^2}\left[-v^2+v_\Delta^2+(v^2-3v_\Delta^2)c_{2\alpha}+2v_\Delta v_\phi s_{2\alpha}\right]m_{H^+}^2\notag\\
&+\frac{2s_\alpha}{v^2+2v_\Delta^2}\left[v^2s_\alpha-2v_\Delta (v_\phi c_\alpha +v_\Delta s_\alpha)\right]m_A^2\notag\\
&+\frac{s_{2\alpha}}{8v_\Delta v_\phi}\left[v^2-(v^2-4v_\Delta^2)c_{2\alpha}-3v_\Delta v_\phi s_{2\alpha}
\right]m_H^2\notag\\
&-\frac{1}{16}\left[9 +\frac{8v^2c_\alpha s_\alpha^3}{v_\phi v_\Delta}+ 
\left(-4c_{2\alpha}+3c_{4\alpha}+\frac{4v_\Delta}{v_\phi}s_{4\alpha}\right)\right]m_h^2\Bigg\},\\
\lambda_{AAAA}&=\frac{v^4-8v^2v_\Delta^2+12v_\Delta^4}{8v_\Delta^2(v^2+2v_\Delta^2)^2}m_A^2\notag\\
&-\frac{v^6-6v^4v_\Delta^2+12v^2v_\Delta^4+8v_\Delta^6}{16v_\Delta^2v_\phi^2(v^2+2v_\Delta^2)^2}(m_H^2+m_h^2)\notag\\
&\hspace{-10mm}
-\frac{1}{16v_\Delta^2v_\phi^2(v^2+2v_\Delta^2)^2}[(v^6-6v^4v_\Delta^2+12v^2v_\Delta^4-24v_\Delta^6)c_{2\alpha}-8v_\Delta^3v_\phi^3s_{2\alpha}](m_H^2-m_h^2),\\
\lambda_{AAAG^0}&=\frac{1}{v_\Delta(v^2+2v_\Delta^2)^2}\Bigg\{
(v^2-4v_\Delta^2)v_\phi m_A^2+\frac{-v^4+4v^2v_\Delta^2}{2v_\phi}(m_H^2+m_h^2)\notag\\
&+\frac{1}{2v_\phi}\left[
-(v^4-4v^2v_\Delta^2+8v_\Delta^4)c_{2\alpha}-v_\Delta v_\phi (v^2-6v_\Delta^2)s_{2\alpha}
\right](m_H^2-m_h^2)\Bigg\},\\
\lambda_{AAG^0G^0}&=\frac{1}{2(v^2+2v_\Delta^2)^2}\Bigg\{
(5v^2-14v_\Delta^2)m_A^2-3(v^2-v_\Delta^2)(m_H^2+m_h^2)\notag\\
&+\frac{1}{4}\left[-12(v^2-3v_\Delta^2)c_{2\alpha}+\frac{1}{v_\phi v_\Delta}(v^4-20v^2v_\Delta^2+52 v_\Delta^4)s_{2\alpha}\right](m_H^2-m_h^2)
\Bigg\},\\
\lambda_{AG^0G^0G^0}&=\frac{1}{2(v^2+2v_\Delta^2)^2}\Bigg\{
4v_\Delta v_\phi m_A^2-3v_\Delta v_\phi (m_H^2+m_h^2)
+\left[-5v_\Delta v_\phi c_{2\alpha}+(v^2-6v_\Delta^2)s_{2\alpha}\right](m_H^2-m_h^2)
\Bigg\},
\end{align}

\begin{align}
\lambda_{AAHH}&=
\frac{m_A^2}{8v_\Delta^2(v^2+2v_\Delta^2)^2}
[v^4-6v^2v_\Delta^2+(v^4-2v^2v_\Delta^2-8v_\Delta^4)c_{2\alpha}]\notag\\
&-\frac{m_H^2}{32v_\Delta^2(v^4-4v_\Delta^4)}[2(v^4-4v^2v_\Delta^2+8v_\Delta^4)(1+c_{2\alpha}^2)+4v^2c_{2\alpha}(v^2-4v_\Delta^2)\notag\\
&-\frac{2v_\Delta}{v_\phi}(v^4-4v_\Delta^4)s_{2\alpha}+\frac{v_\Delta}{v_\phi}(v^4-8v^2v_\Delta^2+12v_\Delta^4)s_{4\alpha}]\notag\\
&-\frac{s_{2\alpha}m_h^2}{16v_\Delta v_\phi^3 (v^2+2v_\Delta^2)}
[v^4-4v_\Delta^4-(v^4-8v^2v_\Delta^2+12v_\Delta^4)c_{2\alpha}+\frac{v_\phi}{v_\Delta}(v^4-4v^2v_\Delta^2+8v_\Delta^4)s_{2\alpha}],\\
\lambda_{AAHh}&=
\frac{s_{2\alpha}}{4v_\Delta^2}\Bigg\{
\frac{v^2-4v_\Delta^2}{v^2+2v_\Delta^2}m_A^2
-\frac{v^2(v^2-4v_\Delta^2)}{2v_\phi^2(v^2+2v_\Delta^2)}(m_H^2+m_h^2)\notag\\
&-\frac{m_H^2-m_h^2}{2v_\phi^2(v^2+2v_\Delta^2)}
\left[(v^4-4v^2v_\Delta^2+8v_\Delta^4)c_{2\alpha}+\frac{v_\Delta}{v_\phi}(v^4-8v^2v_\Delta^2+12v_\Delta^4)s_{2\alpha}\right]
\Bigg\},\\
\lambda_{AAhh}&=\frac{v^4}{8v_\Delta^2(v^2+2v_\Delta^2)^2}
\left[1-\frac{6v_\Delta^2}{v^2}-\left(1-\frac{2v_\Delta^2}{v^2}-\frac{8v_\Delta^4}{v^4}\right)c_{2\alpha}\right]m_A^2\notag\\
&\hspace{-10mm}+\frac{s_{2\alpha}v^4}{16v_\Delta v_\phi^3(v^2+2v_\Delta^2)}\left[1-\frac{4v_\Delta^4}{v^4}+\left(1-\frac{8v_\Delta^2}{v^2}+\frac{12v_\Delta^4}{v^4}\right)c_{2\alpha}-\frac{v_\phi}{v_\Delta}\left(1-\frac{4v_\Delta^2}{v^2}+\frac{8v_\Delta^4}{v^4}\right)s_{2\alpha}\right]m_H^2\notag\\
&-\frac{v^4}{16v_\Delta^2(v^4-4v_\Delta^4)}\Big[\left(1-\frac{4v_\Delta^2}{v^2}+\frac{8v_\Delta^4}{v^4}\right)(1+c_{2\alpha}^2)
-2c_{2\alpha}\left(1-\frac{4v_\Delta^2}{v^2}\right)\notag\\
&+\frac{v_\Delta}{v_\phi}s_{2\alpha}\left(1-\frac{4v_\Delta^4}{v^4}\right)+\frac{v_\Delta}{v_\phi}\left(1-\frac{8v_\Delta^2}{v^2}+\frac{12v_\Delta^4}{v^4}\right)s_{2\alpha}c_{2\alpha}\Big]m_h^2,\\
\lambda_{AG^0HH}&=\frac{1}{2v_\Delta v_\phi (v^2+2v_\Delta^2)}
\Bigg\{
\frac{v_\phi^4+(v^4-4v_\Delta^4)c_{2\alpha}}{v^2+2v_\Delta^2}m_A^2\notag\\
&+\frac{1}{4}\left[-3v^2+9v_\Delta^2+4(-v^2+v_\Delta^2)c_{2\alpha}-(v^2-3v_\Delta^2)c_{4\alpha}-2v_\Delta v_\phi s_{4\alpha} \right]m_H^2\notag\\
&+\frac{s_{2\alpha}}{2}\left[2v_\Delta v_\phi c_{2\alpha}-(v^2-3v_\Delta^2)s_{2\alpha}\right]m_h^2
\Bigg\},\\
\lambda_{AG^0hh}&=
\frac{1}{2v_\Delta v_\phi(v^2+2v_\Delta^2)}\Bigg\{
\frac{v_\phi^4-(v^4-4v_\Delta^4)c_{2\alpha}}{v^2+2v_\Delta^2}m_A^2
+\frac{s_{2\alpha}}{2}\left[2v_\Delta v_\phi c_{2\alpha}-(v^2-3v_\Delta^2)s_{2\alpha}\right]m_H^2\notag\\
&+\frac{1}{4}\left[-3v^2+9v_\Delta^2+4(v^2-v_\Delta^2)c_{2\alpha}-(v^2-3v_\Delta^2)c_{4\alpha}-2v_\Delta v_\phi s_{4\alpha}\right]m_h^2
\Bigg\},\\
\lambda_{G^0G^0Hh}&=\frac{s_{2\alpha}}{2(v^2+2v_\Delta^2)}
\left[m_A^2-\frac{3}{4}(m_H^2+m_h^2)-\frac{1}{4}\left(5c_{2\alpha}-\frac{v^2-6v_\Delta^2}{v_\phi v_\Delta}s_{2\alpha}\right)(m_H^2-m_h^2)\right],\\
\lambda_{G^0G^0hh}&=
\frac{1}{4(v^2+2v_\Delta^2)}\Bigg\{
\frac{1}{v^2+2v_\Delta^2}
\left[
v^2-6v_\Delta^2-(v^2+2v_\Delta^2)c_{2\alpha}
\right]m_A^2\notag\\
&+\frac{s_{2\alpha}}{4v_\Delta v_\phi}
\left[v^2+2v_\Delta^2-(v^2-6v_\Delta^2)c_{2\alpha}
-5v_\Delta v_\phi s_{2\alpha}\right]m_H^2\notag\\
&-\frac{1}{8v_\Delta v_\phi}
\left[
8v^2c_\alpha s_\alpha^3
+v_\Delta v_\phi (15-12 c_{2\alpha}+5c_{4\alpha})
+v_\Delta^2(4s_{2\alpha}+6s_{4\alpha})
\right]m_h^2
\Bigg\},
\end{align}

\begin{align}
\lambda_{HHHH}&=
\frac{1}{8v_\Delta^2(v^2+2v_\Delta^2)}(v_\phi^2c_\alpha^4-v_\Delta^2s_{2\alpha}^2)m_A^2\notag\\
&-\frac{1}{8v_\Delta^2v_\phi^2}\left[v_\phi^2c_\alpha^6-\frac{2v^4v_\Delta}{v_\phi(v^2+2v_\Delta^2)}c_\alpha^3s_\alpha^3+
v_\Delta^2s_\alpha^6+\frac{v_\Delta^5}{v_\phi(v^2+2v_\Delta^2)}s_{2\alpha}^3\right]m_H^2\notag\\
&-\frac{s_{2\alpha}^2}{64v_\Delta^2 v_\phi^2}\left[v^2-v_\Delta^2+(v^2-3v_\Delta^2)c_{2\alpha}+2v_\Delta v_\phi s_{2\alpha}\right]m_h^2,\\
\lambda_{HHHh}&=\frac{s_{2\alpha}}{8v_\Delta^2}\Bigg\{
\frac{v_\phi^2+(v^2+2v_\Delta^2)c_{2\alpha}}{v^2+2v_\Delta^2}m_A^2\notag\\
&-\frac{1}{4v_\phi^2}
\left[3v^2-9v_\Delta^2+4(v^2-v_\Delta^2)c_{2\alpha}+(v^2-3v_\Delta^2)c_{4\alpha}+2v_\Delta v_\phi s_{4\alpha}\right]m_H^2\notag\\
&-\frac{s_{2\alpha}}{2v_\phi^2}[-2v_\Delta v_\phi c_{2\alpha}+(v^2-3v_\Delta^2)s_{2\alpha}]m_h^2\Bigg\},\\
\lambda_{HHhh}&=\frac{v^2}{32v_\Delta^2(v^2+2v_\Delta^2)}\left[3-\frac{10v_\Delta^2}{v^2}-3\left(1+\frac{2v_\Delta^2}{v^2}\right)c_{4\alpha}\right]m_A^2
\notag\\
&+\frac{v^2s_{2\alpha}}{32v_\Delta v_\phi^3}\Bigg\{\left[1-\frac{2v_\Delta^2}{v^2}+
\frac{3v_\phi^2}{v^2}c_{4\alpha}+\left(-\frac{3v_\phi}{2v_\Delta}+\frac{9v_\Delta v_\phi}{2v^2}\right)s_{4\alpha}\right](m_H^2-m_h^2)\notag\\
&-\frac{3v_\phi}{v_\Delta}s_{2\alpha}\left(1-\frac{v_\Delta^2}{v^2}\right)
(m_H^2+m_h^2)\Bigg\},\\
\lambda_{Hhhh}&=\frac{s_{2\alpha}v^2}{4v_\Delta^2(v^2+2v_\Delta^2)}\left(s_\alpha^2-\frac{2v_\Delta^2}{v^2}c_\alpha^2\right)m_A^2\notag\\
&-\frac{s_{2\alpha}^2v^2}{16v_\Delta^2v_\phi^2}\left[\left(1-\frac{3v_\Delta^2}{v^2}\right)s_{2\alpha}-\frac{2v_\Delta v_\phi}{v^2}c_{2\alpha}\right]m_H^2\notag\\
&-\frac{s_{2\alpha}v^2}{16 v_\Delta^2 v_\phi^2}\left[(1+c_{2\alpha}^2)\left(1-\frac{3v_\Delta^2}{v^2}\right)-2\left(1-\frac{v_\Delta^2}{v^2}\right)c_{2\alpha}+\frac{v_\Delta v_\phi}{v^2}s_{4\alpha}\right]m_h^2,\\
\lambda_{hhhh}&=\frac{v^2}{8v_\Delta^2(v^2+2v_\Delta^2)}\left(\frac{v_\phi^2}{v^2}s_\alpha^4-\frac{v_\Delta^2}{v^2}s_{2\alpha}^2\right)m_A^2\notag\\
&-\frac{s_{2\alpha}^2v^2}{32v_\Delta^2 v_\phi^2}\left[s_\alpha^2\left(1-\frac{v_\Delta^2}{v^2}\right)+\frac{v_\Delta^2}{v^2}c_{2\alpha}
-\frac{v_\Delta v_\phi}{v^2}s_{2\alpha}\right]m_H^2\notag\\
&+\frac{1}{16}\left[-\frac{2}{v_\phi^2}c_\alpha^6-\frac{1}{v_\phi^3 v_\Delta}\left(\frac{v^2}{2}-v_\Delta^2\right)s_{2\alpha}^3-\frac{2}{v_\Delta^2}s_\alpha^6\right]m_h^2. 
\end{align}

\section{1PI diagram contributions}

In this Appendix, we summarize the formulae for the 
1PI diagram contributions to the one-, two- and three-point functions in terms of the Passarino-Veltman functions~\cite{PV}. 
The formulae for these 1PI diagram contributions are separately presented 
into the fermion-loop, scalar boson-loop and gauge-boson loop contributions. 
Each contribution is denoted by the subscript of $F$ for the fermion-loop, that of $S$ for the scalar boson-loop and 
that of $V$ for the gauge boson-loop. 
The gauge boson loop contributions are calculated in the 't Hooft-Feynman gauge.  
In the calculation of some physical observables such as $\Delta r$, 
divergent parts in each contribution are cancelled within a corresponding contribution. 
For instance, the divergent part from the scalar boson loop contribution is cancelled by the 
scalar boson loop contribution. 

We use the following definitions of the $A$, $B$ and $C$ functions according to the Passarino and Veltman~\cite{PV} 
as  
\begin{align}
A(m) &=\int \frac{d^Dk}{i\pi^2} \frac{1}{k^2-m^2}, \\
B_0;B^\mu;B^{\mu\nu}(p^2,m_1,m_2) &=\int \frac{d^Dk}{i\pi^2} \frac{1;k^\mu;k^\mu k^\nu}{(k^2-m_1^2)[(k+p)^2-m_2^2]}, \\
C_0;C^\mu;C^{\mu\nu}(p_1^2,p_2^2,q^2,m_1,m_2,m_3) &=\int \frac{d^Dk}{i\pi^2} \frac{1;k^\mu;k^\mu k^\nu}{(k^2-m_1^2)[(k+p_1)^2-m_2^2]
[(k+p_1+p_2)^2-m_3^2]},\notag\\
&\text{with }q^2 = (p_1+p_2)^2, 
\end{align}
where $D=4-2\epsilon$. 
The vector and tensor functions of $B$ and $C$ functions are rewitten in terms of the following form factors by
\begin{align}
B^\mu  &= p^\mu B_1,\\
B^{\mu\nu}&= p^\mu p^\nu B_{21}+g^{\mu\nu}B_{22},\\
C^\mu &= p_1^\mu C_{11}+p_2^\mu C_{12} ,\\
C^{\mu\nu} &= p_1^\mu p_1^\nu C_{21}+p_2^\mu p_2^\nu C_{22} + (p_1^\mu p_2^\nu + p_1^\nu p_2^\mu)C_{23} +g^{\mu\nu} C_{24}. 
\end{align}

\subsection{One-point functions}


The 1PI diagram contributions to the one-point function are calculated by 
\begin{align}
&T_{h,F}^{\text{1PI}}=-\frac{4m_f^2 N_c^f}{16\pi^2}\frac{c_\alpha}{v_\phi}A(m_f),\\
&T_{H,F}^{\text{1PI}}=+\frac{4m_f^2 N_c^f}{16\pi^2}\frac{s_\alpha}{v_\phi}A(m_f), 
\end{align}
\begin{align}
T_{h,S}^{\text{1PI}}&=-\frac{1}{16\pi^2}[\lambda_{H^{++}H^{--}h}A(m_{H^{++}})+\lambda_{H^{+}H^{-}h}A(m_{H^+})
+\lambda_{AAh}A(m_A)\notag\\
&+\lambda_{HHh}A(m_H)+3\lambda_{hhh}A(m_h)],\\
T_{H,S}^{\text{1PI}}&=-\frac{1}{16\pi^2}[\lambda_{H^{++}H^{--}H}A(m_{H^{++}})+\lambda_{H^{+}H^{-}H}A(m_{H^+})\notag\\
&+\lambda_{AAH}A(m_A)+3\lambda_{HHH}A(m_H)+\lambda_{Hhh}A(m_h)],
\end{align}
\begin{align}
T_{h,V}^{\text{1PI}}&=\frac{1}{16\pi^2}\Bigg[-\lambda_{G^+G^-h}A(m_{G^+})-\lambda_{G^0G^0h}A(m_{G^0})\notag\\
&+gm_W(c_\beta c_\alpha +\sqrt{2}s_\beta s_\alpha)DA(m_W)+\frac{g_Zm_Z}{2}(c_{\beta'} c_\alpha +2s_{\beta'} s_\alpha)DA(m_Z)\notag\\
&-gm_W(c_\beta c_\alpha +\sqrt{2}s_\beta s_\alpha)A(m_{c^+})-\frac{g_Zm_Z}{2}(c_{\beta'} c_\alpha +2s_{\beta'} s_\alpha)A(m_{c_Z})
\Bigg],\\
T_{H,V}^{\text{1PI}}&=\frac{1}{16\pi^2}\Bigg[-\lambda_{G^+G^-H}A(m_{G^+})-\lambda_{G^0G^0H}A(m_{G^0})\notag\\
&+gm_W(-c_\beta s_\alpha +\sqrt{2}s_\beta c_\alpha)DA(m_W)+\frac{g_Zm_Z}{2}(-c_{\beta'} s_\alpha +2s_{\beta'} c_\alpha)DA(m_Z)\notag\\
&-gm_W(-c_\beta s_\alpha +\sqrt{2}s_\beta c_\alpha)A(m_{c^+})-\frac{g_Zm_Z}{2}(-c_{\beta'} s_\alpha +2s_{\beta'} c_\alpha)A(m_{c_Z})
\Bigg], 
\end{align}
where $m_{G^+}$ ($m_{c^+}$) and $m_{G^0}$ ($m_{c_Z}$) are the masses of the NG bosons $G^\pm$ and $G^0$ 
(ghost fields $c^\pm$ and $c_Z$ ), respectively. 
In the 't Hooft-Feynman gauge, these masses are the same as the corresponding gauge boson masses, i.e., 
$m_{G^+}=m_{c^+}=m_W$ and $m_{G^0}=m_{c_Z}=m_Z$.


\subsection{Two-point functions}

The 1PI diagram contributions to the scalar boson two point functions are calculated as  
\begin{align}
\Pi_{hh}^{\text{1PI}}(p^2)_F&=-\frac{4m_f^2 N_c^f}{16\pi^2}
\frac{c_\alpha^2}{v_\phi^2}\left[A(m_f)+\left(2m_f^2-\frac{p^2}{2}\right)B_0(p^2,m_f,m_f)\right],\\
\Pi_{HH}^{\text{1PI}}(p^2)_F&=-\frac{4m_f^2 N_c^f}{16\pi^2}
\frac{s_\alpha^2}{v_\phi^2}\left[A(m_f)+\left(2m_f^2-\frac{p^2}{2}\right)B_0(p^2,m_f,m_f)\right],\\
\Pi_{Hh}^{\text{1PI}}(p^2)_F&=\frac{4m_f^2 N_c^f}{16\pi^2}
\frac{c_\alpha s_\alpha}{v_\phi^2}\left[A(m_f)+\left(2m_f^2-\frac{p^2}{2}\right)B_0(p^2,m_f,m_f)\right],\\
\Pi_{AA}^{\text{1PI}}(p^2)_F&=-\frac{4m_f^2 N_c^f}{16\pi^2}
\frac{s_{\beta^\prime}^2}{v_\phi^2}\left[A(m_f)-\frac{p^2}{2}B_0(p^2,m_f,m_f)\right],\\
\Pi_{AG}^{\text{1PI}}(p^2)_F&=+\frac{2m_f^2 N_c^f}{16\pi^2}
\frac{s_{2\beta^\prime}}{v_\phi^2}\left[A(m_f)-\frac{p^2}{2}B_0(p^2,m_f,m_f)\right], 
\end{align}
\begin{align}
&\Pi_{hh}^{\text{1PI}}(p^2)_S=
\frac{1}{16\pi^2}[-2\lambda_{H^{++}H^{--}hh}A(m_{H^{++}})-2\lambda_{H^{+}H^{-}hh}A(m_{H^+})
\notag\\&
-2\lambda_{AAhh}A(m_A)-2\lambda_{HHhh}A(m_H)+12\lambda_{hhhh}A(m_h)
\notag\\
&+\lambda_{H^{++}H^{--}h}^2B_0(p^2,m_{H^{++}},m_{H^{++}})\notag\\
&+\lambda_{H^{+}H^{-}h}^2B_0(p^2,m_{H^+},m_{H^+})
+\lambda_{G^{+}G^{-}h}^2B_0(p^2,m_{G^+},m_{G^+})+2\lambda_{H^{+}G^{-}h}^2B_0(p^2,m_{H^+},m_{G^+})\notag\\
&+2\lambda_{AAh}^2B_0(p^2,m_A,m_A)+2\lambda_{G^0G^0h}^2B_0(p^2,m_{G^0},m_{G^0})
+\lambda_{AG^0h}^2B_0(p^2,m_A,m_{G^0})\notag\\
&+2\lambda_{HHh}^2B_0(p^2,m_H,m_H)
+18\lambda_{hhh}^2B_0(p^2,m_h,m_h)
+4\lambda_{Hhh}^2B_0(p^2,m_h,m_H)],\\
&\Pi_{HH}^{\text{1PI}}(p^2)_S=
\frac{1}{16\pi^2}[-2\lambda_{H^{++}H^{--}HH}A(m_{H^{++}})-2\lambda_{H^{+}H^{-}HH}A(m_{H^+})\notag\\
&-2\lambda_{HHAA}A(m_A)+12\lambda_{HHHH}A(m_H)-2\lambda_{HHhh}A(m_h)]\notag\\
&+\lambda_{H^{++}H^{--}H}^2B_0(p^2,m_{H^{++}},m_{H^{++}})\notag\\
&+\lambda_{H^{+}H^{-}H}^2B_0(p^2,m_{H^+},m_{H^+})
+\lambda_{G^{+}G^{-}H}^2B_0(p^2,m_{G^+},m_{G^+})+2\lambda_{H^{+}G^{-}H}^2B_0(p^2,m_{H^+},m_{G^+})\notag\\
&+2\lambda_{AAH}^2B_0(p^2,m_A,m_A)+2\lambda_{G^0G^0H}^2B_0(p^2,m_{G^0},m_{G^0})
+\lambda_{AG^0H}^2B_0(p^2,m_A,m_{G^0})\notag\\
&+18\lambda_{HHH}^2B_0(p^2,m_H,m_H)
+2\lambda_{Hhh}^2B_0(p^2,m_h,m_h)
+4\lambda_{HHh}^2B_0(p^2,m_h,m_H)],\\
&\Pi_{Hh}^{\text{1PI}}(p^2)_S=
\frac{1}{16\pi^2}[-\lambda_{H^{++}H^{--}Hh}A(m_{H^{++}})
-\lambda_{H^{+}H^{-}Hh}A(m_{H^{+}})\notag\\
&-\lambda_{AAHh}A(m_A)-3\lambda_{HHHh}A(m_H)-3\lambda_{Hhhh}A(m_h)
\notag\\
&+\lambda_{H^{++}H^{--}h}\lambda_{H^{++}H^{--}H} B_0(p^2,m_{H^{++}},m_{H^{++}})
+\lambda_{H^{+}H^{-}h}\lambda_{H^{+}H^{-}H}B_0(p^2,m_{H^+},m_{H^+})\notag\\
&+\lambda_{G^{+}G^{-}h}\lambda_{G^{+}G^{-}H}B_0(p^2,m_{G^+},m_{G^+})
+2\lambda_{H^{+}G^{-}h}\lambda_{H^{+}G^{-}H}B_0(p^2,m_{H^+},m_{G^+})\notag\\
&+2\lambda_{AAh}\lambda_{AAH}B_0(p^2,m_A,m_A)+2\lambda_{hG^0G^0}\lambda_{G^0G^0H}B_0(p^2,m_{G^0},m_{G^0})\notag\\
&+\lambda_{AG^0h}\lambda_{AG^0H}B_0(p^2,m_A,m_{G^0})+6\lambda_{HHh}\lambda_{HHH}B_0(p^2,m_H,m_H)\notag\\
&+6\lambda_{hhh}\lambda_{Hhh}B_0(p^2,m_h,m_h)+4\lambda_{Hhh}\lambda_{HHh}B_0(p^2,m_H,m_h)],\\
&\Pi_{AA}^{\text{1PI}}(p^2)_S=
-\frac{1}{16\pi^2}[2\lambda_{H^{++}H^{--}AA}A(m_{H^{++}})
+2\lambda_{H^{+}H^{-}AA}A(m_{H^{+}})\notag\\
&+12\lambda_{AAAA}A(m_A)+2\lambda_{AAHH}A(m_H)+2\lambda_{AAhh}A(m_h)
]\notag\\
&+\frac{1}{16\pi^2}
[2\lambda_{H^+G^-A}\lambda_{H^+G^-A}^* B_0(p^2,m_{H^{+}},m_{G^+})
+4\lambda_{AAh}^2B_0(p^2,m_A,m_h)\notag\\
&+4\lambda_{AAH}^2B_0(p^2,m_A,m_H)
+\lambda_{AG^0h}^2B_0(p^2,m_h,m_{G^0})+\lambda_{AG^0H}^2B_0(p^2,m_H,m_{G^0})],
\end{align}
\begin{align}
&\Pi_{AG}^{\text{1PI}}(p^2)_S=
-\frac{1}{16\pi^2}[\lambda_{H^{++}H^{--}AG^0}A(m_{H^{++}})
+\lambda_{H^{+}H^{-}AG^0}A(m_{H^{+}})\notag\\
&+3\lambda_{AAAG^0}A(m_A)+\lambda_{AG^0HH}A(m_H)+\lambda_{AG^0hh}A(m_h)
]\notag\\
&+\frac{2}{16\pi^2}
[\lambda_{H^+G^-A}\lambda_{H^+G^-G^0}^* B_0(p^2,m_{H^{+}},m_{G^+})\notag\\
&+\lambda_{AAh}\lambda_{AG^0h}B_0(p^2,m_A,m_h)+\lambda_{AAH}\lambda_{AG^0H}B_0(p^2,m_A,m_H)\notag\\
&+\lambda_{AG^0h}\lambda_{G^0G^0h}B_0(p^2,m_{G^0},m_h)+\lambda_{AG^0H}\lambda_{G^0G^0H}B_0(p^2,m_{G^0},m_H),\\
&\Pi_{H^+H^-}^{\text{1PI}}(p^2)_S=
-\frac{1}{16\pi^2}[\lambda_{H^{++}H^{--}H^{+}H^{-}}A(m_{H^{++}})
+4\lambda_{H^{+}H^{-}H^{+}H^{-}}A(m_{H^{+}})\notag\\
&+\lambda_{H^{+}H^{-}AA}A(m_A)+\lambda_{H^{+}H^{-}HH}A(m_H)+\lambda_{H^{+}H^{-}hh}A(m_h)
]\notag\\
&+\frac{1}{16\pi^2}
[4\lambda_{H^{++}H^-H^-}^2 B_0(p^2,m_{H^{++}},m_{H^+})
+\lambda_{H^{++}H^-G^-}^2 B_0(p^2,m_{H^{++}},m_{G^+})\notag\\
&+\lambda_{H^+H^-H}^2 B_0(p^2,m_{H^{+}},m_H)+\lambda_{H^+H^-h}^2 B_0(p^2,m_{H^{+}},m_h)\notag\\
&+\lambda_{H^+G^-H}^2 B_0(p^2,m_H,m_{G^+})+\lambda_{H^+G^-A}\lambda_{H^+G^-A}^* B_0(p^2,m_A,m_{G^+})\notag\\
&+\lambda_{H^+G^-h}^2 B_0(p^2,m_h,m_{G^+})+\lambda_{H^+G^-G^0}\lambda_{H^+G^-G^0}^* B_0(p^2,m_{G^0},m_{G^+})],\\
&\Pi_{H^{++}H^{--}}^{\text{1PI}}(p^2)_S=
-\frac{1}{16\pi^2}[4\lambda_{H^{++}H^{--}H^{++}H^{--}}A(m_{H^{++}})
+\lambda_{H^{++}H^{--}H^{+}H^{-}}A(m_{H^{+}})\notag\\
&+\lambda_{H^{++}H^{--}AA}A(m_A)+\lambda_{H^{++}H^{--}HH}A(m_H)+\lambda_{H^{++}H^{--}hh}A(m_h)
]\notag\\
&+\frac{1}{16\pi^2}
[2\lambda_{H^{++}H^-H^-}^2 B_0(p^2,m_{H^{+}},m_{H^+})
+2\lambda_{H^{++}G^-G^-}^2 B_0(p^2,m_{G^+},m_{G^+})\notag\\
&+\lambda_{H^{++}H^-G^-}^2 B_0(p^2,m_{H^{+}},m_{G^+})\notag\\
&+\lambda_{H^{++}H^{--}H}^2 B_0(p^2,m_{H^{++}},m_H)+\lambda_{H^{++}H^{--}h}^2 B_0(p^2,m_{H^{++}},m_h)],\\
&\Pi_{hh}^{\text{1PI}}(p^2)_V
=\frac{1}{16\pi^2}\Bigg\{
g^2m_W^2(c_\beta c_\alpha +\sqrt{2}s_\beta s_\alpha)^2DB_0(p^2,m_W,m_W)
+\frac{g^2}{4}(3-c_{2\alpha})DA(m_W)\notag\\
&-\frac{g^2}{2}(c_\alpha c_\beta+\sqrt{2}s_\alpha s_\beta)^2[2A(m_{W})-A(m_{G^+})+(2m_{G^+}^2-m_W^2+2p^2)B_0(p^2,m_W,m_{G^+})]\notag\\
&-\frac{g^2}{2}(-c_\alpha s_\beta+\sqrt{2}s_\alpha c_\beta)^2[2A(m_{W})-A(m_{H^+})+(2m_{H^+}^2-m_W^2+2p^2)B_0(p^2,m_W,m_{H^+})]\notag\\
&-2\lambda_{G^+G^-hh}A(m_{G^+})
-\frac{g^2m_W^2}{2}(c_{\beta}c_\alpha+\sqrt{2}s_{\beta}s_\alpha)^2B_0(p^2,m_{c^+},m_{c^+})\notag\\
&+\frac{g_Z^2m_Z^2}{2}(c_{\beta'} c_\alpha +2s_{\beta'} s_\alpha)^2DB_0(p^2,m_Z,m_Z)
+\frac{g_Z^2}{8}(5-3c_{2\alpha})DA(m_Z)\notag\\
&-\frac{g_Z^2}{4}(c_\alpha c_{\beta'}+2s_\alpha s_{\beta'})^2[2A(m_{Z})-A(m_{G^0})+(2m_{G^0}^2-m_Z^2+2p^2)B_0(p^2,m_Z,m_{G^0})]\notag\\
&-\frac{g_Z^2}{4}(-c_\alpha s_{\beta'}+2s_\alpha c_{\beta'})^2[2A(m_{Z})-A(m_A)+(2m_{A}^2-m_Z^2+2p^2)B_0(p^2,m_Z,m_A)]\notag\\
&-2\lambda_{G^0G^0hh}A(m_{G^0})
-\frac{g_Z^2m_Z^2}{4}(c_{\beta'}c_\alpha+2s_{\beta'}s_\alpha)^2B_0(p^2,m_{c_Z},m_{c_Z})\Bigg\},
\end{align}
\begin{align}
&\Pi_{HH}^{\text{1PI}}(p^2)_V
=\frac{1}{16\pi^2}\Bigg\{
g^2m_W^2(-c_\beta s_\alpha +\sqrt{2}s_\beta c_\alpha)^2DB_0(p^2,m_W,m_W)
+\frac{g^2}{4}(3+c_{2\alpha})DA(m_W)\notag\\
&-\frac{g^2}{2}(-s_\alpha c_\beta+\sqrt{2}c_\alpha s_\beta)^2[2A(m_{W})-A(m_{G^+})+(2m_{G^+}^2-m_W^2+2p^2)B_0(p^2,m_W,m_{G^+})]\notag\\
&-\frac{g^2}{2}(s_\alpha s_\beta+\sqrt{2}c_\alpha c_\beta)^2[2A(m_{W})-A(m_{H^+})+(2m_{H^+}^2-m_W^2+2p^2)B_0(p^2,m_W,m_{H^+})]\notag\\
&-2\lambda_{G^+G^-HH}A(m_{G^+})
-\frac{g^2m_W^2}{2}(-c_{\beta}s_\alpha+\sqrt{2}s_{\beta}c_\alpha)^2B_0(p^2,m_{c^+},m_{c^+})\notag\\
&+\frac{g_Z^2m_Z^2}{2}(-c_{\beta'} s_\alpha +2s_{\beta'} c_\alpha)^2DB_0(p^2,m_Z,m_Z)
+\frac{g_Z^2}{8}(5+3c_{2\alpha})DA(m_Z)\notag\\
&-\frac{g_Z^2}{4}(-s_\alpha c_{\beta'}+2c_\alpha s_{\beta'})^2[2A(m_{Z})-A(m_{G^0})+(2m_{G^0}^2-m_Z^2+2p^2)B_0(p^2,m_Z,m_{G^0})]\notag\\
&-\frac{g_Z^2}{4}(2c_\alpha c_{\beta'}+s_\alpha s_{\beta'})^2[2A(m_{Z})-A(m_A)+(2m_{A}^2-m_Z^2+2p^2)B_0(p^2,m_Z,m_A)]\notag\\
&-2\lambda_{G^0G^0HH}A(m_{G^0})
-\frac{g_Z^2m_Z^2}{4}(-c_{\beta'}s_\alpha+2s_{\beta'}c_\alpha)^2B_0(p^2,m_{c_Z},m_{c_Z})\Bigg\},\\
&\Pi_{Hh}^{\text{1PI}}(p^2)_V
=\frac{1}{16\pi^2}\Bigg\{
g^2m_W^2(c_\beta c_\alpha +\sqrt{2}s_\beta s_\alpha)(-c_\beta s_\alpha +\sqrt{2}s_\beta c_\alpha)DB_0(p^2,m_W,m_W)
+\frac{g^2}{4}s_{2\alpha}DA(m_W)\notag\\
&-\frac{g^2}{2}(c_\alpha c_\beta+\sqrt{2}s_\alpha s_\beta)(-s_\alpha c_\beta+\sqrt{2}c_\alpha s_\beta)
[2A(m_{W})-A(m_{G^+})+(2m_{G^+}^2-m_W^2+2p^2)B_0(p^2,m_W,m_{G^+})]\notag\\
&-\frac{g^2}{2}(-c_\alpha s_\beta+\sqrt{2}s_\alpha c_\beta)(s_\alpha s_\beta+\sqrt{2}c_\alpha c_\beta)
[2A(m_{W})-A(m_{H^+})+(2m_{H^+}^2-m_W^2+2p^2)B_0(p^2,m_W,m_{H^+})]\notag\\
&-\lambda_{G^+G^-Hh}A(m_{G^+})
-\frac{g^2m_W^2}{2}(c_{\beta}c_\alpha+\sqrt{2}s_{\beta}s_\alpha)(-c_{\beta}s_\alpha+\sqrt{2}s_{\beta}c_\alpha)
B_0(p^2,m_{c^+},m_{c^+})\notag\\
&+\frac{g_Z^2m_Z^2}{2}(c_{\beta'} c_\alpha +2s_{\beta'} s_\alpha)(-c_{\beta'} s_\alpha +2s_{\beta'} c_\alpha)
DB_0(p^2,m_Z,m_Z)
+\frac{3g_Z^2}{8}s_{2\alpha}DA(m_Z)\notag\\
&-\frac{g_Z^2}{4}(c_\alpha c_{\beta'}+2s_\alpha s_{\beta'})(-s_\alpha c_{\beta'}+2c_\alpha s_{\beta'})
[2A(m_{Z})-A(m_{G^0})+(2m_{G^0}^2-m_Z^2+2p^2)B_0(p^2,m_Z,m_{G^0})]\notag\\
&-\frac{g_Z^2}{4}(-c_\alpha s_{\beta'}+2s_\alpha c_{\beta'})(s_\alpha s_{\beta'}+2c_\alpha c_{\beta'})
[2A(m_{Z})-A(m_A)+(2m_{A}^2-m_Z^2+2p^2)B_0(p^2,m_Z,m_A)]\notag\\
&-\lambda_{G^0G^0Hh}A(m_{G^0})
-\frac{g_Z^2m_Z^2}{4}(c_{\beta'}c_\alpha+2s_{\beta'}s_\alpha)(-c_{\beta'}s_\alpha+2s_{\beta'}c_\alpha)
B_0(p^2,m_{c_Z},m_{c_Z})\Bigg\},
\end{align}
\begin{align}
&\Pi_{AA}^{\text{1PI}}(p^2)_V=\frac{1}{16\pi^2}\Bigg\{
\frac{g^2}{4}(3+c_{2\beta'})DA(m_W)-2\lambda_{G^+G^-AA}A(m_{G^+})\notag\\
&-\frac{g^2}{2}(-c_\beta s_{\beta'}+\sqrt{2}s_\beta c_{\beta'})^2[2A(m_{W})-A(m_{G^+})+(2m_{G^+}^2-m_W^2+2p^2)B_0(p^2,m_W,m_{G^+})]\notag\\
&-\frac{g^2}{2}(s_\beta s_{\beta'}+\sqrt{2}c_\beta c_{\beta'})^2[2A(m_{W})-A(m_{H^+})+(2m_{H^+}^2-m_W^2+2p^2)B_0(p^2,m_W,m_{H^+})]\notag\\
&+\frac{g^2m_W^2}{2}(-c_{\beta}s_{\beta'}+\sqrt{2}s_{\beta}c_{\beta'})^2B_0(p^2,m_{c^+},m_{c^+})\notag\\
&+\frac{g_Z^2}{8}(5+3c_{2\beta'})DA(m_Z)-2\lambda_{AAG^0G^0}A(m_{G^0})\notag\\
&-\frac{g_Z^2}{4}(-c_\alpha s_{\beta'}+2s_\alpha c_{\beta'})^2[2A(m_{Z})-A(m_{h})+(2m_{h}^2-m_Z^2+2p^2)B_0(p^2,m_Z,m_{h})]\notag\\
&-\frac{g_Z^2}{4}(s_\alpha s_{\beta'}+2c_\alpha c_{\beta'})^2[2A(m_{Z})-A(m_H)+(2m_{H}^2-m_Z^2+2p^2)B_0(p^2,m_Z,m_H)]\Bigg\}, \\
&\Pi_{AG}^{\text{1PI}}(p^2)_V=\frac{1}{16\pi^2}\Bigg\{\frac{g^2}{4}s_{2\beta'}DA(m_W)+\frac{3}{8}g_Z^2s_{2\beta'}DA(m_Z)\notag\\
&-\frac{g^2}{2}(s_\beta s_{\beta'}+\sqrt{2}c_\beta c_{\beta'})(-s_\beta c_{\beta'}+\sqrt{2}c_\beta s_{\beta'})
[2A(m_{W})-A(m_{H^+})+(2m_{H^+}^2-m_W^2+2p^2)B_0(p^2,m_W,m_{H^+})]\notag\\
&-\frac{g_Z^2}{4}(-c_\alpha s_{\beta'}+2s_\alpha c_{\beta'})(c_\alpha c_{\beta'}+2s_\alpha s_{\beta'})
[2A(m_Z)-A(m_h)+(2m_h^2-m_Z^2+2p^2)B_0(p^2,m_Z,m_h)]\notag\\
&-\frac{g_Z^2}{4}(2c_\alpha c_{\beta'}+s_\alpha s_{\beta'})(-s_\alpha c_{\beta'}+2c_\alpha s_{\beta'})
[2A(m_Z)-A(m_H)+(2m_H^2-m_Z^2+2p^2)B_0(p^2,m_Z,m_H)]\notag\\
&-3\lambda_{AG^0G^0G^0}A(m_{G^0})-\lambda_{G^+G^-AG^0}A(m_{G^+})\Bigg\}. 
\end{align}
The 1PI diagram contributions to the gauge boson two point functions are calculated as follows. 
The fermion-loop contributions are  
\begin{align}
\Pi_{WW}^{\text{1PI}}(p^2)_F&=\frac{g^2}{16\pi^2}N_c^f\Big[-B_4+2p^2B_3\Big](p^2,m_f,m_{f'}),\\
\Pi_{ZZ}^{\text{1PI}}(p^2)_F&=\frac{g_Z^2}{16\pi^2}N_c^f\Big[2p^2(4s_W^4Q_f^2-4s_W^2Q_fI_f+2I_f^2)B_3-2I_f^2m_f^2B_0\Big](p^2,m_f,m_f),\\
\Pi_{\gamma\gamma}^{\text{1PI}}(p^2)_F&=\frac{e^2}{16\pi^2}N_c^fQ_f^2\Big[8p^2B_3\Big](p^2,m_f,m_f),\\
\Pi_{Z\gamma}^{\text{1PI}}(p^2)_F&=-\frac{eg_Z}{16\pi^2}N_c^f\Big[2p^2(-4s_W^2Q_f^2+2I_fQ_f)B_3\Big](p^2,m_f,m_f), 
\end{align}
where $B_3(p^2,m_1,m_2)=-B_1(p^2,m_1,m_2)-B_{21}(p^2,m_1,m_2)$ and $B_4(p^2,m_1,m_2)=-m_1^2B_1(p^2,m_2,m_1)-m_2^2B_1(p^2,m_1,m_2)$~\cite{HHKM}. 

The scalar-boson loop contirbutions are
\begin{align}
&\Pi_{WW}^{\text{1PI}}(p^2)_S
=\frac{1}{16\pi^2}\frac{g^2}{4}\Big[4c_{\beta}^2B_5(p^2,m_{H^{++}},m_{H^+})+4s_{\beta}^2B_5(p^2,m_{H^{++}},m_{G^+})\notag\\
&+(c_\alpha s_{\beta}-\sqrt{2}s_\alpha c_{\beta})^2B_5(p^2,m_{H^+},m_h)
+(c_\alpha c_{\beta}+\sqrt{2}s_\alpha s_{\beta})^2B_5(p^2,m_{G^+},m_h)\notag\\
&+(s_\alpha s_{\beta}+\sqrt{2}c_\alpha c_{\beta})^2B_5(p^2,m_{H^+},m_H)
+(s_\alpha c_{\beta}-\sqrt{2}c_\alpha s_{\beta})^2B_5(p^2,m_{G^+},m_H)\notag\\
&+(s_{\beta'} s_{\beta}+\sqrt{2}c_{\beta'} c_{\beta})^2B_5(p^2,m_{H^+},m_A)
+(s_{\beta'} c_{\beta}-\sqrt{2}c_{\beta'} s_{\beta})^2B_5(p^2,m_{G^+},m_A)\notag\\
&+(-c_{\beta'} s_{\beta}+\sqrt{2}s_{\beta'} c_{\beta})^2B_5(p^2,m_{H^+},m_Z)
+(c_{\beta'} c_{\beta}+\sqrt{2}s_{\beta'} s_{\beta})^2B_5(p^2,m_{G^+},m_{G^0}) \Big],\\
&\Pi_{ZZ}^{\text{1PI}}(p^2)_S
=\frac{1}{16\pi^2}\frac{g_Z^2}{4}\Big[4(c_W^2-s_W^2)^2B_5(p^2,m_{H^{++}},m_{H^{++}})
+(c_W^2-s_W^2-c_\beta^2)^2B_5(p^2,m_{H^{+}},m_{H^{+}})\notag\\
&+(c_W^2-s_W^2-s_\beta^2)^2B_5(p^2,m_{G^+},m_{G^+})
+2s_\beta^2c_\beta^2B_5(p^2,m_{H^+},m_{G^+})\notag\\
&+(2c_\alpha c_{\beta'}+s_\alpha s_{\beta'})^2B_5(p^2,m_H,m_A)
+(2s_\alpha c_{\beta'}-c_\alpha s_{\beta'})^2B_5(p^2,m_h,m_A)\notag\\
&+(s_\alpha c_{\beta'}-2c_\alpha s_{\beta'})^2B_5(p^2,m_H,m_{G^0})
+(c_\alpha c_{\beta'}+2s_\alpha s_{\beta'})^2B_5(p^2,m_h,m_{G^0})
\Big],\\
&\Pi_{\gamma\gamma}^{\text{1PI}}(p^2)_S=\frac{e^2}{16\pi^2}\Big[
4B_5(p^2,m_{H^{++}},m_{H^{++}})+B_5(p^2,m_{H^+},m_{H^+})+B_5(p^2,m_{G^+},m_{G^+})\Big],
\end{align}
\begin{align}
&\Pi_{Z\gamma }^{\text{1PI}}(p^2)_S
=-\frac{eg_Z}{16\pi^2}\Big[
2(c_W^2-s_W^2)B_5(p^2,m_{H^{++}},m_{H^{++}})\notag\\
&+\frac{1}{2}(c_W^2-s_W^2-c_{\beta}^2)B_5(p^2,m_{H^+},m_{H^+})
+\frac{1}{2}(c_W^2-s_W^2-s_{\beta_\pm}^2)B_5(p^2,m_{G^+},m_{G^+})\Big], 
\end{align}
where $B_5(p^2,m_1,m_2)=A(m_1)+A(m_2)-4B_{22}(p^2,m_1,m_2)$~\cite{HHKM}. 
The gauge boson loop contributions are
\begin{align}
\overline{\Pi}_{WW}^{\text{1PI}}(p^2)_V
&=\Pi_{WW}^{\text{1PI}}(p^2)_V-\frac{4g^2}{16\pi^2}(p^2-m_W^2)[c_W^2B_0(p^2,m_Z,m_W)+s_W^2B_0(p^2,0,m_W)],\notag\\
\overline{\Pi}_{ZZ}^{\text{1PI}}(p^2)_V
&=\Pi_{ZZ}^{\text{1PI}}(p^2)_V-\frac{4g_Z^2}{16\pi^2}c_W^4(p^2-m_Z^2)B_0(p^2,m_W,m_W),\notag\\
\overline{\Pi}_{\gamma\gamma}^{\text{1PI}}(p^2)_V
&=\Pi_{\gamma\gamma}^{\text{1PI}}(p^2)_V-\frac{4e^2}{16\pi^2}p^2B_0(p^2,m_W,m_W),\notag\\
\overline{\Pi}_{Z\gamma }^{\text{1PI}}(p^2)_V
&=\Pi_{Z\gamma }^{\text{1PI}}(p^2)_V+\frac{4eg_Z}{16\pi^2}c_W^2\left(p^2-\frac{1}{2}m_Z^2 \right)B_0(p^2,m_W,m_W),\label{bar}
\end{align}
where $\overline{\Pi}_{XY}^{\text{1PI}}(p^2)_V$ functions are the gauge invariant two point functions while 
$\Pi_{XY}^{\text{1PI}}(p^2)_V$ functions are the amplitude calculated in the 't Hooft-Feynman gauge. 
The second term of the right-hand side in Eq.~(\ref{bar}) corresponds to the pinch-terms~\cite{pinch} which 
are introduced to maintain the gauge invariance of the gauge boson two point functions.  
The $\Pi_{XY}^{\text{1PI}}(p^2)_V$ functions are calculated as 
\begin{align}
&\Pi_{WW}^{\text{1PI}}(p^2)_V=\frac{g^2}{16\pi^2}\Bigg\{
m_W^2\Big[
\left(c_\beta c_\alpha+\sqrt{2}s_\beta s_\alpha\right)^2
B_0(p^2,m_h,m_W)+\left(c_\beta s_\alpha-\sqrt{2}s_\beta c_\alpha\right)^2B_0(p^2,m_H,m_W)\notag\\
&+4s_\beta^2B_0(p^2,m_{H^{++}},m_W)
+\frac{c_\beta^2s_\beta^2}{c_W^2}B_0(p^2,m_{H^+},m_Z)
+s_W^2B_0(p^2,m_{G^+},0)
+\frac{(s_W^2+s_\beta^2)^2}{c_W^2}B_0(p^2,m_{G^+},m_Z)
\Big]\notag\\
&-c_W^2\left[(6D-8)B_{22}+p^2(2B_{21}+2B_1+5B_0)\right](p^2,m_Z,m_W)
+(D-1)\left[c_W^2A(m_Z)+A(m_W)\right]\notag\\
&-s_W^2\left[(6D-8)B_{22}+p^2(2B_{21}+2B_1+5B_0)\right](p^2,0,m_W)\Bigg\},\\
&\Pi_{ZZ}^{\text{1PI}}(p^2)_V=\frac{g_Z^2}{16\pi^2}\Bigg\{
m_Z^2\Big[(c_{\beta'}c_\alpha+2s_{\beta'}s_\alpha)^2B_0(p^2,m_h,m_Z)+
(c_{\beta'}s_\alpha-2s_{\beta'}c_\alpha)^2B_0(p^2,m_H,m_Z)\Big]\notag\\
&+m_W^2\Big[2c_\beta^2s_\beta^2B_0(p^2,m_{H^+},m_W)
+2\left(s_W^2+s_\beta^2\right)^2B_0(p^2,m_{G^+},m_W)\Big]\notag\\
&-c_W^4\left[(6D-8)B_{22}+p^2(2B_{21}+2B_1+5B_0)\right](p^2,m_W,m_W)
+2(D-1)c_W^4A(m_W)\Bigg\},\\
&\Pi_{\gamma\gamma}^{\text{1PI}}(p^2)_V=
-\frac{e^2}{16\pi^2}\Big[(6D-8)B_{22}(p^2,m_W,m_W)+p^2(2B_{21}+2B_1+5B_0)(p^2,m_W,m_W)\notag\\
&-2(D-1)A(m_W)-2m_W^2B_0(p^2,m_{G^+},m_W)\Big],\\
&\Pi_{Z\gamma }^{\text{1PI}}(p^2)_V=
+\frac{eg_Z}{16\pi^2}\Big[c_W^2(6D-8)B_{22}(p^2,m_W,m_W)+c_W^2p^2(2B_{21}+2B_1+5B_0)(p^2,m_W,m_W)\notag\\
&-2c_W^2(D-1)A(m_W)+2m_W^2(s_W^2+s_\beta^2)B_0(p^2,m_{G^+},m_W)\Big].
\end{align}

\subsection{Three-point functions}
In this subsection, we use the shortened notation for the three-point function of the Passarino-Veltman function as 
$C_i(m_1,m_2,m_3)\equiv C_i(p_1^2,p_2^2,q^2,m_1,m_2,m_3)$. 
The 1PI diagram contributions to the $hhh$ vertex can be expressed as 
a function of the incoming momenta $p_1$ and $p_2$ and the outgoing momentum $q=p_1+p_2$ as
\begin{align}
&\Gamma_{hhh}^{\text{1PI}}(p_1^2,p_2^2,q^2)_F=
-\frac{8m_f^4 N_c^f}{16\pi^2}
\frac{c_\alpha^3}{v_\phi^3}
\Big[B_0(p_1^2,m_f,m_f)+B_0(p_2^2,m_f,m_f)+B_0(q^2,m_f,m_f)\notag\\
&\quad\quad\quad\quad+(4m_f^2-q^2+p_1\cdot p_2)C_0(m_f,m_f,m_f)\Big],
\end{align}
\begin{align}
&\Gamma_{hhh}^{\text{1PI}}(p_1^2,p_2^2,q^2)_S=\frac{1}{16\pi^2}\Big\{\notag\\
&+2\lambda_{H^{++}H^{--}h}\lambda_{H^{++}H^{--}hh}[
B_0(p_1^2,m_{H^{++}},m_{H^{++}})
+B_0(p_2^2,m_{H^{++}},m_{H^{++}})+B_0(q^2,m_{H^{++}},m_{H^{++}})]\notag\\
&+2\lambda_{H^{+}H^{-}h}\lambda_{H^{+}H^{-}hh}[B_0(p_1^2,m_{H^{+}},m_{H^{+}})
+B_0(p_2^2,m_{H^{+}},m_{H^{+}})+B_0(q^2,m_{H^{+}},m_{H^{+}})]\notag\\
&+2\lambda_{hG^{+}G^{-}}\lambda_{hhG^{+}G^{-}}[B_0(p_1^2,m_{W},m_{W})
+B_0(p_2^2,m_{W},m_{W})+B_0(q^2,m_{W},m_{W})]\notag\\
&+4\lambda_{H^{+}G^{-}h}\lambda_{H^{+}G^{-}hh}[B_0(p_1^2,m_{H^{+}},m_{W})
+B_0(p_2^2,m_{H^{+}},m_{W})+B_0(q^2,m_{H^{+}},m_{W})]\notag\\
&+4\lambda_{AAh}\lambda_{AAhh}
[B_0(p_1^2,m_{A},m_{A})+B_0(p_2^2,m_{A},m_{A})+B_0(q^2,m_{A},m_{A})]\notag\\
&+4\lambda_{G^0G^0h}\lambda_{G^0G^0hh}
[B_0(p_1^2,m_{Z},m_{Z})+B_0(p_2^2,m_{Z},m_{Z})+B_0(q^2,m_{Z},m_{Z})]\notag\\
&+2\lambda_{AG^0h}\lambda_{AG^0hh}
[B_0(p_1^2,m_{A},m_{Z})+B_0(p_2^2,m_{A},m_{Z})+B_0(q^2,m_{A},m_{Z})]\notag\\
&+4\lambda_{HHh}\lambda_{HHhh}
[B_0(p_1^2,m_{H},m_{H})+B_0(p_2^2,m_{H},m_{H})+B_0(q^2,m_{H},m_{H})]\notag\\
&+12\lambda_{Hhh}\lambda_{Hhhh}[B_0(p_1^2,m_h,m_H)+B_0(p_2^2,m_h,m_H)+B_0(q^2,m_h,m_H)]\notag\\
&+72\lambda_{hhh}\lambda_{hhhh}[B_0(p_1^2,m_h,m_h)+B_0(p_2^2,m_h,m_h)+B_0(q^2,m_h,m_h)]
\Big]\Big\}\notag\\
&-\frac{1}{16\pi^2}\Big\{
2\lambda_{H^{++}H^{--}h}^3C_0(m_{H^{++}},m_{H^{++}},m_{H^{++}})+2\lambda_{H^+H^-h}^3C_0(m_{H^+},m_{H^+},m_{H^+})\notag\\
&+2\lambda_{G^{+}G^{-}h}^3C_0(m_W,m_W,m_W)+8\lambda_{G^0G^0h}^3C_0(m_Z,m_Z,m_Z)\notag\\
&+8\lambda_{AAh}^3C_0(m_A,m_A,m_A)
+8\lambda_{HHh}^3C_0(m_H,m_H,m_H)
+216\lambda_{hhh}^3C_0(m_h,m_h,m_h)\notag\\
&+2\lambda_{H^+H^-h}\lambda_{H^+G^-h}^2
[C_0(m_{G^+},m_{H^+},m_{H^+})+C_0(m_{H^+},m_{W},m_{H^+})+C_0(m_{H^+},m_{H^+},m_{W})]\notag\\
&+2\lambda_{G^+G^-h}\lambda_{H^+G^-h}^2
[C_0(m_{H^+},m_W,m_W)+C_0(m_{G^+},m_{H^+},m_{W})+C_0(m_{W},m_{W},m_{H^+})]\notag\\
&+2\lambda_{AAh}\lambda_{AG^0h}^2
[C_0(m_Z,m_A,m_A)+C_0(m_A,m_Z,m_A)+C_0(m_A,m_A,m_Z)]\notag\\
&+2\lambda_{G^0G^0h}\lambda_{AG^0h}^2
[C_0(m_A,m_Z,m_Z)+C_0(m_Z,m_A,m_Z)+C_0(m_Z,m_Z,m_A)]\notag\\
&+8\lambda_{HHh}\lambda_{Hhh}^2
[C_0(m_h,m_H,m_H)+C_0(m_H,m_H,m_h)+C_0(m_H,m_h,m_H)]\notag\\
&+24\lambda_{hhh}\lambda_{Hhh}^2
[C_0(m_h,m_h,m_H)+C_0(m_H,m_h,m_h)+C_0(m_h,m_H,m_h)]\Big\},
\end{align}
\begin{align}
&\Gamma_{hhh}^{\text{1PI}}(p_1^2,p_2^2,q^2)_V=\frac{1}{16\pi^2}\Bigg[\notag\\
&+\frac{3g^3}{4}m_W(c_\beta c_\alpha+\sqrt{2}s_\beta s_\alpha)(3-c_{2\alpha})DB_0(q^2,m_W,m_W)
+2g^3m_W^3(c_\beta c_\alpha+\sqrt{2}s_\beta s_\alpha)^3DC_0(m_W,m_W,m_W)\notag\\
&-\frac{g^3}{2}m_W(c_\beta c_\alpha+\sqrt{2}s_\beta s_\alpha)^3
C_{SVV}^{hhh}(m_{G^+},m_W,m_W)\notag\\
&-\frac{g^3}{2}m_W(c_\beta c_\alpha+\sqrt{2}s_\beta s_\alpha)(-s_\beta c_\alpha+\sqrt{2}c_\beta s_\alpha)^2
C_{SVV}^{hhh}(m_{H^+},m_W,m_W)\notag\\
&+\frac{g^2}{2}\lambda_{G^+G^-h}(c_\beta c_\alpha+\sqrt{2}s_\beta s_\alpha)^2C_{VSS}^{hhh}(m_W,m_{G^+},m_{G^+})
\notag\\
&+\frac{g^2}{2}\lambda_{H^+H^-h}(-s_\beta c_\alpha+\sqrt{2}c_\beta s_\alpha)^2C_{VSS}^{hhh}(m_W,m_{H^+},m_{H^+})
\notag\\
&+\frac{g^2}{2}\lambda_{H^+G^-h}(c_\beta c_\alpha+\sqrt{2}s_\beta s_\alpha)(-s_\beta c_\alpha+\sqrt{2}c_\beta s_\alpha)
[C_{VSS}^{hhh}(m_W,m_{G^+},m_{H^+})+C_{VSS}(m_W,m_{H^+},m_{G^+})]
\notag\\
&-\frac{g^3m_W^3}{2}(c_\beta c_\alpha+\sqrt{2}s_\beta s_\alpha)^3C_0(m_{c^+},m_{c^+},m_{c^+})\notag\\
&+\frac{3g_Z^3m_Z}{8}(c_{\beta'} c_\alpha+2s_{\beta'} s_\alpha)(5-3c_{2\alpha})DB_0(q^2,m_Z,m_Z)
+g_Z^3m_Z^3(c_{\beta'} c_\alpha+2s_{\beta'} s_\alpha)^3DC_0(m_Z,m_Z,m_Z)\notag\\
&-\frac{g_Z^3m_Z}{4}(c_{\beta'} c_\alpha+2s_{\beta'} s_\alpha)^3C_{SVV}^{hhh}(m_{G^0},m_Z,m_Z)\notag\\
&-\frac{g_Z^3m_Z}{4}(c_{\beta'} c_\alpha+2s_{\beta'} s_\alpha)(-s_{\beta'} c_\alpha+2c_{\beta'} s_\alpha)^2
C_{SVV}^{hhh}(m_{A},m_Z,m_Z)\notag\\
&+\frac{g_Z^2}{2}\lambda_{G^0G^0h}(c_{\beta'} c_\alpha+2s_{\beta'} s_\alpha)^2
C_{VSS}^{hhh}(m_Z,m_{G^0},m_{G^0})
+\frac{g_Z^2}{2}\lambda_{AAh}(-s_{\beta'} c_\alpha+2c_{\beta'} s_\alpha)^2
C_{VSS}^{hhh}(m_Z,m_{A},m_{A})\notag\\
&+\frac{g_Z^2}{4}\lambda_{AG^0h}(c_{\beta'} c_\alpha+2s_{\beta'} s_\alpha)(-s_{\beta'} c_\alpha+2c_{\beta'} s_\alpha)
[C_{VSS}^{hhh}(m_Z,m_{A},m_{G^0})+C_{VSS}^{hhh}(m_Z,m_{G^0},m_{A})]\notag\\
&-\frac{g_Z^3m_Z^3}{4}(c_{\beta'} c_\alpha+2s_{\beta'} s_\alpha)^3C_0(m_{c_Z},m_{c_Z},m_{c_Z})\Bigg],
\end{align}
where we define
\begin{align}
&C_{SVV}^{hhh}(m_1,m_2,m_3)\equiv \notag\\
&\Big[p_1^2C_{21}+p_2^2C_{22}+2p_1p_2C_{23}+DC_{24}
-(q+p_1)(p_1C_{11}+p_2C_{12})+qp_1C_0
\Big](m_1,m_2,m_3)\notag\\
&+\Big[p_1^2C_{21}+p_2^2C_{22}+2p_1p_2C_{23}+DC_{24}
+(3p_1-p_2)(p_1C_{11}+p_2C_{12})+2p_1(p_1-p_2)C_0
\Big](m_3,m_1,m_2)\notag\\
&+\Big[p_1^2C_{21}+p_2^2C_{22}+2p_1p_2C_{23}+DC_{24}
+(3p_1+4p_2)(p_1C_{11}+p_2C_{12})+2q(q+p_2)C_0
\Big](m_2,m_3,m_1),\\
&C_{VSS}^{hhh}(m_V,m_S,m_S)\equiv \notag\\
&\Big[p_1^2C_{21}+p_2^2C_{22}+2p_1p_2C_{23}+DC_{24}
+(4p_1+2p_2)(p_1C_{11}+p_2C_{12})+4p_1\cdot q C_0
\Big](m_V,m_S,m_S)\notag\\
&\hspace{-5mm}+\Big[p_1^2C_{21}+p_2^2C_{22}+2p_1p_2C_{23}+DC_{24}
+2p_2(p_1C_{11}+p_2C_{12})-p_1(p_1+2p_2)C_0
\Big](m_S,m_V,m_S)\notag\\
&\hspace{-5mm}+\Big[p_1^2C_{21}+p_2^2C_{22}+2p_1p_2C_{23}+DC_{24}
-2p_2(p_1C_{11}+p_2C_{12})-q(p_1-p_2)C_0
\Big](m_S,m_S,m_V).
\end{align}

The 1PI diagram contributions to the form factors of the $hZZ$ and $hWW$ vertices which are defined in Eq.~(\ref{form_factor}) are 
calculated as
\begin{align}
&M_{1,\text{1PI}}^{hZZ}(p_1^2,p_2^2,q^2)_F=-\frac{32m_f^2m_Z^2N_c^fc_\alpha}{v_\phi(v^2+2v_\Delta^2)}
\frac{1}{16\pi^2}
\Bigg\{-\frac{1}{4}(I_f^2-2s_W^2I_fQ_f+2s_W^4Q_f^2)\notag\\
&\Big[B_0(p_1^2,m_f,m_f)+B_0(p_2^2,m_f,m_f)+2B_0(q^2,m_f,m_f)
&\hspace{30mm}+(4m_f^2-p_1^2-p_2^2)C_0(m_f,m_f,m_f)-8C_{24}(m_f,m_f,m_f)\Big]\notag\\
&+\frac{s_W^2}{2}(-I_fQ_f+s_W^2Q_f^2)\Big[B_0(p_2^2,m_f,m_f)+B_0(p_1^2,m_f,m_f)
+(4m_f^2-q^2)C_0(m_f,m_f,m_f)\Big]\Bigg\},\\
&M_{2,\text{1PI}}^{hZZ}(p_1^2,p_2^2,q^2)_F=-\frac{32m_f^2m_Z^4N_c^fc_\alpha}{v_\phi(v^2+2v_\Delta^2)}
\frac{1}{16\pi^2}\notag\\
&\times\Big[\frac{1}{2}(I_f^2-2s_W^2I_fQ_f+2s_W^4Q_f^2)\Big(4C_{23}+3C_{12}+C_{11}+C_0\Big)+s_W^2(-I_fQ_f+s_W^2Q_f^2)\Big(C_{12}-C_{11}\Big)\Big]\notag\\
&(m_f,m_f,m_f),\\
&M_{3,\text{1PI}}^{hZZ}(p_1^2,p_2^2,q^2)_F=-\frac{32m_f^2m_Z^4N_c^fc_\alpha}{v_\phi(v^2+2v_\Delta^2)}
\frac{1}{16\pi^2}
\frac{I_f}{2}(-I_f+2s_W^2Q_f)(C_{11}+C_{12}+C_0)(m_f,m_f,m_f).
\end{align}

\begin{align}
&M_{1,\text{1PI}}^{hWW}(p_1^2,p_2^2,q^2)_F=\frac{4m_W^2m_t^2N_c^fc_\alpha}{v_\phi v^2}\frac{1}{16\pi^2}
\Bigg[\frac{1}{2}B_0(p_2^2,m_t,m_b)+B_0(q^2,m_t,m_t)+\frac{1}{2}B_0(p_1^2,m_t,m_b)\notag\\
&-4C_{24}(p_1^2,p_2^2,q^2,m_t,m_b,m_t)
+\frac{1}{2}(2m_t^2+2m_b^2-p_1^2-p_2^2)C_0(m_t,m_b,m_t)\Bigg]+(m_t\leftrightarrow m_b),\\
&M_{2,\text{1PI}}^{hWW}(p_1^2,p_2^2,q^2)_F=\frac{-4m_W^4m_t^2N_c^fc_\alpha}{v_\phi v^2}\frac{1}{16\pi^2}
\left(4C_{23}+3C_{12}+C_{11}+C_0\right)(m_t,m_b,m_t)+(m_t\leftrightarrow m_b),\\
&M_{3,\text{1PI}}^{hWW}(p_1^2,p_2^2,q^2)_F= \frac{-4m_W^4m_t^2N_c^fc_\alpha}{v_\phi v^2}\frac{1}{16\pi^2}\left(C_{11}+C_{12}+C_0\right)(m_t,m_b,m_t)
+(m_t\leftrightarrow m_b).
\end{align}

\begin{align}
&M_{1,\text{1PI}}^{hZZ}(p_1^2,p_2^2,q^2)_S=-\frac{16 m_Z^2}{v^2+2v_\Delta^2}\frac{1}{16\pi^2}\Bigg\{
2\lambda_{H^{++}H^{--}h}(c_W^2-s_W^2)^2C_{24}(m_{H^{++}},m_{H^{++}},m_{H^{++}})\notag\\
&+\frac{1}{2}\lambda_{H^{+}H^{-}h}(c_W^2-s_W^2-c_\beta^2)^2C_{24}(m_{H^{+}},m_{H^{+}},m_{H^{+}})\notag\\
&+\frac{1}{8}\lambda_{H^{+}H^{-}h}s_{2\beta}^2C_{24}(m_{H^{+}},m_{G^+},m_{H^{+}})
+\frac{1}{8}\lambda_{G^{+}G^{-}h}s_{2\beta}^2C_{24}(m_{G^+},m_{H^+},m_{G^+})\notag\\
&-\frac{1}{4}\lambda_{H^{+}G^{-}h}s_{2\beta}
(c_W^2-s_W^2-c_\beta^2)[C_{24}(m_{H^{+}},m_{H^+},m_{G^+})+C_{24}(m_{G^+},m_{H^+},m_{H^+})]\notag\\
&-\frac{1}{4}\lambda_{H^{+}G^{-}h}s_{2\beta}
(c_W^2-s_W^2-s_\beta^2)[C_{24}(m_{H^{+}},m_{G^+},m_{G^+})+C_{24}(m_{G^+},m_{G^+},m_{H^+})]\notag\\
&+\frac{1}{2}\lambda_{G^+G^{-}h}(c_W^2-s_W^2-s_\beta^2)^2C_{24}(m_{G^+},m_{G^+},m_{G^+})\notag\\
&+\frac{1}{2}\lambda_{AAh}(2c_\alpha c_{\beta'}+s_\alpha s_{\beta'})^2C_{24}(m_A,m_H,m_A)
+\frac{1}{2}\lambda_{AAh}(c_\alpha s_{\beta'}-2s_\alpha c_{\beta'})^2C_{24}(m_A,m_h,m_A)\notag\\
&+\frac{1}{2}\lambda_{HHh}(2c_\alpha c_{\beta'}+s_\alpha s_{\beta'})^2C_{24}(m_H,m_A,m_H)
+\frac{3}{2}\lambda_{hhh}(c_\alpha s_{\beta'}-2s_\alpha c_{\beta'})^2C_{24}(m_h,m_A,m_h)\notag\\
&-\frac{1}{2}\lambda_{Hhh}(2c_\alpha c_{\beta'}+s_\alpha s_{\beta'})(c_\alpha s_{\beta'}-2s_\alpha c_{\beta'})
\left[C_{24}(m_H,m_A,m_h)+C_{24}(m_h,m_A,m_H)\right]\notag\\
&-\frac{1}{4}\lambda_{AG^0h}(2c_\alpha c_{\beta'}+s_\alpha s_{\beta'})(s_\alpha c_{\beta'}-2c_\alpha s_{\beta'})\left[C_{24}(m_A,m_H,m_{G^0})+C_{24}(m_{G^0},m_H,m_A)\right]\notag\\
&-\frac{1}{4}\lambda_{AG^0h}(c_\alpha s_{\beta'}-2s_\alpha c_{\beta'})(c_\alpha c_{\beta'}+2s_\alpha s_{\beta'})\left[C_{24}(m_A,m_h,m_{G^0})+C_{24}(m_{G^0},m_h,m_A)\right]\notag\\
&+\frac{1}{2}\lambda_{G^0G^0h}(s_\alpha c_{\beta'}-2c_\alpha s_{\beta'})^2C_{24}(m_{G^0},m_H,m_{G^0})
+\frac{1}{2}\lambda_{G^0G^0h}(c_\alpha c_{\beta'}+2s_\alpha s_{\beta'})^2C_{24}(m_{G^0},m_h,m_{G^0})\notag\\
&+\frac{1}{2}\lambda_{HHh}(s_\alpha c_{\beta'}-2c_\alpha s_{\beta'})^2C_{24}(m_H,m_{G^0},m_H)
+\frac{3}{2}\lambda_{hhh}(c_\alpha c_{\beta'}+2s_\alpha s_{\beta'})^2C_{24}(m_h,m_{G^0},m_h)\notag\\
&-\frac{1}{2}\lambda_{Hhh}(s_\alpha c_{\beta'}-2c_\alpha s_{\beta'})(c_\alpha c_{\beta'}+2s_\alpha s_{\beta'})\left[C_{24}(m_H,m_{G^0},m_h)+C_{24}(m_h,m_{G^0},m_H)\right]\Bigg\}\notag\\
&+\frac{4 m_Z^2}{v^2+2v_\Delta^2}\frac{1}{16\pi^2}\Bigg\{
2\lambda_{H^{++}H^{--}h}(c_W^2-s_W^2)^2B_0(q^2,m_{H^{++}},m_{H^{++}})\notag\\
&+\frac{1}{4}\lambda_{H^{+}H^{-}h}(2+c_{4W}-4c_{2W}c_{\beta}^2+c_{2\beta})B_0(q^2,m_{H^{+}},m_{H^{+}})\notag\\
&+\frac{1}{4}\lambda_{G^{+}G^{-}h}(2+c_{4W}-4c_{2W}s_{\beta}^2-c_{2\beta})B_0(q^2,m_{G^+},m_{G^+})
+\frac{1}{2}\lambda_{H^{+}G^{-}h}s_{2\beta}(1-2c_{2W})B_0(q^2,m_{H^{+}},m_{G^+})\notag\\
&+\frac{1}{4}\lambda_{AAh}(5+3c_{2\beta'})B_0(q^2,m_A,m_A)
+\frac{1}{4}\lambda_{G^0G^0h}(5-3c_{2\beta'})B_0(q^2,m_{G^0},m_{G^0})\notag\\
&+\frac{1}{4}\lambda_{HHh}(5+3c_{2\alpha})B_0(q^2,m_H,m_H)
+\frac{3}{4}\lambda_{hhh}(5-3c_{2\alpha})B_0(q^2,m_h,m_h)\notag\\
&
+\frac{3}{4}\lambda_{AG^0h}s_{2\beta'}B_0(q^2,m_A,m_{G^0})
+\frac{3}{2}\lambda_{Hhh}s_{2\alpha}B_0(q^2,m_h,m_H)\Bigg\}, 
\end{align}
\begin{align}
&M_{2,\text{1PI}}^{hZZ}(p_1^2,p_2^2,q^2)_S=-\frac{16 m_Z^4}{v^2+2v_\Delta^2}\frac{1}{16\pi^2}\Bigg\{
2\lambda_{H^{++}H^{--}h}(c_W^2-s_W^2)^2C_{1223}(m_{H^{++}},m_{H^{++}},m_{H^{++}})\notag\\
&+\frac{1}{2}\lambda_{H^{+}H^{-}h}(c_W^2-s_W^2-c_\beta^2)^2C_{1223}(m_{H^{+}},m_{H^{+}},m_{H^{+}})\notag\\
&+\frac{1}{8}\lambda_{H^{+}H^{-}h}s_{2\beta}^2C_{1223}(m_{H^{+}},m_{W},m_{H^{+}})
+\frac{1}{8}\lambda_{G^{+}G^{-}h}s_{2\beta}^2C_{1223}(m_{W},m_{H^+},m_{W})\notag\\
&-\frac{1}{4}\lambda_{H^{+}G^{-}h}s_{2\beta}
(c_W^2-s_W^2-c_\beta^2)[C_{1223}(m_{H^{+}},m_{H^+},m_{W})+C_{1223}(m_{W},m_{H^+},m_{H^+})]\notag\\
&-\frac{1}{4}\lambda_{H^{+}G^{-}h}s_{2\beta}
(c_W^2-s_W^2-s_\beta^2)[C_{1223}(m_{H^{+}},m_{W},m_{W})+C_{1223}(m_{W},m_{W},m_{H^+})]\notag\\
&+\frac{1}{2}\lambda_{G^+G^{-}h}(c_W^2-s_W^2-s_\beta^2)^2C_{1223}(m_W,m_W,m_W)\notag\\
&+\frac{1}{2}\lambda_{AAh}(2c_\alpha c_{\beta'}+s_\alpha s_{\beta'})^2C_{1223}(m_A,m_H,m_A)
+\frac{1}{2}\lambda_{AAh}(c_\alpha s_{\beta'}-2s_\alpha c_{\beta'})^2C_{1223}(m_A,m_h,m_A)\notag\\
&+\frac{1}{2}\lambda_{HHh}(2c_\alpha c_{\beta'}+s_\alpha s_{\beta'})^2C_{1223}(m_H,m_A,m_H)
+\frac{3}{2}\lambda_{hhh}(c_\alpha s_{\beta'}-2s_\alpha c_{\beta'})^2C_{1223}(m_h,m_A,m_h)\notag\\
&-\frac{1}{2}\lambda_{Hhh}(2c_\alpha c_{\beta'}+s_\alpha s_{\beta'})
(c_\alpha s_{\beta'}-2s_\alpha c_{\beta'})\left[C_{1223}(m_H,m_A,m_h)+C_{1223}(m_h,m_A,m_H)\right]\notag\\
&-\frac{1}{4}\lambda_{AG^0h}(2c_\alpha c_{\beta'}+s_\alpha s_{\beta'})(s_\alpha c_{\beta'}-2c_\alpha s_{\beta'})\left[C_{1223}(m_A,m_H,m_Z)+C_{1223}(m_Z,m_H,m_A)\right]\notag\\
&-\frac{1}{4}\lambda_{AG^0h}(c_\alpha s_{\beta'}-2s_\alpha c_{\beta'})(c_\alpha c_{\beta'}+2s_\alpha s_{\beta'})\left[C_{1223}(m_A,m_h,m_Z)+C_{1223}(m_Z,m_h,m_A)\right]\notag\\
&+\frac{1}{2}\lambda_{G^0G^0h}(s_\alpha c_{\beta'}-2c_\alpha s_{\beta'})^2C_{1223}(m_Z,m_H,m_Z)
+\frac{1}{2}\lambda_{G^0G^0h}(c_\alpha c_{\beta'}+2s_\alpha s_{\beta'})^2C_{1223}(m_Z,m_h,m_Z)\notag\\
&+\frac{1}{2}\lambda_{HHh}(s_\alpha c_{\beta'}-2c_\alpha s_{\beta'})^2C_{1223}(m_H,m_Z,m_H)
+\frac{3}{2}\lambda_{hhh}(c_\alpha c_{\beta'}+2s_\alpha s_{\beta'})^2C_{1223}(m_h,m_Z,m_h)\notag\\
&-\frac{1}{2}\lambda_{Hhh}(s_\alpha c_{\beta'}-2c_\alpha s_{\beta'})
(c_\alpha c_{\beta'}+2s_\alpha s_{\beta'})\left[C_{1223}(m_H,m_Z,m_h)+C_{1223}(m_h,m_Z,m_H)\right]\Bigg\},
\end{align}
\begin{align}
M_{3,\text{1PI}}^{hZZ}(p_1^2,p_2^2,q^2)_S=0, 
\end{align}
\begin{align}
&M_{1,\text{1PI}}^{hWW}(p_1^2,p_2^2,q^2)_S=-\frac{16 m_W^2}{v^2}\frac{1}{16\pi^2}\Big\{\notag\\
&+\lambda_{H^{++}H^{--}h}c_\beta^2C_{24}(m_{H^{++}},m_{H^{+}},m_{H^{++}})
+\lambda_{H^{++}H^{--}h}s_\beta^2C_{24}(m_{H^{++}},m_{G^+},m_{H^{++}})\notag\\
&+\lambda_{H^{+}H^{-}h}c_\beta^2C_{24}(m_{H^{+}},m_{H^{++}},m_{H^{+}})
+\lambda_{G^{+}G^{-}h}s_\beta^2C_{24}(m_{G^+},m_{H^{++}},m_{G^+})\notag\\
&+\lambda_{H^{+}G^{-}h}c_\beta s_\beta[C_{24}(m_{G^+},m_{H^{++}},m_{H^{+}})+C_{24}(m_{H^+},m_{H^{++}},m_{G^+})]\notag\\
&+\frac{1}{2}\lambda_{AAh}\left[\left(s_\beta s_{\beta'}+\sqrt{2}c_\beta c_{\beta'}\right)^2C_{24}(m_A,m_{H^+},m_A)+
\left(-c_\beta s_{\beta'}+\sqrt{2}s_\beta c_{\beta'}\right)^2C_{24}(m_A,m_{G^+},m_A)\right]\notag\\
&+\frac{1}{2}\lambda_{G^0G^0h}\left[\left(-s_\beta c_{\beta'}+\sqrt{2}c_\beta s_{\beta'}\right)^2C_{24}(m_{G^0},m_{H^+},m_{G^0})+
\left(c_\beta c_{\beta'}+\sqrt{2}s_\beta s_{\beta'}\right)^2C_{24}(m_{G^0},m_{G^+},m_{G^0})\right]\notag\\
&+\frac{1}{4}\lambda_{AG^0h}\left(s_\beta s_{\beta'}+\sqrt{2}c_\beta c_{\beta'}\right)\left(-s_\beta c_{\beta'}+\sqrt{2}c_\beta s_{\beta'}\right)[C_{24}(m_A,m_{H^+},m_{G^0})+C_{24}(m_{G^0},m_{H^+},m_A)]\notag\\
&+\frac{1}{4}\lambda_{AG^0h}\left(-c_\beta s_{\beta'}+\sqrt{2}s_\beta c_{\beta'}\right)\left(c_\beta c_{\beta'}+\sqrt{2}s_\beta s_{\beta'}\right)[C_{24}(m_A,m_{G^+},m_{G^0})+C_{24}(m_{G^0},m_{G^+},m_A)]\notag\\
&+\frac{1}{2}\lambda_{HHh}\left[\left(s_\beta s_{\alpha}+\sqrt{2}c_\beta c_{\alpha}\right)^2C_{24}(m_H,m_{H^+},m_H)+
\left(-c_\beta s_{\alpha}+\sqrt{2}s_\beta c_{\alpha}\right)^2C_{24}(m_H,m_{G^+},m_H)\right]\notag\\
&+\frac{3}{2}\lambda_{hhh}\left[\left(-s_\beta c_{\alpha}+\sqrt{2}c_\beta s_{\alpha}\right)^2C_{24}(m_h,m_{H^+},m_h)+
\left(c_\beta c_{\alpha}+\sqrt{2}s_\beta s_{\alpha}\right)^2C_{24}(m_h,m_{G^+},m_h)\right]\notag\\
&+\frac{1}{2}\lambda_{Hhh}\left(s_\beta s_{\alpha}+\sqrt{2}c_\beta c_{\alpha}\right)
\left(-s_\beta c_{\alpha}+\sqrt{2}c_\beta s_{\alpha}\right)[C_{24}(m_H,m_{H^+},m_h)+C_{24}(m_h,m_{H^+},m_H)]\notag\\
&+\frac{1}{2}\lambda_{Hhh}\left(-c_\beta s_{\alpha}+\sqrt{2}s_\beta c_{\alpha}\right)
\left(c_\beta c_{\alpha}+\sqrt{2}s_\beta s_{\alpha}\right)[C_{24}(m_H,m_{G^+},m_h)+C_{24}(m_h,m_{G^+},m_H)]\notag\\
&+\frac{1}{4}\lambda_{H^+H^-h}\left[\left(s_\beta s_{\beta'}+\sqrt{2}c_\beta c_{\beta'}\right)^2C_{24}(m_{H^+},m_A,m_{H^+})+
\left(-s_\beta c_{\beta'}+\sqrt{2}c_\beta s_{\beta'}\right)^2C_{24}(m_{H^+},m_{G^0},m_{H^+})\right]\notag\\
&+\frac{1}{4}\lambda_{H^+H^-h}\left[\left(s_\beta s_{\alpha}+\sqrt{2}c_\beta c_{\alpha}\right)^2C_{24}(m_{H^+},m_H,m_{H^+})+
\left(-s_\beta c_{\alpha}+\sqrt{2}c_\beta s_{\alpha}\right)^2C_{24}(m_{H^+},m_h,m_{H^+})\right]\notag\\
&+\frac{1}{4}\lambda_{G^+G^-h}\left[\left(-c_\beta s_{\beta'}+\sqrt{2}s_\beta c_{\beta'}\right)^2C_{24}(m_{G^+},m_A,m_{G^+})
+\left(c_\beta c_{\beta'}+\sqrt{2}s_\beta s_{\beta'}\right)^2C_{24}(m_{G^+},m_{G^0},m_{G^+})\right]\notag\\
&+\frac{1}{4}\lambda_{G^+G^-h}\left[\left(-c_\beta s_{\alpha}+\sqrt{2}s_\beta c_{\alpha}\right)^2C_{24}(m_{G^+},m_H,m_{G^+})
+\left(c_\beta c_{\alpha}+\sqrt{2}s_\beta s_{\alpha}\right)^2C_{24}(m_{G^+},m_h,m_{G^+})\right]\notag\\
&+\frac{1}{4}\lambda_{H^+G^-h}(s_\beta s_{\beta'}+\sqrt{2}c_\beta c_{\beta'})(-c_\beta s_{\beta'}+\sqrt{2}s_\beta c_{\beta'})
[C_{24}(m_{H^+},m_A,m_{G^+})+C_{24}(m_{G^+},m_A,m_{H^+})]\notag\\
&+\frac{1}{4}\lambda_{H^+G^-h}(-s_\beta c_{\beta'}+\sqrt{2}c_\beta s_{\beta'})(c_\beta c_{\beta'}+\sqrt{2}s_\beta s_{\beta'})[C_{24}(m_{H^+},m_{G^0},m_{G^+})+C_{24}(m_{G^+},m_{G^0},m_{H^+})]\notag\\
&+\frac{1}{4}\lambda_{H^+G^-h}(s_\beta s_{\alpha}+\sqrt{2}c_\beta c_{\alpha})(-c_\beta s_{\alpha}+\sqrt{2}s_\beta c_{\alpha})[C_{24}(m_{H^+},m_H,m_{G^+})+C_{24}(m_{G^+},m_H,m_{H^+})]\notag\\
&+\frac{1}{4}\lambda_{H^+G^-h}(-s_\beta c_{\alpha}+\sqrt{2}c_\beta s_{\alpha})(c_\beta c_{\alpha}+\sqrt{2}s_\beta s_{\alpha})
[C_{24}(m_{H^+},m_h,m_{G^+})+C_{24}(m_{G^+},m_h,m_{H^+})]\Big\}\notag\\
&+\frac{4 m_W^2}{v^2}\frac{1}{16\pi^2}\Big[\lambda_{H^{++}H^{--}h}B_0(q^2,m_{H^{++}},m_{H^{++}})    
+\lambda_{H^{+}H^{-}h}\frac{5+3c_{2\beta}}{4}B_0(q^2,m_{H^{+}},m_{H^{+}})\notag\\
&+\lambda_{G^{+}G^{-}h}\frac{5-3c_{2\beta}}{4}B_0(q^2,m_{G^+},m_{G^+})  
+2\lambda_{H^{+}G^{-}h}\frac{3s_{2\beta}}{4}B_0(q^2,m_{H^{+}},m_{G^+})\notag\\  
&+2\lambda_{AAh}\frac{3+c_{2\beta'}}{8}B_0(q^2,m_{A},m_{A})+2\lambda_{G^{0}G^{0}h}\frac{3-c_{2\beta'}}{8}B_0(q^2,m_{Z},m_{Z})
+\lambda_{AG^{0}h}\frac{s_{2\beta'}}{4}B_0(q^2,m_{A},m_{Z})\notag\\  
&+2\lambda_{HHh}\frac{3+c_{2\alpha}}{8}B_0(q^2,m_{H},m_{H})+6\lambda_{hhh}\frac{3-c_{2\alpha}}{8}B_0(q^2,m_{h},m_{h})
+2\lambda_{Hhh}\frac{s_{2\alpha}}{4}B_0(q^2,m_{H},m_{h})\Big],
\end{align}
\begin{align}
&M_{2,\text{1PI}}^{hWW}(p_1^2,p_2^2,q^2)_S=-\frac{16 m_W^4}{v^2}\frac{1}{16\pi^2}\Bigg\{\notag\\
&+\lambda_{H^{++}H^{--}h}c_\beta^2C_{1223}(m_{H^{++}},m_{H^{+}},m_{H^{++}})
+\lambda_{H^{++}H^{--}h}s_\beta^2C_{1223}(m_{H^{++}},m_{G^+},m_{H^{++}})\notag\\
&+\lambda_{H^{+}H^{-}h}c_\beta^2C_{1223}(m_{H^{+}},m_{H^{++}},m_{H^{+}})
+\lambda_{G^{+}G^{-}h}s_\beta^2C_{1223}(m_{G^+},m_{H^{++}},m_{G^+})\notag\\
&+\lambda_{H^{+}G^{-}h}c_\beta s_\beta[C_{1223}(m_{G^+},m_{H^{++}},m_{H^{+}})+C_{1223}(m_{H^+},m_{H^{++}},m_{G^+})]\notag\\
&+\frac{1}{2}\lambda_{AAh}\left[\left(s_\beta s_{\beta'}+\sqrt{2}c_\beta c_{\beta'}\right)^2C_{1223}(m_A,m_{H^+},m_A)+
\left(-c_\beta s_{\beta'}+\sqrt{2}s_\beta c_{\beta'}\right)^2C_{1223}(m_A,m_{G^+},m_A)\right]\notag\\
&+\frac{1}{2}\lambda_{G^0G^0h}\left[\left(-s_\beta c_{\beta'}+\sqrt{2}c_\beta s_{\beta'}\right)^2C_{1223}(m_{G^0},m_{H^+},m_{G^0})+
\left(c_\beta c_{\beta'}+\sqrt{2}s_\beta s_{\beta'}\right)^2C_{1223}(m_{G^0},m_{G^+},m_{G^0})\right]\notag\\
&+\frac{1}{4}\lambda_{AG^0h}\left(s_\beta s_{\beta'}+\sqrt{2}c_\beta c_{\beta'}\right)\left(-s_\beta c_{\beta'}+\sqrt{2}c_\beta s_{\beta'}\right)[C_{1223}(m_A,m_{H^+},m_{G^0})+C_{1223}(m_{G^0},m_{H^+},m_A)]\notag\\
&+\frac{1}{4}\lambda_{AG^0h}\left(-c_\beta s_{\beta'}+\sqrt{2}s_\beta c_{\beta'}\right)\left(c_\beta c_{\beta'}+\sqrt{2}s_\beta s_{\beta'}\right)[C_{1223}(m_A,m_{G^+},m_{G^0})+C_{1223}(m_{G^0},m_{G^+},m_A)]\notag\\
&+\frac{1}{2}\lambda_{HHh}\left[\left(s_\beta s_{\alpha}+\sqrt{2}c_\beta c_{\alpha}\right)^2C_{1223}(m_H,m_{H^+},m_H)+
\left(-c_\beta s_{\alpha}+\sqrt{2}s_\beta c_{\alpha}\right)^2C_{1223}(m_H,m_{G^+},m_H)\right]\notag\\
&+\frac{3}{2}\lambda_{hhh}\left[\left(-s_\beta c_{\alpha}+\sqrt{2}c_\beta s_{\alpha}\right)^2C_{1223}(m_h,m_{H^+},m_h)+
\left(c_\beta c_{\alpha}+\sqrt{2}s_\beta s_{\alpha}\right)^2C_{1223}(m_h,m_{G^+},m_h)\right]\notag\\
&+\frac{1}{2}\lambda_{Hhh}\left(s_\beta s_{\alpha}+\sqrt{2}c_\beta c_{\alpha}\right)
\left(-s_\beta c_{\alpha}+\sqrt{2}c_\beta s_{\alpha}\right)[C_{1223}(m_H,m_{H^+},m_h)+C_{1223}(m_h,m_{H^+},m_H)]\notag\\
&+\frac{1}{2}\lambda_{Hhh}\left(-c_\beta s_{\alpha}+\sqrt{2}s_\beta c_{\alpha}\right)
\left(c_\beta c_{\alpha}+\sqrt{2}s_\beta s_{\alpha}\right)[C_{1223}(m_H,m_{G^+},m_h)+C_{1223}(m_h,m_{G^+},m_H)]\notag\\
&+\frac{1}{4}\lambda_{H^+H^-h}\left[\left(s_\beta s_{\beta'}+\sqrt{2}c_\beta c_{\beta'}\right)^2C_{1223}(m_{H^+},m_A,m_{H^+})+
\left(\sqrt{2}c_\beta s_{\beta'}-s_\beta c_{\beta'}\right)^2C_{1223}(m_{H^+},m_{G^0},m_{H^+})\right]\notag\\
&+\frac{1}{4}\lambda_{H^+H^-h}\left[\left(s_\beta s_{\alpha}+\sqrt{2}c_\beta c_{\alpha}\right)^2C_{1223}(m_{H^+},m_H,m_{H^+})+
\left(-s_\beta c_{\alpha}+\sqrt{2}c_\beta s_{\alpha}\right)^2C_{1223}(m_{H^+},m_h,m_{H^+})\right]\notag\\
&+\frac{1}{4}\lambda_{G^+G^-h}\left[\left(-c_\beta s_{\beta'}+\sqrt{2}s_\beta c_{\beta'}\right)^2C_{1223}(m_{G^+},m_A,m_{G^+})
+\left(c_\beta c_{\beta'}+\sqrt{2}s_\beta s_{\beta'}\right)^2C_{1223}(m_{G^+},m_{G^0},m_{G^+})\right]\notag\\
&+\frac{1}{4}\lambda_{G^+G^-h}\left[\left(-c_\beta s_{\alpha}+\sqrt{2}s_\beta c_{\alpha}\right)^2C_{1223}(m_{G^+},m_H,m_{G^+})
+\left(c_\beta c_{\alpha}+\sqrt{2}s_\beta s_{\alpha}\right)^2C_{1223}(m_{G^+},m_h,m_{G^+})\right]\notag\\
&+\frac{1}{4}\lambda_{H^+G^-h}\left(s_\beta s_{\beta'}+\sqrt{2}c_\beta c_{\beta'}\right)\left(-c_\beta s_{\beta'}+\sqrt{2}s_\beta c_{\beta'}\right)
[C_{1223}(m_{H^+},m_A,m_{G^+})+C_{1223}(m_{G^+},m_A,m_{H^+})]\notag\\
&+\frac{1}{4}\lambda_{H^+G^-h}\left(-s_\beta c_{\beta'}+\sqrt{2}c_\beta s_{\beta'}\right)\left(c_\beta c_{\beta'}+\sqrt{2}s_\beta s_{\beta'}\right)^2[C_{1223}(m_{H^+},m_{G^0},m_{G^+})+C_{1223}(m_{G^+},m_{G^0},m_{H^+})]\notag\\
&+\frac{1}{4}\lambda_{H^+G^-h}\left(s_\beta s_{\alpha}+\sqrt{2}c_\beta c_{\alpha}\right)\left(-c_\beta s_{\alpha}+\sqrt{2}s_\beta c_{\alpha}\right)[C_{1223}(m_{H^+},m_H,m_{G^+})+C_{1223}(m_{G^+},m_H,m_{H^+})]\notag\\
&+\frac{1}{4}\lambda_{H^+G^-h}\left(-s_\beta c_{\alpha}+\sqrt{2}c_\beta s_{\alpha}\right)\left(c_\beta c_{\alpha}+\sqrt{2}s_\beta s_{\alpha}\right)
[C_{1223}(m_{H^+},m_h,m_{G^+})+C_{1223}(m_{G^+},m_h,m_{H^+})]\Bigg\},\\
&M_{3,\text{1PI}}^{hWW(\text{1PI})}(p_1^2,p_2^2,q^2)_S=0, 
\end{align}

\begin{align}
&M_{1,\text{1PI}}^{hZZ}(p_1^2,p_2^2,q^2)_V=\frac{1}{16\pi^2}\Bigg\{
2g^3m_Wc_W^2(c_\alpha c_\beta+\sqrt{2}s_\alpha s_\beta)C_{VVV}(m_W,m_W,m_W)\notag\\
&+g^3m_W(c_\alpha c_\beta +\sqrt{2}s_\alpha s_\beta)(s_W^2+s_\beta^2)C_{SVV}^{hVV}(m_{G^+},m_W,m_W)\notag\\
&+g^3m_W(-c_\alpha s_\beta +\sqrt{2}s_\alpha c_\beta)s_\beta c_\beta
C_{SVV}^{hVV}(m_{H^+},m_W,m_W)\notag\\
&+g^3m_W(c_\alpha c_\beta +\sqrt{2}s_\alpha s_\beta)(s_W^2+s_\beta^2)C_{VVS}^{hVV}(m_W,m_W,m_{G^+})\notag\\
&+g^3m_W(-c_\alpha s_\beta +\sqrt{2}s_\alpha c_\beta)s_\beta c_\beta C_{VVS}^{hVV}(m_W,m_W,m_{H^+})\notag\\
&-gg_Z^2m_W(c_\alpha c_\beta +\sqrt{2}s_\alpha s_\beta)(c_W^2-s_W^2-s_\beta^2)(s_W^2+s_\beta^2)[C_{24}(m_W,m_{G^+},m_{G^+})
+C_{24}(m_{G^+},m_{G^+},m_W)]\notag\\
&-gg_Z^2m_W(-c_\alpha s_\beta +\sqrt{2}s_\alpha c_\beta)(c_W^2-s_W^2-c_\beta^2)s_\beta c_\beta[C_{24}(m_W,m_{H^+},m_{H^+})
+C_{24}(m_{H^+},m_{H^+},m_W)]\notag\\
&+gg_Z^2m_W(c_\alpha c_\beta +\sqrt{2}s_\alpha s_\beta)s_\beta^2 c_\beta^2[C_{24}(m_W,m_{H^+},m_{G^+})
+C_{24}(m_{H^+},m_{G^+},m_W)]\notag\\
&+gg_Z^2m_W(-c_\alpha s_\beta +\sqrt{2}s_\alpha c_\beta)s_\beta c_\beta(s_W^2+s_\beta^2)[C_{24}(m_W,m_{G^+},m_{H^+})
+C_{24}(m_{G^+},m_{H^+},m_W)]\notag\\
&-2gg_Z^2(s_W^2+s_\beta^2)^2m_W^3(c_\alpha c_\beta +\sqrt{2}s_\alpha s_\beta)C_0(m_W,m_{G^+},m_W)\notag\\
&-2gg_Z^2s_\beta^2c_\beta^2m_W^3(c_\alpha c_\beta +\sqrt{2}s_\alpha s_\beta)C_0(m_W,m_{H^+},m_W)\notag\\
&+2\lambda_{G^+G^-h}g_Z^2m_W^2(s_W^2+s_\beta^2)^2C_0(m_{G^+},m_W,m_{G^+})
+2\lambda_{H^+H^-h}g_Z^2m_W^2s_\beta^2c_\beta^2C_0(m_{H^+},m_W,m_{H^+})\notag\\
&+2\lambda_{H^+G^-h}g_Z^2m_W^2s_\beta c_\beta(s_W^2+s_\beta^2)[C_0(m_{H^+},m_W,m_{G^+})+C_0(m_{G^+},m_W,m_{H^+})]
\notag\\
&-2g^3c_W^2m_W(c_\alpha c_\beta+\sqrt{2}s_\alpha s_\beta)C_{24}(m_{c^+},m_{c^+},m_{c^+})\notag\\
&-g^3c_W^2m_W(c_\alpha c_\beta +\sqrt{2}s_\alpha s_\beta)(2D-2)B_0(q^2,m_W,m_W)\notag\\
&+gg_Z^2m_W[-c_\alpha c_\beta s_W^2 +\sqrt{2}(c_W^2-2)s_\alpha s_\beta](s_W^2+s_\beta^2)
[B_0(p_2^2,m_W,m_{G^+})+B_0(p_1^2,m_{G^+},m_W)]\notag\\
&+gg_Z^2m_W[c_\alpha s_\beta s_W^2 +\sqrt{2}(c_W^2-2)s_\alpha c_\beta]s_\beta c_\beta
[B_0(p_2^2,m_W,m_{H^+})+B_0(p_1^2,m_{H^+},m_W)]\notag\\
&+\frac{g_Z^3}{2}m_Z(c_\alpha c_{\beta'}+2s_\alpha s_{\beta '})^3[C_{24}(m_Z,m_h,m_{G^0})+C_{24}(m_{G^0},m_h,m_Z)]\notag\\
&+\frac{g_Z^3}{2}m_Z(c_\alpha c_{\beta'}+2s_\alpha s_{\beta '})(-c_\alpha s_{\beta'}+2s_\alpha c_{\beta '})^2[C_{24}(m_Z,m_h,m_{A})+C_{24}(m_{A},m_h,m_Z)]\notag\\
&+\frac{g_Z^3}{2}m_Z(c_\alpha c_{\beta'}+2s_\alpha s_{\beta '})(-s_\alpha c_{\beta'}+2c_\alpha s_{\beta '})^2[C_{24}(m_Z,m_H,m_{G^0})+C_{24}(m_{G^0},m_H,m_Z)]\notag\\
&+\frac{g_Z^3}{2}m_Z(-s_\alpha c_{\beta'}+2c_\alpha s_{\beta '})
(2c_\alpha c_{\beta'}+s_\alpha s_{\beta '})
(-c_\alpha s_{\beta'}+2s_\alpha c_{\beta '})
[C_{24}(m_Z,m_H,m_{A})+C_{24}(m_{A},m_H,m_Z)]\notag\\
&-g_Z^3m_Z^3(c_\alpha c_{\beta'}+2s_\alpha s_{\beta '})^3C_0(m_Z,m_h,m_Z)
-g_Z^3m_Z^3(c_\alpha c_{\beta'}+2s_\alpha s_{\beta '})(-s_\alpha c_{\beta'}+2c_\alpha s_{\beta '})^2C_0(m_Z,m_H,m_Z)\notag\\
&+6\lambda_{hhh}g_Z^2m_Z^2(c_\alpha c_{\beta'}+2s_\alpha s_{\beta '})^2C_0(m_h,m_Z,m_h)
+2\lambda_{HHh}g_Z^2m_Z^2(-s_\alpha c_{\beta'}+2c_\alpha s_{\beta '})^2C_0(m_H,m_Z,m_H)\notag\\
&+2\lambda_{Hhh}g_Z^2m_Z^2(c_\alpha c_{\beta'}+2s_\alpha s_{\beta '})(-s_\alpha c_{\beta'}+2c_\alpha s_{\beta '})
[C_0(m_h,m_Z,m_H)+C_0(m_H,m_Z,m_h)]\notag\\
&-\frac{g_Z^3}{4}m_Z(5-3c_{2\alpha})(c_\alpha c_{\beta'}+2s_\alpha s_{\beta '})[B_0(p_1^2,m_h,m_Z)+B_0(p_2^2,m_h,m_Z)]\notag\\
&-\frac{3g_Z^3}{4}m_Zs_{2\alpha}(-s_\alpha c_{\beta'}+2c_\alpha s_{\beta '})[B_0(p_1^2,m_H,m_Z)+B_0(p_2^2,m_H,m_Z)]\Bigg\}, 
\end{align}
\begin{align}
&M_{1,\text{1PI}}^{hWW}(p_1^2,p_2^2,q^2)_V=\frac{1}{16\pi^2}\Bigg\{
g^3c_Wm_Z(c_\alpha c_{\beta'}+2s_\alpha s_{\beta'})C_{VVV}(m_Z,m_W,m_Z)\notag\\
&
+g^3m_W[c_W^2(c_\alpha c_\beta+\sqrt{2}s_\alpha s_\beta)C_{VVV}(m_W,m_Z,m_W)
+s_W^2(c_\alpha c_\beta+\sqrt{2}s_\alpha s_\beta)C_{VVV}(m_W,0,m_W)]\notag\\
&-\frac{1}{2}g^3m_W(c_\alpha c_\beta +\sqrt{2}s_\alpha s_\beta)[
(s_W^2+s_\beta^2)C_{SVV}^{hVV}(m_{G^+},m_Z,m_W)
-s_W^2C_{SVV}^{hVV}(m_{G^+},0,m_W)]\notag\\
&-\frac{1}{2}g^3m_W(-c_\alpha s_\beta +\sqrt{2}s_\alpha c_\beta)s_\beta c_\beta C_{SVV}^{hVV}(m_{H^+},m_Z,m_W)\notag\\
&-\frac{1}{2}g^3m_W(c_\alpha c_\beta +\sqrt{2}s_\alpha s_\beta)[(s_W^2+s_\beta^2)C_{VVS}^{hVV}
(m_W,m_Z,m_{G^+})
-s_W^2C_{VVS}^{hVV}(m_W,0,m_{G^+})]\notag\\
&-\frac{1}{2}g^3m_W(-c_\alpha s_\beta +\sqrt{2}s_\alpha c_\beta)s_\beta c_\beta C_{VVS}^{hVV}
(m_W,m_Z,m_{H^+})\notag\\
&-4g^3s_\beta^2m_W^3(c_\alpha c_{\beta} +\sqrt{2}s_\alpha s_{\beta})C_0(m_W,m_{H^{++}},m_W)
-g_Z^3m_W^3(s_W^2+s_\beta^2)^2(c_\alpha c_{\beta'} +2s_\alpha s_{\beta'})C_0(m_Z,m_{G^+},m_Z)\notag\\
&-g_Z^3m_W^3s_\beta^2c_\beta^2(c_\alpha c_{\beta'} +2s_\alpha s_{\beta'})C_0(m_Z,m_{H^+},m_Z)
-g^3m_W^3(c_\alpha c_{\beta} +\sqrt{2}s_\alpha s_{\beta})^3C_0(m_W,m_{h},m_W)\notag\\
&-g^3m_W^3(c_\alpha c_{\beta} +\sqrt{2}s_\alpha s_{\beta})(-s_\alpha c_{\beta} +\sqrt{2}c_\alpha s_{\beta})^2C_0(m_W,m_H,m_W)\notag\\
&+6\lambda_{hhh}g^2m_W^2(c_\alpha c_\beta+\sqrt{2}s_\alpha s_\beta)^2C_0(m_h,m_W,m_h)\notag\\
&+2\lambda_{HHh}g^2m_W^2(-s_\alpha c_\beta+\sqrt{2}c_\alpha s_\beta)^2C_0(m_H,m_W,m_H)\notag\\
&+2\lambda_{Hhh}g^2m_W^2(c_\alpha c_\beta+\sqrt{2}s_\alpha s_\beta)(-s_\alpha c_\beta+\sqrt{2}c_\alpha s_\beta)
[C_0(m_h,m_W,m_H)+C_0(m_H,m_W,m_h)]\notag\\
&+4\lambda_{H^{++}H^{--}h}g^2m_W^2s_\beta^2C_0(m_{H^{++}},m_W,m_{H^{++}})\notag\\
&+\lambda_{G^+G^-h}g_Z^2m_W^2(s_W^2+s_\beta^2)^2C_0(m_{G^+},m_Z,m_{G^+})+\lambda_{hG^+G^-}g^2s_W^2m_W^2C_0(m_{G^+},0,m_{G^+})\notag\\
&+\lambda_{H^+H^-h}g_Z^2m_W^2s_\beta^2c_\beta^2 C_0(m_{H^+},m_Z,m_{H^+})\notag\\
&+\lambda_{H^+G^-h}g_Z^2m_W^2s_\beta c_\beta(s_W^2+s_\beta^2)[C_0(m_{H^+},m_Z,m_{G^+})+C_0(m_{G^+},m_Z,m_{H^+})]\notag\\
&+2g^3m_Ws_\beta^2(c_\alpha c_\beta+\sqrt{2}s_\alpha s_\beta)[C_{24}(m_{G^+},m_{H^{++}},m_W)+C_{24}(m_W,m_{H^{++}},m_{G^+})]\notag\\
&+2g^3m_Ws_\beta c_\beta(-c_\alpha s_\beta+\sqrt{2}s_\alpha c_\beta)[C_{24}(m_{H^+},m_{H^{++}},m_W)+C_{24}(m_W,m_{H^{++}},m_{H^+})]\notag\\
&+\frac{1}{2}g^3m_W(c_\alpha c_\beta +\sqrt{2}s_\alpha s_\beta)^3[C_{24}(m_W,m_{h},m_{G^+})+C_{24}(m_{G^+},m_{h},m_W)]\notag\\
&+\frac{1}{2}g^3m_W(-s_\alpha c_\beta +\sqrt{2}c_\alpha s_\beta)^2(c_\alpha c_\beta +\sqrt{2}s_\alpha s_\beta)[C_{24}(m_W,m_{H},m_{G^+})+C_{24}(m_{G^+},m_{H},m_W)]\notag\\
&+\frac{1}{2}g^3m_W(c_\alpha c_\beta +\sqrt{2}s_\alpha s_\beta)(-c_\alpha s_\beta +\sqrt{2}s_\alpha c_\beta)^2[C_{24}(m_W,m_{h},m_{H^+})+C_{24}(m_{H^+},m_{h},m_W)]\notag\\
&+\frac{1}{2}g^3m_W(-s_\alpha c_\beta +\sqrt{2}c_\alpha s_\beta)(-c_\alpha s_\beta +\sqrt{2}s_\alpha c_\beta)
(s_\alpha s_\beta +\sqrt{2}c_\alpha c_\beta)[C_{24}(m_W,m_H,m_{H^+})+C_{24}(m_{H^+},m_H,m_W)]\notag\\
&+\frac{1}{2}gg_Z^2m_W(s_W^2+s_\beta^2)(c_\beta c_{\beta'} +\sqrt{2}s_\beta s_{\beta'})
(c_\alpha c_{\beta'} +2s_\alpha s_{\beta'})[C_{24}(m_{G^0},m_{G^+},m_Z)+C_{24}(m_{Z},m_{G^+},m_{G^0})]\notag\\
&+\frac{1}{2}gg_Z^2m_W(s_W^2+s_\beta^2)(-c_\beta s_{\beta'} +\sqrt{2}s_\beta c_{\beta'})
(-c_\alpha s_{\beta'} +2s_\alpha c_{\beta'})[C_{24}(m_A,m_{G^+},m_Z)+C_{24}(m_{Z},m_{G^+},m_A)]\notag\\
&+\frac{1}{2}gg_Z^2m_Ws_\beta c_\beta(-s_\beta c_{\beta'} +\sqrt{2}c_\beta s_{\beta'})
(c_\alpha c_{\beta'} +2s_\alpha s_{\beta'})[C_{24}(m_{G^0},m_{H^+},m_Z)+C_{24}(m_{Z},m_{H^+},m_{G^0})]\notag\\
&+\frac{1}{2}gg_Z^2m_Ws_\beta c_\beta(s_\beta s_{\beta'} +\sqrt{2}c_\beta c_{\beta'})
(-c_\alpha s_{\beta'} +2s_\alpha c_{\beta'})[C_{24}(m_A,m_{H^+},m_Z)+C_{24}(m_{Z},m_{H^+},m_A)]\notag
\end{align}
\begin{align}
&-g^3c_Wm_Z(c_\alpha c_{\beta'}+2s_\alpha s_{\beta'})C_{24}(m_{c_Z},m_{c^+},m_{c_Z})\notag\\
&-g^3c_W^2 m_W(c_\alpha c_{\beta}+\sqrt{2}s_\alpha s_{\beta})C_{24}(m_{c^+},m_{c_Z},m_{c^+})
-g^3s_W^2 m_W(c_\alpha c_{\beta}+\sqrt{2}s_\alpha s_{\beta})C_{24}(m_{c^+},m_{c_\gamma},m_{c^+})\notag\\
&-g^3m_W(c_\alpha c_\beta +\sqrt{2}s_\alpha s_\beta)(D-1)B_0(q^2,m_W,m_W)
-g^3c_Wm_Z(c_\alpha c_{\beta'} +2s_\alpha s_{\beta'})(D-1)B_0(q^2,m_Z,m_Z)\notag\\
&-\frac{4}{\sqrt{2}}g^3m_Ws_\alpha s_\beta[B_0(p_1^2,m_W,m_{H^{++}})+B_0(p_2^2,m_W,m_{H^{++}})]\notag\\
&-\frac{1}{4}g^3(3-c_{2\alpha})m_W(c_\alpha c_\beta  +\sqrt{2}s_\alpha s_\beta)
[B_0(p_1^2,m_W,m_{h})+B_0(p_2^2,m_W,m_h)]\notag\\
&-\frac{1}{4}g^3s_{2\alpha}m_W(-s_\alpha c_\beta +\sqrt{2}c_\alpha s_\beta)
[B_0(p_1^2,m_W,m_H)+B_0(p_2^2,m_W,m_H)]\notag\\
&-\frac{1}{2}gg_Z^2m_W[c_\alpha c_\beta s_W^2 -\sqrt{2}(c_W^2-2)s_\alpha s_\beta](s_W^2+s_\beta^2)
[B_0(p_2^2,m_Z,m_{G^+})+B_0(p_1^2,m_{Z},m_{G^+})]\notag\\
&+\frac{1}{2}gg_Z^2m_W[c_\alpha s_\beta s_W^2 +\sqrt{2}(c_W^2-2)s_\alpha c_\beta]s_\beta c_\beta
[B_0(p_2^2,m_Z,m_{H^+})+B_0(p_1^2,m_{Z},m_{H^+})]\notag\\
&-\frac{1}{2}ge^2m_W(c_\alpha c_\beta +\sqrt{2}s_\alpha s_\beta)
[B_0(p_2^2,0,m_{G^+})+B_0(p_1^2,0,m_{G^+})]\Bigg\},
\end{align}
where $C_{1223}(m_1,m_2,m_3)\equiv C_{12}(m_1,m_2,m_3)+C_{23}(m_1,m_2,m_3)$ and 
\begin{align}
&C_{VVV}^{hVV}(m_1,m_2,m_3)\equiv \notag\\
&\left[6(D-1)C_{24}+p_1^2(2C_{21}+3C_{11}+C_{0})+p_2^2(2C_{22}+C_{12})+p_1\cdot p_2(4C_{23}+3C_{12}+C_{11}-4C_0)\right]\notag\\
&(m_1,m_2,m_3),
\\
&C_{SVV}^{hVV}(m_1,m_2,m_3)\equiv\notag\\
&\left[(D-1)C_{24}+p_1^2(C_{21}-C_0)+p_2^2(C_{22}-2C_{12}+C_0)+2p_1\cdot p_2 (C_{23}-C_{11})\right](m_1,m_2,m_3),\\
&C_{VVS}^{hVV}(m_1,m_2,m_3)\equiv\notag\\
&\left[(D-1)C_{24}+p_1^2(C_{21}+4C_{11}+4C_0)+p_2^2(C_{22}+2C_{12})+2p_1\cdot p_2 (C_{23}+2C_{12}+C_{11}+2C_0)\right]\notag\\
&(m_1,m_2,m_3).
\end{align}

\end{appendix}


\end{document}